\documentclass{aastex}
\usepackage{graphicx}
\usepackage{emulateapj5}
\usepackage{times}
\usepackage{color}



\def\chandra    {{\em Chandra}\/}

\def\xmm        {{\em XMM-Newton}\/}
\def\rosat      {{\em ROSAT}\/}

\def\planck     {{\em Planck}\/}

\def\cluster    {{PLCK G036.7+14.9}\/}

\usepackage{mathptmx}

\begin{document}

\title{\textit{CHANDRA} and \textit{XMM-NEWTON} observations of the merging cluster of galaxies \newline \cluster }

\author{
B.\ Zhang$^{\!}$\altaffilmark{1,2,3},
L. P.\ David$^{\!}$\altaffilmark{3},
C.\ Jones$^{\!}$\altaffilmark{3},
F.\ Andrade-Santos$^{\!}$\altaffilmark{3},
E.\ O'Sullivan$^{\!}$\altaffilmark{3},
H.\ Dahle$^{\!}$\altaffilmark{4},
P.\ E.\ J.\ Nulsen$^{\!}$\altaffilmark{3},
T. E.\ Clarke$^{\!}$\altaffilmark{5},
E.\ Pointecouteau$^{\!}$\altaffilmark{6,7},
G. W.\ Pratt$^{\!}$\altaffilmark{8},
M.\ Arnaud$^{\!}$\altaffilmark{8},
J. M.\ Vrtilek$^{\!}$\altaffilmark{3},
L. \ Ji$^{\!}$\altaffilmark{9},
R. J.\ van Weeren$^{\!}$\altaffilmark{3},
R. P.\ Kraft$^{\!}$\altaffilmark{3},
and X.\ Kong$^{\!}$\altaffilmark{1,2}
}

\smallskip

\affil{\scriptsize 1) Center for Astrophysics, University of Science and Technology of China, Hefei, Anhui, 230026, China; zhb006@mail.ustc.edu.cn, xkong@ustc.edu.cn}
\affil{\scriptsize 2) Key Laboratory for Research in Galaxies and Cosmology, Chinese Academy of Sciences, Hefei, Anhui, 230026, China}
\affil{\scriptsize 3) Harvard-Smithsonian Center for Astrophysics, 60 Garden Street, Cambridge, MA 02138, USA}
\affil{\scriptsize 4) Institute of Theoretical Astrophysics, University of Oslo, P. O. Box 1029, Blindern, N-0315 Oslo, Norway}
\affil{\scriptsize 5) Naval Research Laboratory, Code 7213, Washington, DC 20375, USA}
\affil{\scriptsize 6) Universit\'{e} de Toulouse, UPS-OMP, IRAP, F-31028 Toulouse cedex 4, France}
\affil{\scriptsize 7) CNRS, IRAP, 9 Av. colonel Roche, BP 44346, F-31028 Toulouse cedex 4, France}
\affil{\scriptsize 8) Laboratoire AIM, IRFU/Service d'Astrophysique - CEA/DSM - CNRS - Universit\'{e} Paris Diderot, B\^{a}t. 709, CEA-Saclay, F-91191 Gif-sur-Yvette Cedex, France}
\affil{\scriptsize 9) Purple Mountain Observatory, Chinese Academy of Sciences, Nanjing, Jiangsu, 210008, China}
\shorttitle{X-ray observations of \cluster }
\shortauthors{Zhang et al.}

\begin{abstract}

We present \chandra\ and \xmm\ observations of \cluster\ from the \chandra -\planck\ Legacy Program. The high resolution X-ray observations reveal two close ($\sim$$72''=193$ kpc in projection) subclusters, G036N and G036S, which were not resolved by previous \rosat , optical, or recent \planck\ observations. We perform detailed imaging and spectral analyses and use a simplified model to study the kinematics of this system. The basic picture is that \cluster\ is undergoing a major merger (mass ratio close to unity) between the two massive subclusters, with the merger largely along the line-of-sight ($\sim$$80^\circ$ between the merger axis and the plane of the sky from the simplified model) and probably at an early stage (less than $\sim0.4-0.7$ Gyr since the merger began). G036N hosts a small ($\sim$27 kpc), moderate cool-core (cooling time $t_{\rm cool}\sim2.6-4.7$ Gyr), while G036S has at most a very weak cool-core ($t_{\rm cool}\sim5.7-10.3$ Gyr) in the central $\sim$$40$ kpc region. The difference in core cooling times is unlikely to be caused by the ongoing merger disrupting a pre-existing cool-core in G036S. G036N also hosts an unresolved radio source in the center, which may be heating the gas if the radio source is extended. The total mass of the whole cluster determined from \xmm\ is $\sim(5.9-8.0)\times10^{14}$ $M_{\rm \odot}$, and is $\sim(6.7-9.9)\times10^{14}$ $M_{\rm \odot}$ from \chandra . The \planck\ derived mass, $\sim(5.1-6.0)\times10^{14}$ $M_{\rm \odot}$, is higher than the X-ray measured mass of either subcluster, but is lower than the X-ray measured mass of the whole cluster, due to the fact that \planck\ does not resolve \cluster\ into subclusters and interprets it as a single cluster. This mass discrepancy could induce significant bias to the mass function if such previously unresolved systems are common in the \planck\ cluster sample. High resolution X-ray observations are necessary to identify the fraction of such systems and correct such a bias for the purpose of precision cosmological studies.

\end{abstract}

 \keywords{X-rays: galaxies: clusters ---galaxies: clusters: general --- galaxies: clusters: individual (\cluster ) --- cosmology: observations}

\section{INTRODUCTION}

In the concordance Lambda Cold Dark Matter ($\Lambda$CDM) cosmology, small scale systems form first and subsequently undergo a series of mergers and accretions to form larger objects such as groups and clusters of galaxies (e.g., Springel et al. 2005; Boylan-Kolchin et al. 2009; Kravtsov \& Borgani 2012). Clusters of galaxies are the largest objects whose inner parts have had time to virialize. Larger systems are still forming, with gravity the dominant force governing the behavior of the formation process. On smaller scales, the impact of non-gravitational processes such as radiative cooling, feedback from supernovae or Active Galactic Nuclei (AGNs) becomes significant (e.g., Ponman et al. 1999; Voit et al. 2005). Due to their large volumes and high masses, clusters of galaxies can be used in a number of ways to constrain cosmological parameters (see Allen et al. 2011, and references therein). In particular, the number density of clusters as a function of mass and redshift, i.e., the mass function, is strongly related to the mean matter density $\Omega _{\rm M}$ and the power spectrum amplitude $\sigma _8$ (e.g., Voit 2005; Vikhlinin et al. 2009; Mantz et al. 2010).

The primary link between theory and observation is the mass of the cluster. Observationally, the mass can be obtained by different techniques with different assumptions and limitations. Velocity dispersions of the member galaxies can be used to derive the mass under the assumption of dynamical equilibrium (e.g., Binney \& Tremaine 1987). However, this method suffers from strong projection effects and is observationally expensive. In X-rays, by measuring the gas density and temperature profiles, the mass can be derived assuming hydrostatic equilibrium (e.g., Sarazin 1988), which can underestimate the mass by $10\%-30\%$ due to non-thermal pressure support (e.g., Nagai et al. 2007; Piffaretti \& Valdarnini 2008; Jeltema et al. 2008; Battaglia et al. 2012). Strong and weak lensing offer, in principle, unbiased mass measurements, but strong lensing allows detailed mass distribution modeling primarily in the central regions (e.g., Meneghetti et al. 2013) while weak lensing usually has large intrinsic scatter due to its statistical nature and the results somewhat depend on the fitting procedure (e.g., Meneghetti et al. 2010; Becker \& Kravtsov 2011; Rasia et al. 2012).

The Sunyaev-Zeldovich (SZ) effect (Sunyaev \& Zeldovich 1972), a distortion of the Cosmic Microwave Background (CMB) spectrum produced by inverse Compton scattering of the CMB photons as they travel through the intracluster medium (ICM), has proven to be very efficient at finding massive clusters through three recent SZ surveys: the South Pole Telescope (SPT; e.g., Reichardt et al. 2013), the Atacama Cosmology Telescope (ACT; e.g., Hasselfield et al. 2013), and the \planck\ mission (e.g., Planck Collaboration XXIX 2014). \planck\ is an all-sky survey while SPT and ACT perform deeper observations of smaller solid angles with higher angular resolution. In 2011, \planck\ released an early catalogue of 189 cluster candidates with high reliability based on the first 9 months of data (Planck Collaboration VIII 2011). Recently, this early sample was increased to 1227 cluster candidates, based on 15.5 months of data (Planck Collaboration XXIX 2014). The \chandra -\planck\ Legacy Program\footnote{http://hea-www.cfa.harvard.edu/CHANDRA\_PLANCK\_CLUSTERS/.} is collecting \chandra\ observations for all the $z<0.35$ clusters (165 in total) in the \planck\ early catalogue. Each observation has at least 10,000 source counts to ensure sufficient accuracy in characterizing the clusters' masses, dynamical states, and scaling relations. The ultimate goal is to obtain the local cluster mass function, and compare it to that at higher redshifts (e.g., from SPT and ACT) to constrain cosmological parameters.

In addition, clusters of galaxies are of great interest in an astrophysical context. For dynamically relaxed clusters, the surface brightness usually exhibits a sharp cusp toward the center where the temperature drops below the surrounding region (e.g., Fabian 1994), a distinct feature of cool-core clusters. For these clusters, whose radiative cooling times in the central regions are usually short, some heating mechanisms, such as thermal conduction (e.g., Zakamska \& Narayan 2003; Voigt \& Fabian 2004; Sanderson et al. 2009), buoyantly raising bubbles (e.g., Churazov et al. 2001, 2002; Miraghaei et al. 2014), shocks (e.g., David et al. 2001; Nulsen et al. 2005), sound waves (e.g., Forman et al. 2005; Fabian et al. 2006), and cosmic ray leakage (e.g., Guo \& Oh 2008; Mathews \& Brighenti 2008), are believed to operate in order to offset the cooling, although the details are poorly known (see McNamara \& Nulsen 2007, 2012 for reviews).  Cluster outskirts, which present opportunities to study a range of physical process (see Reiprich et al. 2013 for a review), e.g., deviation from hydrostatic/thermal/ionization equilibrium, gas clumping, and accretion of the warm-hot intergalactic medium (WHIM), are largely unexplored because of their low signal-to-noise ratio. For merging systems, shocks and cold fronts are generated, producing contact discontinuities in the surface brightness and temperature profiles (e.g., Markevitch et al. 2000; Vikhlinin et al. 2001; Markevitch \& Vikhlinin 2007). Turbulence is also expected, which, together with shocks, might be related to the amplification of magnetic fields and acceleration of relativistic particles, as inferred from observations of radio halos and radio relics (see Feretti et al. 2012, and references therein). Cluster merging could also be responsible for the disruption of cool-cores (e.g., Sanderson et al. 2006; Burns et al. 2008). Merging clusters have also provided direct evidence for dark matter (e.g., Clowe et al. 2006) and constraints on the self-interacting dark matter cross-section (e.g., Markevitch et al. 2004; Randall et al. 2008).

In this paper, we present \chandra\ and \xmm\ observations of \cluster\ (CIZA J1804.4+1002) from the \chandra -\planck\ Legacy Program. Our X-ray observations demonstrate that \cluster\ is undergoing a major merger of two close (in projection) yet separated subclusters. \cluster\ was also observed by \rosat\ and in the optical band (Ebeling et al. 2002), but was not resolved into subclusters by either observation. Due to the size-flux degeneracy (Planck Collaboration VIII 2011), \planck\ used the \rosat\ determined position and size from the Meta-Catalogue of X-ray Detected Clusters (MCXC; Piffaretti et al. 2011), so \planck\ actually measures the total flux from the whole cluster instead of the flux from each individual subcluster.

This paper is structured as follows: The observations and data reduction are presented in Section 2. Section 3 provides the imaging analysis, i.e., the morphology and surface brightness profile, while detailed spectral analysis, including the treatments of background and other uncertainties, gas mass, total mass, and core properties can be found in Section 4. We discuss our main results in Section 5, with a summary presented in Section 6. An Appendix is also supplied for the model 2D/3D temperature, gas mass, total mass, and gas mass fraction profiles. We adopt the Hubble constant $H_{0}=70$ km s$^{-1}$ Mpc$^{-1}$, the matter density parameter $\Omega$$_{\rm M}=0.3$, and the dark energy density parameter $\Omega_{\rm \Lambda}=0.7$. At the cluster redshift of 0.1547, the luminosity distance is 737.7 Mpc and 1$''=2.682$ kpc. Uncertainties quoted are $1\sigma$ throughout this paper.

\section{OBSERVATIONS AND DATA REDUCTION}

\subsection{Chandra Data}

\cluster\ was observed by \chandra\ (ObsID 15098) on 2014 February 5 for 9.6 ks with the Advanced CCD Imaging Spectrometer (ACIS-I). The observation was telemetered in VFAINT mode. Standard \chandra\ data analysis was performed with CIAO version 4.6.1 and calibration database version 4.6.1. The CIAO tool \textit{chandra\_repro} was applied to perform initial processing and to obtain a new event file. Point sources were detected by running \textit{wavdetect} and were excluded in further analysis (except for the AGN in the center of the north subcluster, see Sections 3 and 4.4). Light curves in the soft- and hard-band were examined and no background flares were found, so we proceeded with the full exposure. 

The imaging analysis was performed in the $0.7-7.0$ keV band to maximize the signal-to-noise ratio. The period F blank-sky background with proper scaling according to the exposure time and high energy flux was used. The exposure map was weighted by the best fitting thermal model to the spectrum of the central part of the cluster. 

In the spectral analysis, the blank-sky background was used as the baseline background model. We also varied the blank-sky background by $\pm$5\% and added a soft-band adjustment to test how sensitive the results are to the background (Sections 4.4 and 4.6). All spectral analysis was performed in the $0.7-9.0$ keV band, unless otherwise stated.

\subsection{XMM-Newton Data}

\cluster\ was observed by \xmm\ (ObsID 0692931901) on 2013 March 30 for 10.5 ks (EPIC MOS1), 10.6 ks (EPIC MOS2), and 9.7 ks (EPIC PN). Full frame mode was used for the three cameras with medium filters for the EPIC MOS detectors and thin filters for the EPIC PN. Data reduction was done with ESAS\footnote{http://heasarc.gsfc.nasa.gov/docs/xmm/xmmhp\_xmmesas.html.} version 5.6 (Snowden \& Kuntz 2013), which uses SAS version 13.5.0 and calibration files updated to 2014 February 27. The initial data processing was done by running \textit{emchain} and \textit{epchain} with the default setting in ESAS. Out-of-time (OOT) events were also accounted for in the EPIC PN analysis. Background flares were identified and rejected by applying a $1.5\sigma$ clipping method to the high energy count rate histogram. The resulting ``cleaned" exposure times are 6.8 ks (EPIC MOS1), 5.4 ks (EPIC MOS2), and 2.0 ks (EPIC PN), respectively. This shows that the \xmm\ observation was affected by high energy particle induced background. Due to the limited statistics, we combined data from the three cameras in our analysis. Point sources were detected in the $0.5-10.0$ keV band and were excluded from further analysis (except for the AGN in the center of the north subcluster, see Sections 3 and 4.4).

Imaging analysis was performed in the $0.7-2.0$ keV band. The count images, quiescent particle background (QPB) images, residual soft proton (SP) background images (see Section 4.1.2), and exposure maps for the three cameras were generated and combined with the ESAS task \textit{comb}. The QPB and SP images were then subtracted from the combined count image and the resulting image was exposure corrected with the combined exposure map. 

Spectra from the same regions of the three cameras were extracted with ESAS tasks \textit{mos-spectra} and \textit{pn-spectra} and grouped to contain a minimum of 25 counts per bin. We modeled the various components of the background (see Section 4.1.2). Spectra from different regions and from the three cameras were fit simultaneously. The spectral analysis was carried out in the $0.7-10.0$ keV band for EPIC MOS spectra and in the $0.7-7.0$ keV band for EPIC PN spectra.

\section{IMAGING ANALYSIS}

\subsection{Morphology}

The \xmm\ image (Figure 1, left panel) shows that the large scale X-ray emission from \cluster\ is elongated in a northeast-southwest direction. The central region reveals two close ($\sim$72$''=193$ kpc in projection), yet clearly separated  subclusters (G036N in the north and G036S in the south), which suggests that \cluster\ may be undergoing a merger. The \chandra\ image (Figure 1, right panel) reveals a bow shaped gap (with an angle of $\sim$$145^\circ$) between G036N and G036S, anther indication of interaction between the two subclusters. Spectral analysis reveals a hotter region between G036N and G036S, confirming that they are interacting (see Section 4.7). Beyond the interaction region, neither G036N nor G036S deviates significantly from spherical symmetry, suggesting that the merger is probably at an early stage.

Optical images of \cluster\ were obtained using the MOSaic CAmera (MOSCA) at the 2.56 m Nordic Optical Telescope (NOT). Using MOSCA in $2 \times 2$ binned mode yielded a pixel scale of $0\farcs 217 \ {\rm pixel}^{-1}$, and the field-of-view (FOV) is $7\farcm 7 \times 7\farcm 7$. Three individual exposures were made in each of the Sloan Digital Sky Survey (SDSS) $g$-, $r$-, and $i$-band, adding up to total exposure times of $1800$~s, $600$~s, and $600$~s, respectively. The images in the $r$- and $i$-band were smoothed with Gaussian kernels to match the seeing (full width at half maximum, ${\rm FWHM}=1\farcs 45$) of the combined $g$-band image. Object detection and photometry were performed on the combined image for each filter using the SExtractor software (Bertin \& Arnouts 1996) in dual image mode with the $g$-band chosen for the reference image. Total magnitudes were measured by the MAG$\_$AUTO parameter while colors were measured within a $3\farcs 6$ diameter aperture. Photometric zeropoints were calibrated from SDSS photometry of stars in other fields observed using the same setup, and the derived magnitudes were corrected for Galactic extinction using the Schlafly \& Finkbeiner (2011) recalibration of the Schlegel et al.\ (1998) dust maps. The $g-r$ vs.\ $r-i$ color-color diagram of non-stellar objects in the field of \cluster\ is shown in the left panel of Figure 2. A significant overdensity, encompassed by the shaded circle, corresponding to the location of \cluster , is clearly visible. In order to select the red sequence of \cluster , we fit a linear relation to the $g-i$ vs.\ $i$ color-magnitude diagram (right panel of Figure 2) for galaxies within the circle and obtained a best fitting relation represented by the dashed line in the right panel of Figure 2. The color scatter (1$\sigma$) about this relation, illustrated by the dotted lines in the diagram, is 0.072 mag, which is typical of the measured intrinsic scatter around the mean color-magnitude relation for early-type galaxies in rich clusters (e.g., Stanford et al. 1998). Galaxies which fall within twice the value of the scatter from the best fitting color-magnitude relation, down to $i_{AB}\sim21$ mag (at fainter magnitudes, the photometric errors become non-negligible compared to the intrinsic scatter around the color-magnitude relation), are considered to be the red sequence of \cluster\ and are plotted in red in the diagram. The colors of the red sequence are in good agreement with those in Eisenstein et al. (2001) at redshifts close to \cluster . The Brightest Cluster Galaxies (BCGs) of G036N and G036S are located on the red sequence and colored green in particular in both panels of Figure 2. As shown in Figure 3 (left panel), the BCGs of the two subclusters, marked with blue squares ($5''\times5''$), are very close to the X-ray centers (defined as the X-ray centroids), suggesting that the merger is at an early stage and it should be largely along the line-of-sight.

An X-ray AGN was detected in the $2.0-7.0$ keV band at the center of G036N in the \chandra\ image (Figure 1, right panel), while no point source was detected at the center of G036S. Interestingly, the NRAO VLA Sky Survey (NVSS) image (right panel of Figure 3) also reveals a radio source, NVSS J180431+100323, coincident with the center of G036N. This radio source should be hosted by the BCG of G036N. The extent of the radio source is comparable to the angular resolution of the NVSS (${\rm FWHM}=45''$), so it is essentially unresolved. The properties of the AGN and the core regions are discussed in more detail in Section 4.8.

\subsection{Surface Brightness Profile}

Surface brightness profiles for each subcluster were extracted from the background subtracted\footnote{For \chandra , the normalized blank-sky background was subtracted; for \xmm , the QPB and SP background images were subtracted (see Section 2).} and exposure corrected images. We chose a half circular region opposite to the interaction region (see Figure 4) to derive the surface brightness profiles for both G036N and G036S. The azimuthally averaged surface brightness profiles are shown in Figure 5. We first tried to model these with single $\beta$ models (Cavaliere \& Fusco-Femiano 1976), 
\begin{equation}
S(r)=S_0[1+(\frac{r}{r_c})^2]^{-3\beta+1/2}, 
\end{equation}
where $r_c$ is the core radius. We found excess emission in the central $\sim$$10''$ regions for both G036N and G036S, so we excluded the radial bins covering the central $\sim$$10''$ regions and refit the surface brightness profiles. The best fitting $\beta$ models are plotted in Figure 5, with the best fitting parameters given in Table 1. 

As can be seen, $\beta$ models describe the surface brightness profiles quite well beyond the central $\sim$$10''$ for both G036N and G036S, although the reduced $\chi^2$ for \chandra\ is larger. The best fitting core radii are consistent for \chandra\ and \xmm\ while the differences between $\beta$ values are only $1.9\sigma$ for G036N and $1.6\sigma$ for G036S. In the central $\sim$10$''$ regions, both G036N and G036S show excess emission relative to the $\beta$ models. Note that the X-ray AGN at the center of G036N is not excluded in the surface brightness profiles. Compared to the surface brightness profile of G036S, it can be inferred that the contribution from the X-ray AGN to the total flux from G036N is negligible. We further demonstrate this from the spectral analysis in Section 4.4. Given the quality of the data and the goodness of the $\beta$ model beyond the central $\sim$10$''$ regions, we did not try more complicated models. 

We also extracted surface brightness profiles in different wedge-shaped regions, trying to identify surface brightness discontinuities, but the current data do not reveal any unambiguous edges.

\section{SPECTRAL ANALYSIS}

\subsection{Background Treatment}

\subsubsection{Chandra}

The background level is quiescent throughout the \chandra\ observation, so we simply used the period F blank-sky background as our baseline background model. In addition, to check how sensitive the results are to the background variation (with time and/or location), we varied the background level by $\pm$5\% (by changing the BACKSCAL keyword by $\mp$5\%), and added a soft-band adjustment by fitting the spectrum from a (largely) source-free region near the edge of the CCD (Figure 4, right panel) with an unabsorbed thermal model (best fitting $kT=0.23$ keV) allowing the normalization to be negative (e.g., Vikhlinin et al. 2005). We treated the uncertainty in the background subtraction as a systematic uncertainty (see Section 4.6).

\subsubsection{XMM-Newton}

As the \xmm\ data were affected by periods of high energy particle induced background (Section 2.2), there may be residual SP contamination even after flare screening. Thus we decided to model the various components of the background as in Snowden \& Kuntz (2013). The \xmm\ background is very complex (e.g., Nevalainen et al. 2005; Carter \& Read 2007; Leccardi \& Molendi 2008), but can be separated into three components: instrumental background, cosmic background, and solar wind charge exchange (SWCX) background.

The instrumental background contains QPB (quiescent particle background), instrumental lines, and residual SP (soft proton) contamination. The QPB component can be easily subtracted using the filter wheel closed (FWC) data. The instrumental lines are dominated by the EPIC MOS Al K$\alpha$ ($\sim$1.49 keV), EPIC MOS Si K$\alpha$ ($\sim$1.75 keV), EPIC PN Al K$\alpha$ ($\sim$1.49 keV), and (six) EPIC PN Cu ($\sim$8 keV) lines. We set an upper limit of 7.0 keV for EPIC PN spectra, leaving only three instrumental lines, which were modeled as three Gaussians with zero width. The residual SP contamination was modeled as a power law not folded through the instrumental effective areas, with the indices linked together for the EPIC MOS cameras.

The cosmic background was modeled as three components: an unabsorbed thermal component with $kT=0.1$ keV, representing the emission from the Local Hot Bubble (LHB) or heliosphere; an absorbed thermal component, allowing the temperature to vary, representing the emission from the hotter Galactic halo and/or the intergalactic medium; and an absorbed power-law with the index fixed at 1.46 representing the background from unresolved cosmological sources. Both the absorbed and unabsorbed thermal components have zero redshift and solar abundance.

The SWCX background only produces line emission, with the strongest lines being O {\scshape VII} ($\sim$0.57 keV) and O {\scshape VIII} ($\sim$0.65 keV) lines. We set a lower energy cutoff of 0.7 keV, so that the SWCX background can be neglected.

The normalizations of the various components of the background  were determined by simultaneously fitting the spectra from the global cluster regions (see Figure 4). We define seven regions for the global spectral fitting: core (small circle, 35$''$ in radius), cluster (ellipse), cluster minus core, for both G036N and G036S, and a large annulus (480$''$-720$''$ in radii) free of cluster emission. The normalizations of the cosmic background components, together with the normalization of the source spectrum (a single absorbed thermal model), were computed in units of arcmin$^{-2}$. All the normalizations of the background components were left as free parameters in the global spectral fitting, and were fixed in the subsequent spectral analysis (e.g., temperature profile), accounting for the solid angle. The temperature of the absorbed cosmic background component and the indices of the residual SP contamination were also fixed at the best fitting values in the subsequent spectral analysis.

\subsection{Hydrogen Column Density}

Since \cluster\ is located at low galactic latitude ($b=14.9^\circ$), we expect the total Hydrogen column density, $N_H$, to be higher than the Leiden/Argentine/Bonn (LAB) H {\scshape I} survey (Kalberla et al. 2005) value of $N_{\rm HI}=0.087\times10^{22}$ cm$^{-2}$, due to additional absorption by molecular Hydrogen. Indeed, the best fitting $N_H$ obtained by the global spectral fitting is $0.092^{+0.004}_{-0.006}\times10^{22}$ cm$^{-2}$ for \xmm , $0.164^{+0.012}_{-0.013}\times10^{22}$ cm$^{-2}$ for \chandra\ (baseline background model and baseline background model varied by $\pm$5\%), and $0.195^{+0.014}_{-0.007}\times10^{22}$ cm$^{-2}$ if a soft-band adjustment is added to the \chandra\ baseline background model. 

\xmm\ and \chandra\ give significantly different $N_H$, by almost a factor of two. To investigate the origin of this discrepancy, we refit the spectra with energy cutoffs of 0.3 keV, 0.5 keV, 0.7 keV and 1.5 keV. The best fitting $N_H$ values are all consistent for each observatory, but still significantly different between the two observatories. We also tried different abundance tables, and found that while $N_H$ changes with different abundance tables, the ratio between the two observatories remains almost unchanged. Therefore, we conclude that the discrepancy is most likely caused by cross-calibration issues (see Section 4.6 on their effects on the total mass determinations).

To examine whether $N_H$ is different between G036N and G036S, and whether $N_H$ changes with radius, we fit all spectra (Section 4.4) of G036N and G036S with $N_H$ free in each annulus. We found that all the $N_H$ values are consistent within the $1\sigma$ errors for \chandra , and consistent within the $2\sigma$ errors for \xmm . Therefore, in our spectral analysis we linked $N_H$ in different regions together. 

In addition, we estimated the total $N_H$ with the relation\footnote {http://www.swift.ac.uk/analysis/nhtot/index.php.} 
\begin{equation}
N_{\rm H,tot}=N_{\rm HI}+2N_{\rm H_2}=N_{\rm HI}+2N_{\rm H_2, max}(1-e^\frac{N_{\rm HI}E(B-V)}{N_c})^\alpha, 
\end{equation}
where $N_{\rm H_2, max}=7.3\times10^{20}$ cm$^{-2}$, $E(B-V)=0.162$ is the dust extinction, $N_c=3.0\times10^{20}$ cm$^{-2}$, and $\alpha=1.1$ (Willingale et al. 2013). We obtained $N_{\rm H,tot}=0.136\times10^{22}$ cm$^{-2}$. All the results obtained with $N_H$ fixed at this total value are also included for comparison.

To summarize, we used $N_H=0.092\times10^{22}$ cm$^{-2}$ for \xmm , $N_H=0.164\times10^{22}$ cm$^{-2}$ for the \chandra\ baseline background model and the baseline background model varied by $\pm$5\%, and $N_H=0.195\times10^{22}$ cm$^{-2}$ for the \chandra\ baseline background model with a soft-band adjustment. We also included $N_H=0.136\times10^{22}$ cm$^{-2}$ for both observatories for comparison. See Table 2 for the $N_H$ values used for the two observatories and for the different \chandra\ background models. We treated the uncertainty in $N_H$ as a systematic uncertainty (see Section 4.6).

\subsection{Redshift}

The only redshift available for \cluster\ is from the Cluster in the Zone of Avoidance (CIZA) project, which measured a spectroscopic redshift $z=0.1525$ (Ebeling et al. 2002). However, \cluster\ was treated as a single cluster in Ebeling et al. (2002), so we decided to use our best fitting X-ray redshifts for G036N and G036S. 

Table 2 lists the redshifts obtained by \xmm\ and \chandra\ with different \chandra\ background models and different $N_H$ values. As can be seen, a different $N_H$ for the same observatory and the same background model (\chandra ) only has a minor effect on the derived $z$. The obtained best fitting redshifts differ by $\lesssim$2\% and are consistent. However, \xmm\ and \chandra\ give values that differ by $\sim$13\% ($\sim$$3\sigma$) and $\sim$7\% ($\sim$$2\sigma$) for G036N and G036S, respectively, which are probably caused by cross-calibration issues between the two observatories.

We simply averaged the best fitting redshifts over \xmm\ and \chandra\ (baseline background model), and those obtained with different $N_H$ values, to obtain $z_n=0.1501\pm0.0022$ and $z_s=0.1592\pm0.0020$, where (and hereafter) we use the subscripts $n$ and $s$ to distinguish between the north and south subclusters. We assume that the difference, which has a significance of $3\sigma$, between the redshifts of the two subclusters is caused by their relative motion, and the redshift of the cluster is the mean of the two, $0.1547\pm0.0015$, which is very close to the optical redshift of 0.1525. All the distance-relevant quantities are relative to a redshift of 0.1547. In our spectral analysis, the redshift is fixed at the value shown in Table 2 according to the context. The uncertainty in the redshift was treated as a systematic uncertainty (see Section 4.6).

\subsection{Temperature Profile and Deprojection}

Azimuthally averaged temperature profiles were derived in half circular regions opposite to the interaction region (see Figure 4) for G036N and G036S. Spectra were extracted in the 0-35$''$ (0-35$''$), 35$''$-70$''$ (35$''$-70$''$), 70$''$-130$''$ (70$''$-140$''$), 130$''$-260$''$ (140$''$-280$''$), 260$''$-600$''$ (280$''$-600$''$) regions for G036N (G036S) for \xmm , and in the 0-35$''$\footnote{We did not exclude the AGN at the center of G036N, since the bolometric X-ray luminosity ($0.01-100$ keV in practice) of the AGN is only $\sim$1\% compared to that of the first bin, so the presence of the AGN does not affect the temperature of the first bin for G036N.} (0-45$''$), 35$''$-80$''$ (45$''$-100$''$), 80$''$-200$''$ (100$''$-300$''$), 200$''$-450$''$ regions for G036N (G036S) for \chandra . All spectra for each observatory were fit simultaneously. Due to the limited statistics, the abundances in each annulus cannot be constrained, so we fixed them at the global best fitting values.

The temperature profiles for each subcluster and for each set of \{$N_H$, $z$\} values in Table 2, are displayed in Figure 6. In each panel of Figure 6, the temperature profiles for \xmm\ and for \chandra\ with different background models (see Section 4.1.1) are indicated by different symbols. When using the same $N_H$ value, \xmm\ and \chandra\ generally give consistent results, although at large radii ($>200''$), a direct comparison is not straightforward as they do not probe exactly the same region. For G036N, there is an indication of a temperature drop in the central region, although the uncertainties are large. At large radii, the temperature profile declines, dropping to $\sim$1/2 ($\sim$1/4) of the ``peak''  value for \chandra\ (\xmm ). This behavior is qualitatively consistent with large samples of cluster temperature profiles (e.g., Vikhlinin et al. 2005; Pratt et al. 2007; Leccardi \& Molendi 2008). For G036S, \xmm\ exhibits a temperature drop in the innermost bin, and a declining temperature profile at large radii, while \chandra , in the much smaller radial range it covers because the cluster is close to the chip boundary, gives an essentially isothermal temperature profile, but with large uncertainties. Different $N_H$ values and different \chandra\ background models produce the same trend as previously stated for the temperature profiles for both G036N and G036S.

The observationally determined temperature profile is the projected 2D profile, which contains emission from different parts of the cluster along the line-of-sight. We used the functional form, 
\begin{equation}
T_{\rm 3D}(r)=T_0\frac{(\frac{r}{r_{\rm cool}})^{a_{\rm cool}}+\frac{T_{\rm min}}{T_0}}{1+(\frac{r}{r_{\rm cool}})^{a_{\rm cool}}} \frac{(\frac{r}{r_t})^{-a}}{[1+(\frac{r}{r_t})^b]^{c/b}}, 
\end{equation}
proposed by Vikhlinin et al. (2006), to model the 3D temperature profile. Assuming spherical symmetry, this 3D temperature profile is then projected along the line-of-sight, weighted by $n_e^2/T^{3/4}$ (Mazzotta et al. 2004), where $n_e$ is the electron density and $T$ is the 3D temperature, and is fit to the observed 2D temperature profile. As this functional form has many free parameters, it can describe the temperature profile very well. The results are shown in the Appendix.

\subsection{Gas Mass, Total Mass, and Gas Mass Fraction}

For a surface brightness profile described by a $\beta$ model, the electron density is given by 
\begin{equation}
n_e(r)=n_{e,0}[1+(\frac{r}{r_c})^2]^{-3\beta/2}. 
\end{equation}
The central density\footnote{Note that there is excess emission relative to the $\beta$ model (see Figure 5) in the central $\sim$10$''$ region, so the ``central density", $n_{e,0}$, is not equal to the physical central density. We use ``central density" just in terms of the functional form. However, it is still accurate to use $n_{e,0}$ to determine the gas mass at large radius, as the central $\sim$10$''=27$ kpc region only occupies a tiny volume.}, $n_{e,0}$, was determined by fitting the spectrum extracted in a 200$''$ circular region, utilizing the relation between the normalization $K$ of the thermal model and electron number density $n_e$, 
\begin{equation}
K=\frac{10^{-14}}{4\pi[D_A(1+z)]^2}\int{n_en_H}dV, 
\end{equation}
where $D_A$ is the angular diameter distance, $n_H$ is the Hydrogen number density, and the integration is over a cylindrical volume. The gas mass within radius $r$ is then 
\begin{equation}
M_g(r)=\int_0^r{4\pi\mu_en_{e,0}m_p[1+(\frac{r}{r_c})^2]^{-3\beta/2}r^2}dr, 
\end{equation}
where $\mu_e=1.172$ is the mean molecular weight of electrons, and $m_p$ is the mass of a proton.

The total mass within radius $r$ can be derived, under the assumption of hydrostatic equilibrium, from (e.g., Sarazin 1988)

\begin{equation}
M(r)=-\frac{kTr}{G\mu m_p}(\frac{d\ {\rm ln}\ n_e}{d\ {\rm ln}\ r}+\frac{d\ {\rm ln}\ T}{d\ {\rm ln}\ r}), 
\end{equation}
where $k$ is the Boltzmann constant, $G$ is the gravitational constant, $\mu=0.614$ is the mean molecular weight of the total particles, and $T$ is the 3D temperature at radius $r$. Since \cluster\ is undergoing a merger (Sections 3.1 and 4.7), we avoided the interaction region and measured the surface brightness and temperature profiles in the outer halves of both subclusters (see Figure 4). The surface brightness profiles from these regions are well described by $\beta$ models (Figure 5 and Table 1), and the morphologies do not deviate significantly from spherical symmetry (Figure 1), so hydrostatic equilibrium in these regions should be a reasonable approximation.

Uncertainties were estimated from Monte Carlo simulations. The errors of the measured 2D temperature profile, the parameters of the surface brightness profile, and the normalization $K$ of the spectral fit to the 200$''$ region, are assumed to obey Gaussian distributions. For each Monte Carlo simulation, a set of randomly drawn 2D temperature profile, $\beta$ model parameters\footnote{It is worth noting that, strictly speaking, $r_c$ and $\beta$ are not independent (e.g., Henriksen \& Tittley 2002), so the errors of the gas mass obtained in this way are overestimated.}, and normalization $K$ from their respective Gaussian distributions, was used to obtain the gas mass, total mass, and gas mass fraction profiles, using the deprojection method in Section 4.4. This process was repeated 1000 times. In practice, too steep temperature profiles often lead to unphysical mass determinations. We only accept physical solutions with $\rho_{\rm tot}>\rho_{\rm gas}$, where $\rho_{\rm tot}$ and $\rho_{\rm gas}$ are the total and gas densities respectively, and which are convectively stable, $d\ {\rm ln}\ T/d\ {\rm ln}\ \rho_{\rm gas}<2/3$. Finally, we quote the mean of all accepted solutions as the ``best fitting" gas mass, total mass, and gas mass fraction, with the uncertainties as the $1\sigma$ standard deviation of all the accepted solutions.

In particular, we are interested in the gas mass, total mass, and gas mass fraction inside three characteristic radii, $r_\Delta$ ($\Delta=2500, 500, 200$), where $r_\Delta$ is the radius within which the mean matter density of the Universe is $\Delta$ times the critical density at the cluster redshift. The results are listed in Table 3. Since G036S is located near the chip boundary in the \chandra\ FOV (Figure 1), and the \chandra\ temperature profile does not cover the declining region at large radii (Figure 6), we did not measure the mass-related quantities for G036S with \chandra .

\subsection{Systematic Uncertainties}

From previous sections, we see that, due to cross-calibration issues, \xmm\ and \chandra\ give results inconsistent within their $1\sigma$ errors for some parameters which affect the mass determinations. We treat the uncertainties in $N_H$, $z$, other cross-calibration uncertainties between \xmm\ and \chandra , and the uncertainties in \chandra\ background subtraction, as systematic uncertainties. For each set of \{$N_H$, $z$\} in Table 2, new temperature profiles for G036N and G036S were derived. The characteristic radii, gas mass, total mass, and gas mass fraction were then calculated following the method given in Section 4.5. We also tried different background subtraction methods for the \chandra\ data (see Section 4.1.1). The results are presented in Table 3, from which we can draw the following conclusions:

1. Comparing $N_H$ fixed at the total value to the best fitting values, all the quantities differ by less than $1.6\sigma$. The differences between the ``best fitting" values of $r_{\Delta}$, $M_{g,\Delta}$, $M_{\Delta}$, and $f_{\Delta}$ for \xmm\ are $\lesssim8$\%, $\lesssim7$\%, $\lesssim25$\%, and $\lesssim16$\%, while those for \chandra\ are $\lesssim10$\%, $\lesssim8$\%, $\lesssim31$\%, and $\lesssim27$\%, respectively. The larger differences for \chandra\ are caused by the \chandra\ baseline background model with a soft-band adjustment, without which the differences for all four quantities for \chandra\ are reduced by a factor of at least 2.

2. Comparing \xmm\ to \chandra , all quantities differ by no more than $2.5\sigma$. Varying the \chandra\ baseline background by +5\% (back*0.95) gives the closest results to \xmm , with $\lesssim9$\%, $\lesssim6$\%, $\lesssim28$\%, and $\lesssim20$\% differences between the ``best fitting" values of $r_{\Delta}$, $M_{g,\Delta}$, $M_{\Delta}$, and $f_{\Delta}$, while adding a soft-band adjustment to the \chandra\ baseline background gives the largest differences to \xmm , with $\lesssim34$\%, $\lesssim27$\%, $\lesssim154$\%, and $\lesssim46$\% differences for the above four quantities, respectively. 

3. Comparing \chandra\ different background models, all the quantities are consistent within their $1\sigma$ errors. The differences (with respect to the baseline background model) of $r_{\Delta}$, $M_{g,\Delta}$, $M_{\Delta}$, and $f_{\Delta}$ when varying the baseline background model by $\pm$5\% are $\lesssim6$\%, $\lesssim5$\%, $\lesssim15$\%, and $\lesssim10$\%, while adding a soft-band adjustment to the \chandra\ baseline background gives much larger differences, $\lesssim19$\%, $\lesssim19$\%, $\lesssim75$\%, and $\lesssim28$\%, respectively. 

By comparing the results in the above mentioned cases, we can see that adding a soft-band adjustment to the \chandra\ baseline background always produces the largest differences of the ``best fitting" values of $r_{\Delta}$, $M_{g,\Delta}$, $M_{\Delta}$, and $f_{\Delta}$. Actually, adding a soft-band adjustment also increases the best fitting $N_H$ for the global cluster (Table 2), which further affects the temperature measurements. This ``soft-band-adjustment-$N_H$-$kT$" dependence makes the mass determination much more uncertain, which gives rise to the largest differences of the ``best fitting" values for the mass-related quantities in Table 3. Without accurate $N_H$ determination, we consider this model less constraining and do not include it in our further analysis and final results.

To summarize, uncertainties in $N_H$, $z$, other \xmm\ and \chandra\ cross-calibration uncertainties, and \chandra\ background subtraction, do not lead to any significantly different ($\lesssim2.5\sigma$) mass determinations. \xmm\ gives a mass $M_{200}\sim(3.1-4.2)\times10^{14}$ $M_{\rm \odot}$ for G036N and $M_{200}\sim(3.0-4.7)\times10^{14}$ $M_{\rm \odot}$ for G036S, while \chandra\ gives a somewhat higher mass, $M_{200}\sim(3.4-6.7)\times10^{14}$ $M_{\rm \odot}$, for G036N.

\subsection{Is There Shocked Gas between the Subclusters?}

The X-ray morphology of \cluster\ suggests that it is undergoing a merger between G036N and G036S (Section 3.1). To confirm this spectroscopically, we extracted spectra from a 130$''$$\times$35$''$ rectangular region (Region 1, see Figure 7) between G036N and G036S, and from the symmetric regions with respect to their centers (Region 2 for G036N and Region 3 for G036S), and fit them to single thermal models. We found that the temperature in Region 1 is indeed higher (by a factor of $\sim$2 based on the \chandra\ data) than those in Regions 2 and 3, which could be interpreted as shock heating during the merger. 

Using the Rankine-Hugoniot jump conditions, the Mach number, $\mathcal{M}$, is related to the temperature jump as (e.g., Sarazin 2002) 
\begin{equation}
\frac{1}{C}=\frac{2}{\mathcal{M}^2(\gamma +1)}+\frac{\gamma -1}{\gamma +1},
\end{equation}
and
\begin{equation}
\frac{1}{C}=[\frac{1}{4}(\frac{\gamma +1}{\gamma -1})^2(\frac{T_2}{T_1}-1)^2+\frac{T_2}{T_1}]^{1/2}-\frac{1}{2}\frac{\gamma +1}{\gamma -1}(\frac{T_2}{T_1}-1),
\end{equation}
where the subscripts 1 and 2 denote the preshock and postshock gas, $C\equiv \rho_2 / \rho_1$ is the shock compression, and $\gamma =5/3$ is the adiabatic index. If we simply assume that the temperature of the preshock gas is given by the gas temperature in Region 2 (Region 3) and that of the postshock gas is given by the gas temperature in Region 1, using the above relations, we can estimate the Mach number of the shock propagating in G036N (G036S) both for \xmm\ and \chandra\ (see Table 4). Note that due to projection effects, the inferred Mach number is only a lower limit.

\xmm\ suggests a weak shock ($\mathcal{M}\sim1.0-1.6$), while \chandra\ indicates a stronger shock ($\mathcal{M}\sim1.5-2.7$). The difference is caused by, when comparing the same $N_H$, the lower temperature in the interaction region measured from \xmm\ ($7.65-9.19$) keV compared to \chandra\ ($10.52-16.86$ keV). Actually, from the HIghest X-ray FLUx Galaxy Cluster Sample (HIFLUGCS; Reiprich \& B$\ddot{\rm o}$hringer 2002), Schellenberger et al. (2014) found systematic lower temperatures given by \xmm\ compared to \chandra\ for high temperature systems, especially for $>$$8$ keV clusters. When comparing the Mach number obtained with the best fitting $N_H$ values, the large difference in Mach number between \xmm\ and \chandra\ is then caused by higher temperatures in the preshock regions (Regions 2 and 3) given by \xmm . This is because the best fitting \xmm\ $N_H$ is almost a factor of 2 less than that of \chandra\ (Section 4.2), and a smaller $N_H$ increases the temperature, leading to smaller temperature difference in the postshock region, and larger temperature difference in the preshock regions. Thus, cross-calibration issues again manifest themselves in terms of the Mach number discrepancy.

\subsection{Subcluster Cores}

From Section 3.1, we know that G036N has an X-ray AGN in the center. We extracted a spectrum in a circular region with a radius of $3''$, and used the surrounding region ($4''-8''$ annulus) as the background. By fixing the index of the power-law model at 1.7, we acquired a bolometric X-ray luminosity ($0.01-100$ keV in practice) of $(1.45-1.50)\times10^{42}$ erg s$^{-1}$.

The surface brightness profiles (Figure 5) and the temperature profiles (Figure 6) suggest that the gas may be cooling in the central regions of both G036N and G036S. The isobaric cooling time was calculated as
\begin{equation}
t_{\rm cool}=\frac{H}{L_{\rm bol}}=\frac{\frac{5}{2}\frac{M_{\rm gas}}{\mu m_p}kT}{L_{\rm bol}},
\end{equation}
where $H$ is the enthalpy, $L_{\rm bol}$ is the bolometric X-ray luminosity, and $M_{\rm gas}$ is the gas mass. We then compared this to the look back time to $z=1$, which is $t_L=7.7$ Gyr, a representative time for a cluster to relax and develop a cool-core (e.g., B\^{\i}rzan et al. 2004; Hudson et al. 2010). If $t_{\rm cool}<t_L$, the gas may be cooling. Due to the much larger PSF of \xmm , we only used the \chandra\ data to study the cores. From Figure 4, varying the blank-sky background by $\pm$5\% has little effect on the temperature in the central bright part of the subclusters, so we only used the baseline background model to study the cores.

For G036N, the cooling time (single thermal model) in the central $3''-10''$ (full annulus) region is $t_{\rm cool}=2.64-4.71$ Gyr, while that in the outer $10''-35''$ (half annulus to avoid the interaction region) region is $6.82-12.15$ Gyr, so only the gas in the central $\sim$$10''$ ($27$ kpc) region may be cooling. The abundance obtained with a single thermal model is $0.85-2.44$ $Z_{\odot}$, significantly larger than the global abundance  ($0.09-0.47$ $Z_{\odot}$) of G036N, suggesting enrichment from the BCG. A two thermal components model, with the higher temperature component fixed at the cluster temperature and abundance, does not improve the fit and results in essentially zero normalization for the higher temperature component. Adding a cooling flow model to a single thermal model does not improve the fit either, and the mass deposition rate given by the normalization of the cooling flow model is highly uncertain with $\dot{M}_{\rm spec}=2.3-27.6$ $M_{\odot}$ yr$^{-1}$.

For G036S, the cooling time (single thermal model) in the central $0''-15''$ (full annulus) region is $t_{\rm cool}=5.65-10.27$ Gyr, already comparable to or greater than $t_L$, so cooling should be very weak in this region. Adding a cooling flow model, we obtained a $1\sigma$ upper limit on the mass deposition rate of 14.3 $M_{\odot}$ yr$^{-1}$.

\section{DISCUSSION}

\subsection{The Merging Activity}

Previous \rosat\ and optical studies (Ebeling et al. 2002) did not resolve the morphological details of \cluster . With our high resolution \chandra\ and \xmm\ observations, two close (in projection) subclusters, G036N and G036S, were clearly resolved, and spatially resolved spectroscopy could be performed. Spectral analysis produces very similar redshifts for G036N and G036S (Table 2), and reveals a higher temperature in the region between them compared to the symmetric regions away from their centers which could be due to shock heating during the merger (Section 4.7).

The morphologies of G036N and G036S excluding the interaction region do not deviate significantly from spherical symmetry (see Figure 1) and their surface brightness profiles are well described by single $\beta$ models beyond the central $\sim$10$''$ regions (see Figure 5), suggesting that the merger is probably at an early stage so that perturbations generated by the merger have not yet disturbed the outer regions. Another piece of evidence for an early stage of the merger comes from the good spatial correspondence between the X-ray emission centroids of the two subclusters and their BCGs (see left panel of Figure 3). This also favors a merger close to the line-of-sight.

Projection effects hamper our knowledge of the geometry of the merger. The only observables are the projected distance, $d_p=193$ kpc, and the radial velocity difference, $v_r=2356$ km s$^{-1}$, between the two subclusters. To gain some insight into the merger kinematics, we applied the simplified model in Ricker \& Sarazin (2001) (see also Sarazin 2002; Sauvageot et al. 2005) to G036N and G036S. This model treats the two subclusters as point masses, and assumes that they initially expand away from each other in the Hubble flow. Due to their gravitational attraction, they collapse after reaching the greatest separation $d_0$, and finally merge. Let $t_{\rm coal}$ be the age of the universe when core coalesce occurs. The parameters $d_0$ and $t_{\rm coal}$ are related by Kepler's Third Law as
\begin{equation}
d_0\approx [2G(M_n+M_s)]^{1/3}(\frac{t_{\rm coal}}{\pi})^{2/3},
\end{equation}
where $M$ is the total mass (assumed to be $M_{200}$, see Table 3).  Let $d$ be the 3D distance between the two subclusters. The merger velocity $v$ is then 
\begin{equation}
v^2\approx 2G(M_n+M_s)(\frac{1}{d}-\frac{1}{d_0})[1-(\frac{b}{d_0})^2]^{-1}, 
\end{equation}
where $b$ is the impact parameter as defined in Sarazin (2002), and is assumed to be zero in our case since it is usually very small compared to $d_0$ (Sarazin 2002). The parameters $d$ and $v$ are related to the observables by
\begin{equation}
d=\frac{d_p}{\cos \theta},
\end{equation}
\begin{equation}
v=\frac{v_r}{\sin \theta},
\end{equation}
where $\theta$ is the angle between the merger axis and the plane of the sky. On the other hand, since the two subclusters have not attained core coalesce, $t_{\rm coal}$ can be written as 
\begin{equation}
t_{\rm coal}\approx t_{\rm age}+\frac{d}{v}, 
\end{equation}
where $t_{\rm age}=11.52$ Gyr is the age of the Universe at the cluster redshift in our adopted cosmology. To obtain the time since the merger began, $t_{\rm merge}$, i.e., the time from when the two subclusters started physical contact to the present day, we have to treat the two subclusters as extended sources and $t_{\rm merge}$ can be approximated by
\begin{equation}
t_{\rm merge}\approx \frac{r_n + r_s - d}{v}, 
\end{equation}
where $r$ is the virial radius (assumed to be $r_{200}$, see Table 3). Solving Equations $(11)-(16)$, we obtained $\theta \approx80^\circ$, $t_{\rm merge}\approx0.8$ Gyr (see Table 4). Although overly simplified, this model produces consistent results with our expectations, i.e., an early stage of the merger, mostly along the line-of-sight, which can explain the main observational features outlined in the above two paragraphs. Targeted simulations are required to study in detail the dynamics of this system, which is beyond the scope of this paper.

To summarize, \cluster\ is undergoing a major merger by two nearly equal mass ($\sim1:1$) subclusters. The merger should be at an early stage, and should be happening mostly along the line-of-sight. A simplified model suggests that the merger probably began $\sim$0.8 Gyr ago, with an angle between the merger axis and the plane of the sky of $\sim$$80^\circ$.

\subsection{X-ray Derived Mass vs. SZ Derived Mass}

In Sections 4.5 and 4.6, we derived the total masses of G036N and G036S under the assumption of hydrostatic equilibrium of the ICM using the regions not affected by the ongoing merger. \chandra\ does not cover a large enough region to measure the total mass of G036S. If we assume that \chandra\ would give the same mass ratio of G036N to G036S as \xmm , and the relative errors are the same for G036N and G036S for \chandra , we can obtain the total mass of G036S that would be ``measured'' by \chandra . Adding the total X-ray masses of G036N and G036S together, we obtained the total X-ray derived mass (measured inside the X-ray derived $r_{500}$, $r_{X,500}$, see Table 3) of \cluster , which is $M_{X,500}=(5.91-8.00)\times10^{14}$ $M_{\rm \odot}$ for \xmm\ and $M_{X,500}=(6.66-9.85)\times10^{14}$ $M_{\rm \odot}$ for \chandra \footnote{Note that the values quoted are simply the lowest and highest masses for different $N_H$ and different background models (\chandra ) in Table 3, so the uncertainties are the most conservative estimates. This is important when comparing to the SZ derived mass. Also note that the large uncertainties in the X-ray mass are dominated by systematic uncertainties (see Table 3).}.

The integrated SZ signal, $Y_{SZ}$, is proportional to the total electron pressure of the cluster, 
\begin{equation}
D_A^2Y_{SZ}=\frac{\sigma _{\rm T}}{m_ec^2}\int{p_e}dV,
\end{equation}
where $\sigma _{\rm T}$ is the Thomson cross-section, $m_e$ is the mass of an electron, $c$ is the speed of light, and $p_e$ is the electron pressure. Due to the size-flux degeneracy (Planck Collaboration VIII 2011), additional information is needed to refine the blind \planck\ SZ flux measurement. Using the X-ray measured cluster position and size can break the degeneracy, and if the redshift of the cluster is known, it may be used to further break the degeneracy (Planck Collaboration XXIX 2014). After obtaining the refined SZ flux, the total SZ mass, $M_{SZ}$, can be estimated using the Malmquist-bias corrected scaling relation between $Y_{SZ}$ and $M_{SZ}$ (Planck Collaboration XX 2014). With the above method, Planck Collaboration XXIX (2014) published the SZ flux, $Y_{SZ,500}$, and SZ mass, $M_{SZ,500}$, measured inside the SZ $r_{500}$, $r_{SZ,500}$, for 813 confirmed clusters with measured redshifts out of the 1227 all-sky SZ sources using the first 15.5 months of data. For \cluster , $Y_{SZ,500}=0.00234^{+0.00033}_{-0.00032}$ arcmin$^2$ (adopting a redshift of 0.1525) and $M_{SZ,500}=5.54^{+0.42}_{-0.43}$ $(5.11-5.96)$ $\times10^{14}$ $M_{\rm \odot}$.

It is clear that $M_{SZ,500}$ is greater than the X-ray mass of either G036N or G036S (see Table 3), which is fully expected since \planck\ did not resolve the two subclusters so the derived SZ mass is the mass of the whole cluster. However, compared to the X-ray mass of the whole cluster, $M_{X,500}$, especially the \chandra\ ``measurement" (note that \chandra\ actually does not directly measure the total mass of the whole cluster), $M_{SZ,500}$ is smaller, although there is a small overlapping range between $M_{SZ,500}$ and the \xmm\ derived mass because of the large uncertainties in the X-ray mass measurements (however, see Footnote 8). 

In the derivation of the SZ mass, Planck Collaboration XXIX (2014) used the best fitting $Y_{SZ,500}-M_{SZ,500}$ relation of Planck Collaboration XX (2014). Even if we accounted for the uncertainties in that relation, and chose the parameters that maximize or minimize $M_{SZ,500}$, we obtained the most conservative SZ mass estimate, $M_{SZ,500}=(4.99-6.06)$ $\times10^{14}$ $M_{\rm \odot}$. This only slightly increases the overlapping range between the SZ and \xmm\ mass measurements, while the lower SZ mass compared to X-ray mass determinations remains unexplained.

The difference between the X-ray and SZ masses should not be entirely caused by the different extraction regions for the SZ and X-ray mass measurements. In Figure 8, we show $r_{SZ,500}$ used by \planck\ (Planck Collaboration XXIX 2014; Piffaretti et al. 2011; Ebeling et al. 2002) and $r_{X,500}$ used by \xmm\ and \chandra .  The SZ extraction region is larger (by $\sim$$14\%$) than the \xmm\ extraction regions, but the \xmm\ measured mass is higher. The \chandra\ extraction regions are close to ($\sim$$3\%$ smaller than) the SZ extraction region, but the \chandra\ measured mass is even higher. Therefore, while different extraction region is a factor affecting the mass determinations from X-ray and SZ measurements, some other factors may be more important to explain the difference.

As we show in Sections 5.1 and 5.3, the merger is most likely at an early stage. The SZ signal $Y_{SZ}$ relates to the mass $M$ via a power-law, which can be written as $Y_{SZ}=AM^\eta$, where $A$ and $\eta$ can be considered as constants for our purpose. For an early stage of the merger, the total mass of the whole cluster is $M_n + M_s$. However, if the two subclusters are not resolved, as in the case of the \planck\ observation, the mass of the whole cluster determined in this way will be $(M_n^\eta + M_s^\eta )^{1/\eta}$. From Table 3, $M_n\approx M_s$; $\eta=1.79$ as used by Planck Collaboration XXIX (2014). This gives a total mass a factor of $\sim$0.74 smaller than $M_n + M_s$, which is about the difference between the SZ derived mass and the X-ray derived mass. This demonstrates that the mass discrepancy between the SZ and X-ray measurements is caused by the fact that \planck\ does not resolve the two subclusters and interprets the whole system as a single cluster. 

Based on simulations, Kravtsov et al. (2006) propose that the X-ray analogue of $Y_{SZ}$, $Y_{X}=M_{\rm gas}\times T_{X}$,  where $T_X$ is the emission weighted temperature within a certain region, is a low scatter mass proxy and less sensitive to the cluster dynamical state. We computed the total $Y_{X,500}$ by adding those of G036N and G036S, where $T_{X}$ is measured in the $[0.15-0.75]r_{500}$ region and $M_{\rm gas}$ is measured within $r_{500}$ (see Table 5 for details). We list $Y_{X,500}$ in Table 5. $Y_{X,500}$ and $M_{X,500}$ follow the scaling relations presented in Arnaud et al. (2010).

Neglecting gas clumping, $Y_{X}$ relates to $Y_{SZ}$ by (e.g., Arnaud et al. 2010; Rozo et al. 2012; Planck Collaboration XXIX 2014)
\begin{equation}
\frac{D_A^2 Y_{SZ}}{Y_X}=\frac{\sigma _{\rm T}}{m_e c^2}\frac{1}{\mu_e m_p}\frac{<n_eT>}{<n_e>T_X}=C\frac{<n_eT>}{<n_e>T_X}. 
\end{equation}
Therefore, the ratio $D_A^2Y_{SZ}/CY_X$ should be close to 1, although the exact value depends on the structure of the cluster. Hereafter, we will refer $Y_{SZ}/Y_X$ to $D_A^2Y_{SZ}/CY_X$ for simplicity. Previous studies show that $Y_{SZ,500}/Y_{X,500}$ is between 0.8 and 1 (e.g., Arnaud et al. 2010; Rozo et al. 2012; Planck Collaboration XXIX 2014). Our results are consistent with these studies, although with large uncertainties (see Table 5).

In summary, the SZ mass given by \planck\ is higher than the X-ray derived mass by \xmm\ and \chandra\ for either subcluster, but is lower than the X-ray mass of the whole cluster, due to the fact that \planck\ does not resolve the two subclusters and interprets the whole system as a single cluster. The $Y_{SZ}/Y_X$ ratio is consistent with previous studies, although with large errors.

\subsection{Cool-Core, Merger, and Feedback}

The surface brightness profiles show excess emission in the central $\sim$$10''$ regions relative to single $\beta$ models for both G036N and G036S (see Figure5), and the temperature profiles show indications of temperature drops toward their centers (see Figure 6). The cooling time of the central $3''-10''$ region for G036N is $2.64-4.71$ Gyr, and beyond this region, the cooling time is longer than $7.7$ Gyr. The cooling time of the $0-15''$ region for G036S is $5.65-10.27$ Gyr, comparable to or longer than $7.7$ Gyr (see Section 4.8 for details). Therefore, G036N has a small ($\sim$$10''=27$ kpc), moderate cool-core, while G036S has at most a very weak cool-core within the central $\sim$$15''=40$ kpc region. The spectroscopically derived mass deposition rate (see Section 4.8) confirms this scenario.

It is interesting to ask what causes the difference between the cores of the two subclusters. One possibility is that the ongoing merger has disrupted a pre-existing cool-core in G036S (e.g., Sanderson et al. 2006; Burns et al. 2008; Rossetti \& Molendi 2010; but see, e.g., Poole et al. 2008). From Table 4, the merger began $\sim0.78-0.80$ Gyr ago based on the simplified merger model (Section 5.1), while the time for the shock to propagate to the center of G036N (G036S), $t_{r,n}$ ($t_{r,s}$), $t_r=r_{200}/\mathcal{M}c_{\rm sound}$, is $0.65-1.12$ ($0.67-1.05$) Gyr obtained from \xmm , and is $0.43-0.74$ ($0.41-0.67$) Gyr from \chandra , respectively. Taken at face value, \chandra\ favors a scenario in which the shock has passed the cores of the two subclusters, while the situation for \xmm\ is less clear. 

To further examine this question, we divided the central $3''-10''$ annular region of G036N into two equal area half annular regions, with the symmetry axis perpendicular to the line connecting the centers of the two subclusters. We compared the derived temperatures in the two half annular and full annular regions, finding them to be consistent within their $1\sigma$ errors. We then varied the outer radius, i.e., $3''-8''$, $3''-12''$, $3''-15''$, and $3''-20''$ (maximum outer radius avoiding the interaction region, see Figure 7), performed the same analysis, and found that all the temperatures are consistent within their $1\sigma$ errors (although in some cases the temperature in the half annular region near the interaction region was poorly constrained). Two possibilities can explain this: (i) the shock has not reached the central $\sim$$10''$ region of G036N; (ii) the shock has passed the central $\sim$$10''$ region of G036N, but the core is dense enough not to be destroyed. We now show that possibility (ii) is less likely. We first applied the same method to G036S, but the data quality is poorer and all the temperatures are consistent given the large uncertainties. Notice that $M_{200}$, $r_{200}$, $\mathcal{M}$, $c_{\rm sound}$, and $t_r$ for G036N and G036S are all similar (Tables 3 and 4), so we expect that the shock travels similar distances in G036N and G036S at any given time. As a result, if the shock has reached the core of G036N, it should also have reached the core of G036S. The pressure ratio, $p_n/p_s=0.61-1.16$, where $p_n$ and $p_s$ are the mean pressures in the central $3''-10''$ region of G036N and in the central $0-15''$ region of G036S respectively, although with large uncertainty, is consistent with 1. This implies that, if possibility (ii) is correct, the core of G036S should also have survived the merger induced shock, in contradiction with observations. Thus, the consistent temperatures in the two half annular regions and in their parent regions in the core of G036N imply that the shock probably has not reached the core, so the lack of gas cooling in the core of G036S is unlikely caused by the merger disrupting a pre-existing cool-core in G036S. This is also consistent with our previous expectation of an early stage of the merger, suggesting that the simplified model in Section 5.1 overestimates the time since the merger began. The actual time since the merger began is probably less than $\sim0.4-0.7$ Gyr (see Table 4).

For cool-core clusters, some heating mechanisms must be taking place to prevent the gas from catactrophic cooling (see Fabian 1994 and references therein). Interestingly, the BCG of G036N hosts a radio source (Section 3.1), with a flux density of 19.7 mJy and a 1.4 GHz luminosity ($L_{1.4\ \rm GHz}$) of $1.32\times10^{24}$ W Hz$^{-1}$ after the $K$-correction (Singal et al. 2011) assuming a radio spectral index of $0.8$\footnote{The spectral index $\alpha$ is defined as $S_\nu \propto \nu^{-\alpha}$, where $S_\nu$ is the flux density at frequency $\nu$.} (Condon et al. 1998). The extent of the radio source is comparable to the angular resolution of the NVSS (${\rm FWHM}=45''$, $\sim$120 kpc at the cluster redshift, see right panel of Figure 3). If it is an extended source, i.e., with radio lobes filled with relativistic particles and magnetic fields (e.g., B\^{\i}rzan et al. 2008), there may be undetected cavities, which are generally thought to play a key role in the ``radio mode'' feedback (e.g., Bower et al. 2006; Croton et al. 2006; Sijacki et al. 2007) in clusters of galaxies (see McNamara \& Nulsen 2007, 2012; Fabian 2012; and references therein). Using the scaling relation in Cavagnolo et al. (2010), we estimated the cavity power, $P_{\rm cav}\sim(0.8-1.4)\times10^{43}$ erg s$^{-1}$, based on the 1.4 GHz luminosity. The X-ray luminosity of the cool-core is $(3.3-4.9)\times10^{43}$ erg s$^{-1}$, $2-6$ times larger than $P_{\rm cav}$. Studies of large samples show that although in some cases the cavity power is smaller than the X-ray luminosity, on average it is energetic enough to balance the cooling (e.g., Rafferty et al. 2006; B\^{\i}rzan et al. 2008; Dunn \& Fabian 2008; Cavagnolo et al. 2010). Higher resolution radio data are required to confirm the extended nature of the radio source, and deeper \chandra\ observations are also necessary to identify any cavities in the core of G036N. Sun (2009) studied a sample of 161 BCGs and 74 strong radio AGNs ($L_{1.4\ \rm GHz}>10^{24}$ W Hz$^{-1}$) in 152 nearby ($z<0.11$) clusters and groups using \chandra\ archive data. He found that all BCGs with $L_{1.4\ \rm GHz}>2\times10^{23}$ W Hz$^{-1}$ in the sample have cool-cores. Although with a higher redshift, G036N is also consistent with his finding.

To conclude, G036N hosts a small, moderate cool-core, while G036S has at most a very weak cool-core. This difference is unlikely to be caused by the ongoing merger. G036N also hosts a central radio source, which may be heating the gas if the radio source is extended. Examination of the temperature variations in the core of G036N also suggests that the simplified merger dynamical model presented in Section 5.1 overestimates the time since the merger began, which should be less than $\sim0.4-0.7$ Gyr.

\section{SUMMARY}

We present \xmm\ and \chandra\ observations of \cluster\ from the \chandra -\planck\ Legacy Program. The X-ray images reveal two close (in projection), yet clearly separated subclusters, G036N and G036S (Figure 1); spectral analysis yields similar redshifts for the two subclusters (Section 4.3) and a higher temperature between them (Section 4.7), confirming that they are interacting. Excluding the interaction region, the morphologies of G036N and G036S do not deviate significantly from spherical symmetry (Figure 1) and the surface brightness profiles can be well described by single $\beta$ models beyond the central $\sim$$10''$ regions (Figure 5 and Table 1), suggesting that the merger is still at an early stage. The BCGs of the two subclusters are very close to the X-ray centers (Figure 3), consistent with an early stage of the merger, and in favor of a merger mostly along the line-of-sight. A simplified dynamical model suggests that the merger began about 0.8 Gyr ago, with an angle between the merger axis and the plane of the sky of $\sim80^\circ$ (Section 5.1), in accordance with our expectations.

Surface brightness profiles (Figure 5) and temperature profiles (Figure 6) indicate that the gas may be cooling in the cores of both subclusters. By calculating the cooling time, we conclude that G036N hosts a small, moderate cool-core, while G036S has at most a very weak cool-core. Based on the temperature variations in the core of G036N, we suggest that the difference in core cooling times between G036N and G036S is unlikely to be caused by the ongoing merger (Section 5.3). The temperature variations also indicate that the simplified model (Section 5.1) overestimates the time since the merger began, which is probably less than $\sim0.4-0.7$ Gyr. Interestingly, the BCG of G036N hosts an unresolved radio source (Figure 3). If the radio source is extended, the ``radio mode" AGN feedback may be taking place in the core of G036N. Using the 1.4 GHz luminosity, we estimate a cavity power about $2-6$ times smaller than the X-ray luminosity of the core of G036N. Higher resolution radio data and deeper X-ray data are necessary to study the nature of the radio source and the physical processes in the core, e.g., jet/ICM interaction.

Since the merger is probably at an early stage, we apply the hydrostatic equation to the ICM in the regions opposite to the interaction region to obtain the total mass (Section 4.5). We treat the uncertainties in Hydrogen column density (Section 4.2), redshift (Section 4.3), \chandra\ background subtraction methods (Section 4.1.2), and other \xmm\ and \chandra\ cross-calibration uncertainties as systematic uncertainties (Section 4.6). We find that \chandra\ gives a slightly higher total mass than \xmm , although the difference is not significant (Section 4.6). We also compare the X-ray mass to the SZ mass (Section 5.2). The SZ mass is higher than the X-ray mass of either subcluster, but is lower than the X-ray mass of the whole cluster, which is caused by the fact that \planck\ does not resolve \cluster\ into two subclusters and interprets it as a single cluster. 

Despite the short exposure times of the \chandra\ and \xmm\ observations and large systematic uncertainties, we were able to infer the basic merger kinematics and some useful time scales based on the morphologies and core properties, from which we further estimate the total masses of the constituent subclusters and the whole cluster. They are all self-consistent. Obviously, deeper X-ray data and multiwavelength observations as well as targeted simulations are required to study this interesting system in greater detail. 

As the angular resolution of \planck\ is substantially coarser than those of \chandra\ and \xmm , it cannot resolve closely interacting subclusters or close cluster pairs. Because of the size-flux degeneracy, \planck\ has to use the X-ray determined position and radius to refine the SZ flux measurement (e.g., Planck Collaboration XXIX 2014). Much of this information comes from MCXC, which collects cluster parameters from publicly available \rosat\ All-Sky Survey based and serendipitous based catalogues (Piffaretti et al. 2011). These catalogues usually consist of observations too shallow to resolve clusters in much detail. Consequently, clusters with small projected separations might not be resolved and are treated as a single cluster, which is then adopted by MCXC and \planck . If such close clusters occupy a considerable fraction, the consequence is that the \planck\ all-sky survey will overestimate the number of massive clusters. Using this biased mass function to constrain cosmological parameter will result in lower $\Omega _{\rm M}$ and higher $\sigma _8$. Thus, higher resolution X-ray observations by \chandra\ and \xmm\ of the \planck\ cluster sample are necessary to identify the fraction of previously unresolved clusters like \cluster\ and correct for this bias before applying the \planck\ SZ mass function to constrain cosmological parameters.

\acknowledgments

We thank Nabila Aghanim, Iacopo Bartalucci, James G. Bartlett, Hans B$\ddot{\rm o}$hringer, Stefano Borgani, Shea Brown, Gayoung Chon, Eugene Churazov, Jessica Democles, Daniel Eisenstein, Chiara Ferrari, William Forman, Simona Giacintucci, Charles Lawrence, Marceau Limousin, Giulia Macario, Douspis Marian, Pasquale Mazzotta, Jean-Baptiste Melin, Stephen S. Murray, Elena Pierpaoli, Rashid A. Sunyaev, Alexey Vikhlinin for reading an early version of the manuscript and Harald Ebeling for communications on \cluster . We thank the referee for useful comments and suggestions. 
The scientific results reported in this article are based on observations made by the \chandra\ X-ray Observatory (ObsID 15098), which is operated by the Smithsonian Astrophysical Observatory (SAO) for and on behalf of the National Aeronautics Space Administration (NASA) under contract NAS8-03060, and \xmm\ (ObsID 0692931901), an ESA science mission with instruments and contributions directly funded by ESA Member States and NASA. We also present results based on observations made with the Nordic Optical Telescope, operated by the Nordic Optical Telescope Scientific Association at the Observatorio del Roque de los Muchachos, La Palma, Spain, of the Instituto de Astrofisica de Canarias.
B.Z. acknowledges the SAO predoctoral fellowship program and financial support by China Scholarship Council. L.P.D is supported in part by NASA grant GO2-13146X. C.J. acknowledges support from the \chandra\ X-ray Center.  L.J. is supported by the 100 Talents program of Chinese Academy of Sciences (CAS) and the Key Laboratory of Dark Matter and Space Astronomy of CAS. X.K. is supported by the National Natural Science Foundation of China (NSFC, Nos.11225315, 1320101002, 11433005, and 11421303), the Strategic Priority Research Program "The Emergence of Cosmological Structures" of the Chinese Academy of Sciences (No. XDB09000000), the Specialized Research Fund for the Doctoral Program of Higher Education (SRFDP, No. 20123402110037), and the Chinese National 973 Fundamental Science Programs (973 program) (2015CB857004).

{\it Facilities:} \facility{CXO (ACIS)}, \facility{XMM (EPIC)}, \facility{NOT (MOSCA)}.

\clearpage

%%%%%%%%%%%%%%%%%%%%%%%%%%%%%% 
\begin{table}
\begin{center}
{\caption{Best Fitting $\beta$ Model Parameters of the Surface Brightness Profiles}
\vspace{0.5cm}
\begin{tabular}{cccccc} \hline \hline 
 Region & Facility & $\beta$ & $r_c$ (arcsec)   & $r_c$ (kpc) & $\chi^2/\rm {d.o.f}$ \\ \hline 
 G036N & \xmm\        & $0.6418\pm{0.0060}$ & $14.57\pm{1.36}$ & $39.08\pm{3.65}$ & $118.8/94$  \\
       & \chandra\  & $0.6635\pm{0.0095}$ & $14.65\pm{1.22}$ & $39.29\pm{3.27}$ & $128.6/54$  \\

&&&&& \\

 G036S & \xmm\        & $0.6850\pm{0.0078}$ & $23.65\pm{1.34}$ & $63.43\pm{3.59}$ & $121.3/94$   \\
       & \chandra\  & $0.7104\pm{0.0134}$ & $22.43\pm{1.66}$ & $60.16\pm{4.45}$ & $123.4/44$  \\
 \hline \hline
\end{tabular}
}
\end{center}
\end{table}
%%%%%%%%%%%%%%%%%%%%%%%%%%%%%

%%%%%%%%%%%%%%%%%%%%%%%%%%%%%% 
\begin{table}
\begin{center}
{\caption{Best Fitting Hydrogen Column Density and Redshift}
\vspace{0.5cm}
\begin{tabular}{cccc} \hline \hline 
 Method & Parameter$^a$ & G036N & G036S  \\ \hline 
 \xmm\                              & $N_H$$^b$   & 0.136 ($0.092^{+0.004}_{-0.006}$) & ...   \\
                                  & $z$     & $0.1415^{+0.0025}_{-0.0054}$ ($0.1406^{+0.0039}_{-0.0029}$) &   $0.1535^{+0.0058}_{-0.0051}$ ($0.1554^{+0.0043}_{-0.0059}$) \\
&&& \\
 \chandra\ baseline +             & $N_H$$^b$   & 0.136 ($0.164^{+0.012}_{-0.013}$) & ...     \\
  back*0.95 + back*1.05$^c$      & $z$     &  $0.1590^{+0.0057}_{-0.0034}$ ($0.1594^{+0.0050}_{-0.0066}$) &   $0.1639^{+0.0018}_{-0.0029}$ ($0.1640^{+0.0035}_{-0.0015}$)  \\
&&& \\
\chandra\ with soft$^d$               & $N_H$$^b$   & 0.136 ($0.195^{+0.014}_{-0.007}$) & ...     \\
                                  & $z$     & $0.1602^{+0.0052}_{-0.0054}$ ($0.1569^{+0.0042}_{-0.0050}$) &   $0.1620^{+0.0048}_{-0.0047}$ ($0.1638^{+0.0027}_{-0.0049}$)   \\

 \hline \hline
\end{tabular}
}
\tablenotetext{a}{Obtained by global spectral fitting with the regions shown in Figure 4.}
\tablenotetext{b}{Hydrogen column density, in units of $10^{22}$ cm$^{-2}$. The value without errors was obtained from the relation given in Willingale et al. (2013), while the values with errors are the best fitting values from our analysis. $N_H$ for G036S is always linked to that for G036N (see Section 4.2).}
\tablenotetext{c}{\chandra\ baseline background model and varying \chandra\ baseline background model by $\pm$5\%. Varying the background level by $\pm$5\% only affects regions with faint source emission. Since the regions used for the global spectral fitting cover the bright part of the cluster, we only used the baseline background to obtain the best fitting $N_H$ and $z$. In deriving the temperature profiles, we used these $N_H$ and $z$ for the three \chandra\ background models.}
\tablenotetext{d}{\chandra\ baseline background model with a soft-band adjustment.}
\end{center}
\end{table}
%%%%%%%%%%%%%%%%%%%%%%%%%%%%%

%%%%%%%%%%%%%%%%%%%%%%%%%%%%%
\begin{table*}
\small
\begin{center}
\caption{Radius, Gas mass, Total mass, and Gas Mass Fraction}
\vspace{0.5cm}
\begin{tabular}{c c c c c c }
\hline  \hline                                                            
Region & Method  &   Parameter$^a$     &        $\Delta=2500$              &       $\Delta=500$                 &         $\Delta=200$                \\
\hline

G036N  &                    & $r_\Delta$       & $0.52\pm0.01 (0.55\pm0.02)$  & $0.99\pm0.03 (1.03\pm0.03)$   & $1.38\pm0.04 (1.43\pm0.04)$      \\   
&                           & $M_{g,\Delta}$   & $1.42\pm0.28 (1.51\pm0.32)$  & $3.00\pm0.62 (3.12\pm0.70)$   & $4.33\pm0.91 (4.49\pm1.02)$      \\
&  \xmm                       & $M_{\Delta}$     & $2.25\pm0.18 (2.69\pm0.29)$  & $3.20\pm0.28 (3.60\pm0.29)$   & $3.46\pm0.32 (3.85\pm0.32)$      \\
&                           & $f_{\Delta}$     & $6.32\pm1.41 (5.63\pm1.38)$  & $9.42\pm2.21 (8.69\pm2.16)$ & $12.57\pm3.02 (11.71\pm2.97)$    \\
&  &&&& \\

       &                    & $r_\Delta$       & $0.55\pm0.03 (0.53\pm0.03)$  & $1.10\pm0.06 (1.07\pm0.05)$   & $1.56\pm0.08 (1.51\pm0.07)$      \\   
&  \chandra\                & $M_{g,\Delta}$   & $1.51\pm0.32 (1.48\pm0.31)$  & $3.22\pm0.69 (3.19\pm0.68)$   & $4.65\pm0.99 (4.60\pm0.99)$      \\
&  baseline$^{b}$                 & $M_{\Delta}$     & $2.77\pm0.46 (2.52\pm0.38)$  & $4.44\pm0.67 (4.08\pm0.59)$   & $5.02\pm0.73 (4.61\pm0.65)$      \\
&                           & $f_{\Delta}$     & $5.49\pm1.29 (5.92\pm1.36)$  & $7.29\pm1.78 (7.83\pm1.89)$   & $9.31\pm2.32 (10.02\pm2.49)$    \\
&  &&&& \\

       &                    & $r_\Delta$       & $0.54\pm0.04 (0.54\pm0.03)$  & $1.20\pm0.10 (1.15\pm0.06)$   & $1.85\pm0.23 (1.66\pm0.10)$      \\   
&  \chandra\                & $M_{g,\Delta}$   & $1.45\pm0.33 (1.52\pm0.35)$  & $3.47\pm0.77 (3.46\pm0.80)$   & $5.51\pm1.33 (5.09\pm1.21)$      \\
&  with soft$^c$                & $M_{\Delta}$     & $2.59\pm0.65 (2.63\pm0.41)$  & $5.75\pm1.47 (5.04\pm0.73)$   & $8.80\pm3.55 (6.07\pm1.11)$      \\
&                           & $f_{\Delta}$     & $5.71\pm1.53 (5.81\pm1.40)$  & $6.21\pm1.76 (6.91\pm1.75)$   & $6.70\pm2.38 (8.49\pm2.35)$    \\
&  &&&& \\

       &                    & $r_\Delta$       & $0.54\pm0.03 (0.52\pm0.03)$  & $1.06\pm0.06 (1.03\pm0.05)$   & $1.50\pm0.08 (1.45\pm0.07)$      \\   
&  \chandra\                & $M_{g,\Delta}$   & $1.46\pm0.31 (1.42\pm0.30)$  & $3.09\pm0.66 (3.03\pm0.65)$   & $4.46\pm0.94 (4.37\pm0.93)$      \\
&  back*0.95$^{d}$                & $M_{\Delta}$     & $2.55\pm0.41 (2.30\pm0.34)$  & $3.96\pm0.64 (3.57\pm0.53)$   & $4.43\pm0.69 (4.01\pm0.58)$      \\
&                           & $f_{\Delta}$     & $5.77\pm1.37 (6.22\pm1.43)$  & $7.89\pm1.99 (8.50\pm2.09)$   & $10.12\pm2.61 (10.94\pm2.75)$    \\
&  &&&& \\

       &                    & $r_\Delta$       & $0.52\pm0.02 (0.54\pm0.03)$  & $1.10\pm0.05 (1.08\pm0.07)$   & $1.63\pm0.09 (1.55\pm0.09)$      \\   
&  \chandra\                & $M_{g,\Delta}$   & $1.43\pm0.31 (1.49\pm0.35)$  & $3.21\pm0.74 (3.22\pm0.77)$   & $4.88\pm1.16 (4.70\pm1.12)$      \\
&  back*1.05$^{e}$                & $M_{\Delta}$     & $2.36\pm0.27 (2.56\pm0.44)$  & $4.37\pm0.55 (4.20\pm0.75)$   & $5.74\pm0.98 (4.92\pm0.81)$      \\
&                           & $f_{\Delta}$     & $6.06\pm1.41 (5.87\pm1.47)$  & $7.40\pm1.81 (7.69\pm2.08)$   & $8.60\pm2.28 (9.59\pm2.62)$    \\
&  &&&& \\

G036S$^f$  &                    & $r_\Delta$       & $0.52\pm0.02 (0.56\pm0.02)$  & $0.98\pm0.03 (1.06\pm0.04)$   & $1.37\pm0.05 (1.48\pm0.06)$      \\   
&                           & $M_{g,\Delta}$   & $1.68\pm0.33 (1.79\pm0.31)$  & $3.36\pm0.67 (3.56\pm0.63)$   & $4.74\pm0.94 (5.02\pm0.90)$      \\
&  \xmm                       & $M_{\Delta}$     & $2.36\pm0.23 (2.95\pm0.32)$  & $3.13\pm0.31 (3.90\pm0.41)$   & $3.38\pm0.34 (4.24\pm0.47)$      \\
&                           & $f_{\Delta}$     & $7.10\pm1.42 (6.07\pm1.13)$  & $10.72\pm2.27 (9.12\pm1.77)$  & $14.03\pm3.04 (11.85\pm2.39)$    \\
\hline  \hline 
\end{tabular}
\tablenotetext{a}{$r_\Delta$, $M_{g,\Delta}$, $M_{\Delta}$, and $f_{\Delta}$ are the radius, gas mass, total mass, and gas mass fraction, measured inside a radius where the overdensity within that radius is $\Delta$ times the critical density of the Universe at the cluster redshift, while the units are Mpc, $10^{13}$ $M_{\rm \odot}$, $10^{14}$ $M_{\rm \odot}$, and \%, respectively. The values outside and inside of the parentheses were obtained with $N_H$ fixed at the total and best-fitting values (Table 2), respectively.}
\tablenotetext{b}{\chandra\ baseline background model, i.e., the blank-sky background.}
\tablenotetext{c}{\chandra\ baseline background model with a soft-band adjustment.}
\tablenotetext{d}{\chandra\ baseline background model varied by $+5\%$.}
\tablenotetext{e}{\chandra\ baseline background model varied by $-5\%$.}
\tablenotetext{f}{As G036S is close to the chip boundary in the \chandra\ FOV, we can only measure the temperature profile up to $\sim$$0.8r_{500}$. In this relatively small region, \chandra\ gives an essentially isothermal temperature profile, unlike the obviously declining temperature profile revealed by \xmm\ at large radii where \chandra\ do not cover.  Thus we did not measure mass-related quantities for G036S with \chandra .}
\end{center}
\end{table*}
%%%%%%%%%%%%%%%%%%%%%%%%%%%%%

%%%%%%%%%%%%%%%%%%%%%%%%%%%%%% 
\begin{table}
\begin{center}
{\caption{Characteristics of the Shock and the Merger}
\vspace{0.5cm}
\small
\begin{tabular}{cccccc} \hline \hline 
 Region  &   Parameter$^a$      & \xmm\  & \chandra\ baseline$^b$ & \chandra\ back*0.95$^c$ & \chandra\ back*1.05$^d$  \\ \hline 
              &  $\mathcal{M}$$^e$   & $1.16\pm0.21$ $(1.19\pm0.30)$                         & $2.01\pm0.46$ $(1.91\pm0.38)$     & $2.01\pm0.45$ $(1.90\pm0.38)$ & $2.02\pm0.46$ $(1.91\pm0.38)$   \\
G036N   & $c_{\rm sound}$$^f$ &  $1246^{+54}_{-46}$ ($1353^{+68}_{-54}$) & $1323^{+91}_{-71}$ ($1267^{+76}_{-71}$) & $1290^{+85}_{-74}$ ($1232^{+77}_{-68}$) & $1349^{+103}_{-69}$ ($1295^{+84}_{-69}$) \\
              & $t_r$$^g$   &  $0.94\pm0.18$ ($0.87\pm0.22$) &  $0.57\pm0.14$ ($0.61\pm0.13$)  & $0.57\pm0.14$ ($0.61\pm0.13$) & $0.59\pm0.14$ ($0.61\pm0.13$) \\

&&&&& \\

             & $\mathcal{M}$$^e$    & $1.25\pm0.19$ $(1.28\pm0.28)$                         & $2.27\pm0.44$ $(2.15\pm0.36)$    & $2.26\pm0.43$ $(2.14\pm0.35)$ & $2.28\pm0.44$  $(2.17\pm0.36)$   \\
G036S  & $c_{\rm sound}$$^f$ &  $1224^{+54}_{-41}$ ($1318^{+44}_{-62}$) & $1295^{+81}_{-61}$ ($1244^{+64}_{-60}$) & $1271^{+63}_{-62}$ ($1217^{+65}_{-53}$) & $1317^{+98}_{-60}$ ($1266^{+63}_{-61}$) \\
              & $t_r$$^g$   &  $0.88\pm0.14$ ($0.86\pm0.19$) &  $0.52\pm0.11$ ($0.57\pm0.10$)  & $0.51\pm0.10$ ($0.56\pm0.10$) & $0.53\pm0.11$ ($0.57\pm0.10$) \\
&&&&& \\

...           & $t_{\rm merge}$$^h$   &  0.78 (0.79) &  0.80 (0.79)  & 0.80 (0.80) & 0.80 (0.80) \\
              & $\theta$$^i$   &  76 (78) &  80 (80)  & 79 (79) & 81 (81) \\
 \hline \hline
\end{tabular}
}
\tablenotetext{a}{Values outside and inside of the parentheses were obtained with $N_H$ fixed at the total and best fitting values (Table 2), respectively.}
\tablenotetext{b}{\chandra\ baseline background model.}
\tablenotetext{c}{\chandra\ baseline background model varied by $+5\%$.}
\tablenotetext{d}{\chandra\ baseline background model varied by $-5\%$.}
\tablenotetext{e}{Mach number obtained using the temperatures in Region 1 and Region 2 (Region 3) in Figure 7 as the temperatures of the postshock and preshock gas in G036N (G036S).}
\tablenotetext{f}{Sound speed, in units of km s$^{-1}$, with the temperature measured in the $[0.15-0.75]r_{500}$ region (Table 5).}
\tablenotetext{g}{Time for the shock to propagate to the subcluster center in units of Gyr. For G036S, as \chandra\ does not cover large enough region to measure the mass, hence $r_{200}$, we assumed that the ratios of $r_{200}$ for G036N and G036S are the same for \xmm\ and \chandra\ and that the relative errors are the same for both subclusters for \chandra , to obtain $r_{200}$ of G036S for \chandra .}
\tablenotetext{h}{Time in Gyr since the merger began, obtained from the simplified model in Section 5.1.}
\tablenotetext{i}{Angle in degree between the merger axis (assuming zero impact parameter) and the plane of the sky obtained from the simplified model in Section 5.1.}
\end{center}
\end{table}
%%%%%%%%%%%%%%%%%%%%%%%%%%%%%

%%%%%%%%%%%%%%%%%%%%%%%%%%%%%% 
\begin{table}
\begin{center}
{\caption{Temperature, $Y_{X,500}$, and $Y_{SZ,500}/Y_{X,500}$}
\vspace{0.5cm}
\small
\begin{tabular}{cccccc} \hline \hline 
 Region  &   Parameter$^a$      & \xmm\  & \chandra\ baseline$^b$ & \chandra\ back*0.95$^c$ & \chandra\ back*1.05$^d$  \\ \hline 
 G036N             &  $kT$$^e$   & $5.97^{+0.52}_{-0.44}$ $(7.04^{+0.71}_{-0.56})$                         & $6.73^{+0.93}_{-0.72}$ $(6.17^{+0.74}_{-0.69})$     & $6.40^{+0.84}_{-0.73}$ $(5.84^{+0.73}_{-0.64})$ & $7.00^{+1.07}_{-0.72}$ $(6.45^{+0.84}_{-0.69})$   \\
   & $Y_{X,500}$$^f$ &  $1.82\pm0.43$ ($2.13\pm0.54$) & $2.17\pm0.52$ ($2.02\pm0.48$) & $2.06\pm0.50$ ($1.91\pm0.46$) & $2.26\pm0.58$ ($2.11\pm0.54$) \\

&&&&& \\ 

 G036S            & $kT$$^e$    & $5.76^{+0.51}_{-0.39}$ $(6.68^{+0.63}_{-0.45})$                         & $6.45^{+0.81}_{-0.61}$ $(5.95^{+0.61}_{-0.57})$    & $6.21^{+0.62}_{-0.61}$ $(5.70^{+0.61}_{-0.50})$ & $6.67^{+0.99}_{-0.61}$  $(6.16^{+0.61}_{-0.59})$   \\
  & $Y_{X,500}$$^f$ &  $1.97\pm0.41$ ($2.24\pm0.42$) & $2.11\pm0.48$ ($1.95\pm0.43$) & $1.99\pm0.44$ ($1.86\pm0.41$) & $2.15\pm0.50$ ($2.02\pm0.45$) \\

&&&&& \\ 
Whole              & $Y_{SZ,500}/Y_{X,500}$$^g$   &  $1.11\pm0.35$ ($0.96\pm0.30$) &  $0.98\pm0.33$ ($1.06\pm0.35$)  & $1.04\pm0.34$ ($1.12\pm0.37$) & $0.96\pm0.33$ ($1.02\pm0.35$) \\

 \hline \hline
\end{tabular}
}
\tablenotetext{a}{Values outside and inside of the parentheses were obtained with $N_H$ fixed at the total and best fitting values (Table 2), respectively.}   
\tablenotetext{b}{\chandra\ baseline background model.}
\tablenotetext{c}{\chandra\ baseline background model varied by $+5\%$.}
\tablenotetext{d}{\chandra\ baseline background model varied by $-5\%$.}
\tablenotetext{e}{Temperature in keV measured in the $[0.15-0.75]r_{500}$ region. For G036N, we adopted $r_{500}=1.0$ (1.1) Mpc for \xmm\ (\chandra ), while for G036S, $r_{500}=1.0$ (1.0) Mpc for \xmm\ (\chandra ), for simplicity.}
\tablenotetext{f}{Product of temperature and gas mass, where temperature was measured in the $[0.15-0.75]r_{500}$ region and gas mass is that within $r_{500}$. $Y_{X,500}$ is in $10^{14}$ $M_{\rm \odot}$ keV.}
\tablenotetext{g}{See Section 5.2 for the definition of this ratio. $Y_{SZ,500}$ has been scaled to a redshift of 0.1547.}
\end{center}
\end{table}
%%%%%%%%%%%%%%%%%%%%%%%%%%%%%

\clearpage

%%%%%%%%%%%%%%%%%%%%%%%%%%%%%
\begin{figure*}[hbt!]
\centerline{
\includegraphics[width=0.5\textwidth]{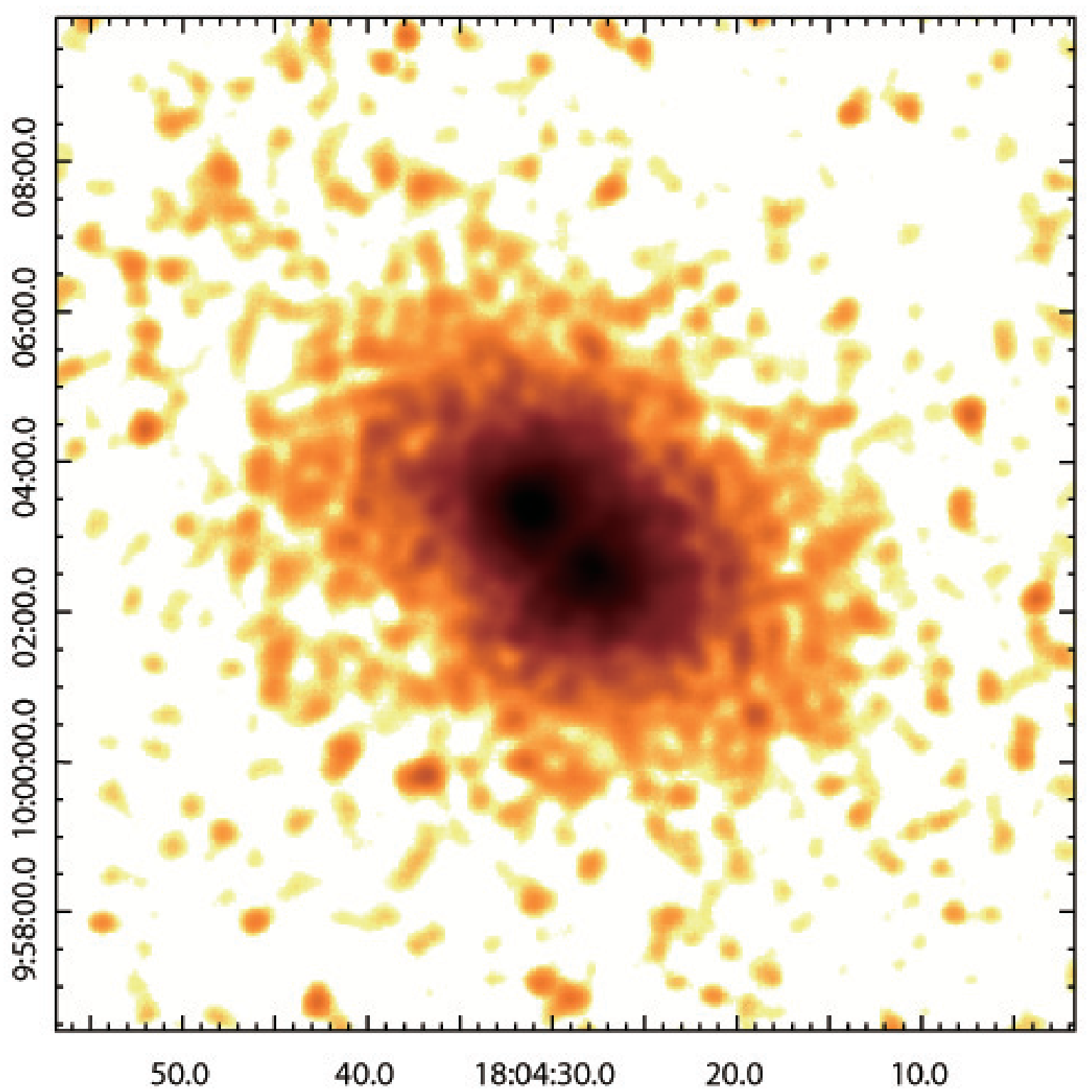}
\includegraphics[width=0.5\textwidth]{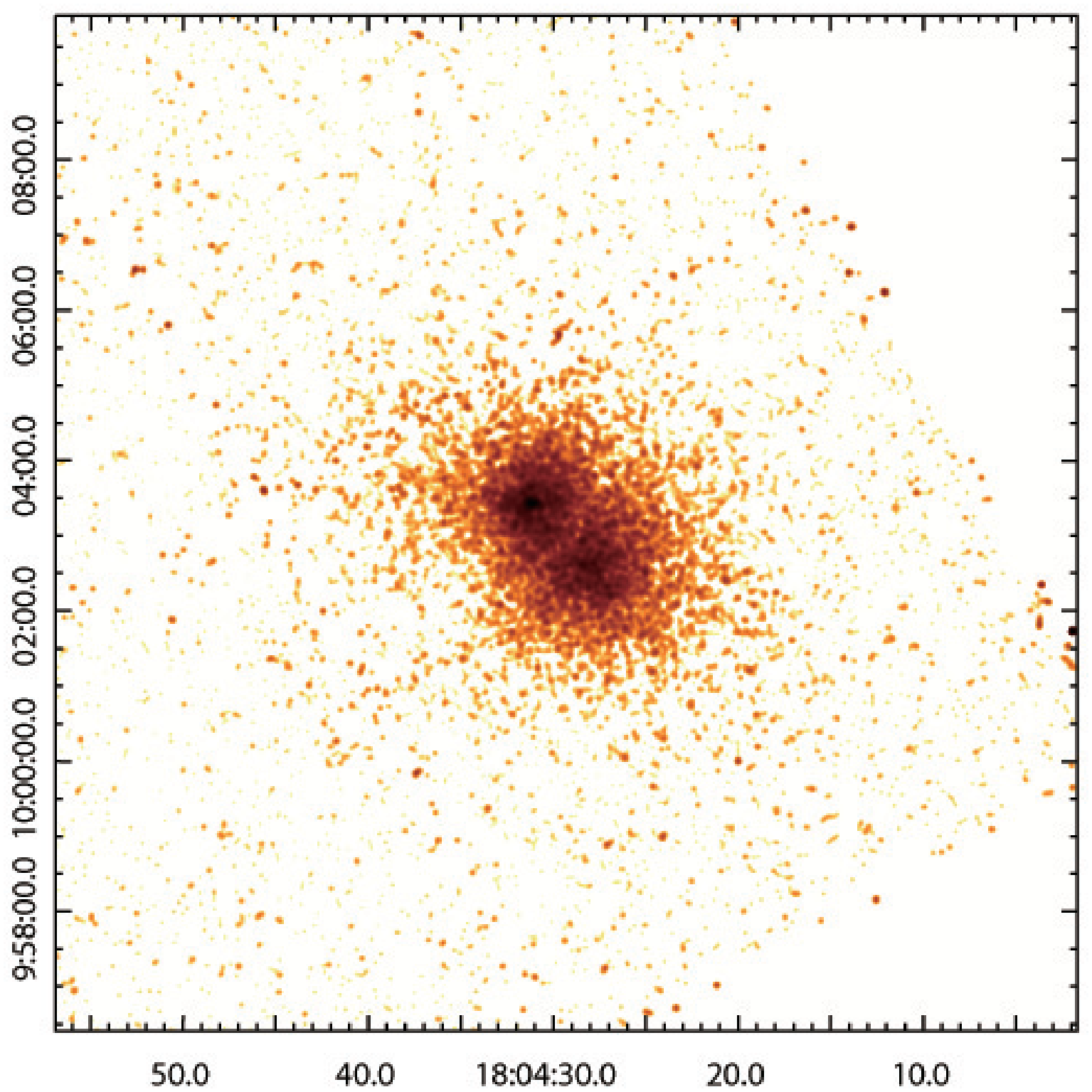}}
\caption{X-ray images (2.2 Mpc $\times$ 2.1 Mpc) of \cluster . North is up and East is to the left. \textit{Left}: \xmm\ EPIC (quiescent particle and residual soft proton) background subtracted and exposure corrected image in the $0.7-2.0$ keV band, smoothed with a $15''$ Gaussian kernel. Each pixel has a size of $2\farcs5\times2\farcs5$. The large scale X-ray emission is elongated in a northeast-southwest direction, suggesting that \cluster\ may be undergoing a merger. Two subclusters, G036N in the north and G036S in the south, were also resolved, although the separation between them is quite small ($\sim$$72''=193$ kpc in projection). Outside the interaction region between the two subclusters, the morphologies of G036N and G036S do not deviate significantly from spherical symmetry. \textit{Right}: \chandra\ $0.7-7.0$ keV background subtracted and exposure corrected image, smoothed with a $3\farcs94$ Gaussian kernel. The image is not binned; i.e., 1 pixel has a size of $0\farcs492\times0\farcs492$. The large scale X-ray emission exhibits the same behavior as \xmm\ revealed, with more details of the two subclusters resolved. A point source is visible at the center of G036N, which is an AGN. A bow shaped gap (with an angle of $\sim$$145^\circ$), is seen between G036N and G036S, another indication of interaction between G036N and G036S. 
}
\end{figure*}
%%%%%%%%%%%%%%%%%%%%%%%%%%%%%

%%%%%%%%%%%%%%%%%%%%%%%%%%%%%%%%%%%%%%%%%%%%%%%%%%%%%%%%%%%%%%%%%%%%%%%%%%
\begin{figure}
\centerline{\includegraphics[height=0.8\linewidth, angle=-90]{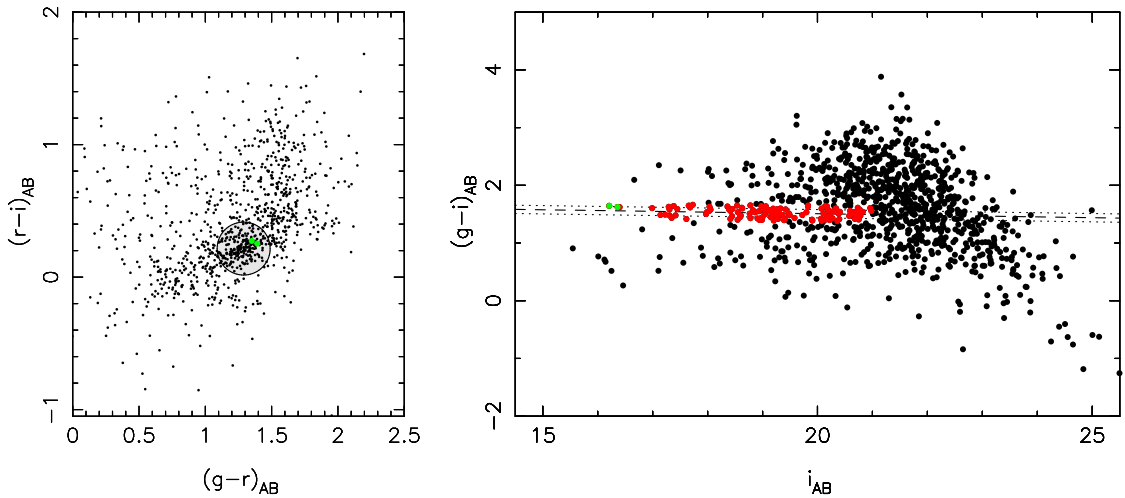}}
\vspace{0.5cm}
\scriptsize
  \caption{\textit{Left}: Color-color ($g-r$ vs.\ $r-i$) diagram of non-stellar objects in the FOV ($7\farcm 7 \times 7\farcm 7$) of MOSCA. The shaded circle indicates the region from which galaxies were selected to derive a linear fit to the color-magnitude relation (right panel) of the cluster galaxies. \textit{Right}: Color-magnitude ($g-i$ vs.\ $i$) diagram of non-stellar objects in the same FOV as in the left panel. The dashed line is the best fitting color-magnitude relation for galaxies within the circle shown in the left panel, while the scatter (0.072 mag, 1$\sigma$) about the best fitting relation is indicated by the dotted lines. Galaxies within twice the measured scatter around the best fitting relation, down to $i_{AB}\sim21$ mag, are represented by red dots and are considered to be the red sequence of the cluster. The BCGs (marked in Figure 3, left panel) of the two subclusters are located on the red sequence and colored green in particular in both panels.} 
\end{figure}
%%%%%%%%%%%%%%%%%%%%%%%%%%%%%%%%%%%%%%%%%%%%%%%%%%%%%%%%%%%%%%%%%%%%%%%%%%

%%%%%%%%%%%%%%%%%%%%%%%%%%%%%
\begin{figure*}[hbt!]
\centerline{
\includegraphics[width=0.5\textwidth]{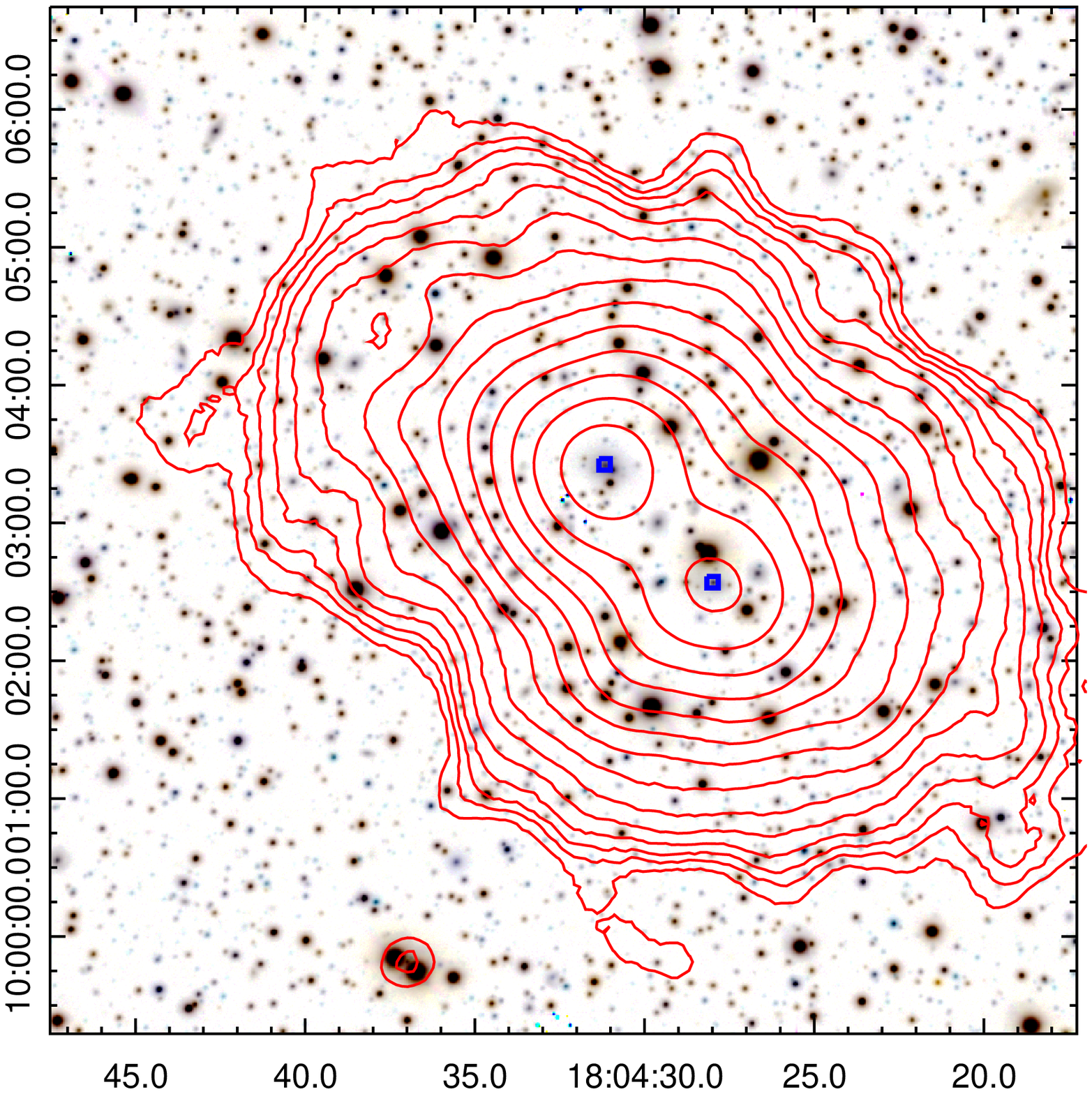}
\includegraphics[width=0.5\textwidth]{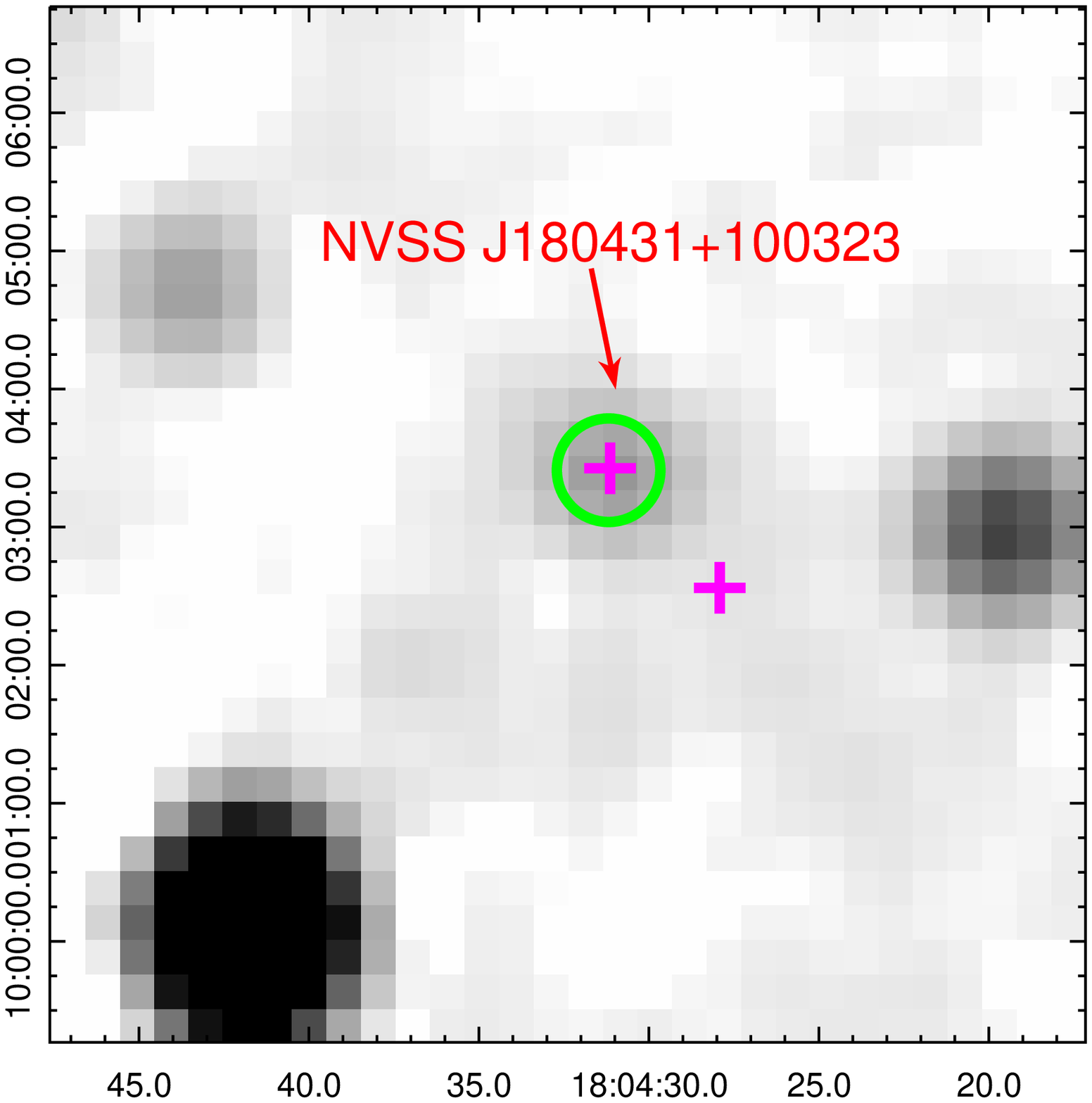}}
\caption{\textit{Left}: Optical mosaic image (1.2 Mpc $\times$ 1.2 Mpc) with the \xmm\ $0.7-2.0$ keV contours overlaid. The two blue squares ($5''\times5''$) mark the locations of the two BCGs identified in Figure 2. As can be seen, the two BCGs are very close to the X-ray centroids (defined to be the X-ray centers) of G036N and G036S. \textit{Right}: NVSS 1.4 GHz radio image, same size as the left panel. The magenta crosses mark the X-ray centers of G036N and G036S. While there is a radio source (name labeled) at the location of G036N, no radio emission was detected at the position of G036S. The radio source should be hosted by the BCG of G036N. The green circle shows the angular resolution of the NVSS (${\rm FWHM}=45''$, 121 kpc at the cluster redshift). The extent of the radio source is comparable to the angular resolution of the NVSS.
}
\end{figure*}
%%%%%%%%%%%%%%%%%%%%%%%%%%%%%

%%%%%%%%%%%%%%%%%%%%%%%%%%%%%
\begin{figure*}[hbt!]
\centerline{
\includegraphics[width=0.5\textwidth]{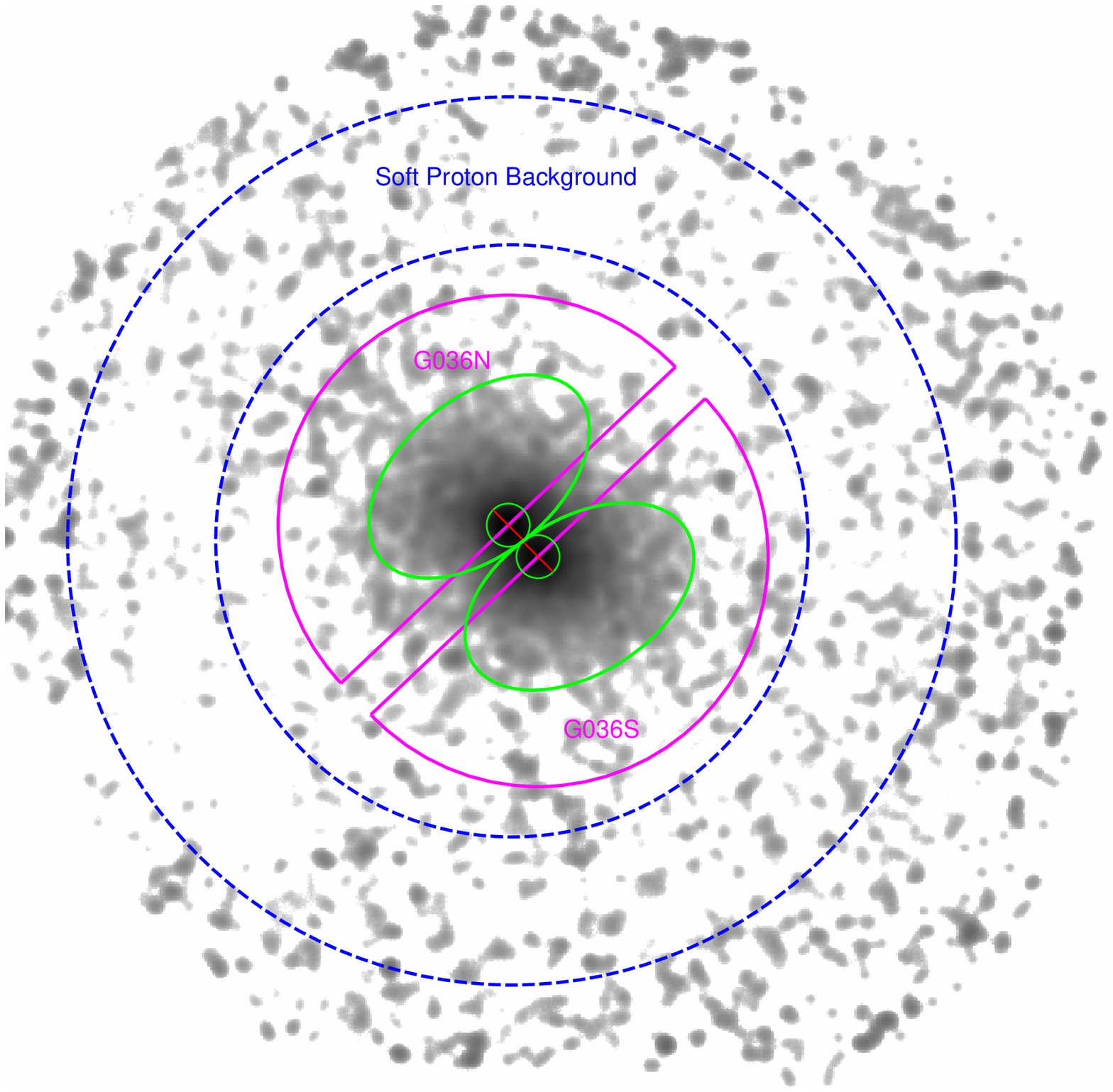}
\includegraphics[width=0.5\textwidth]{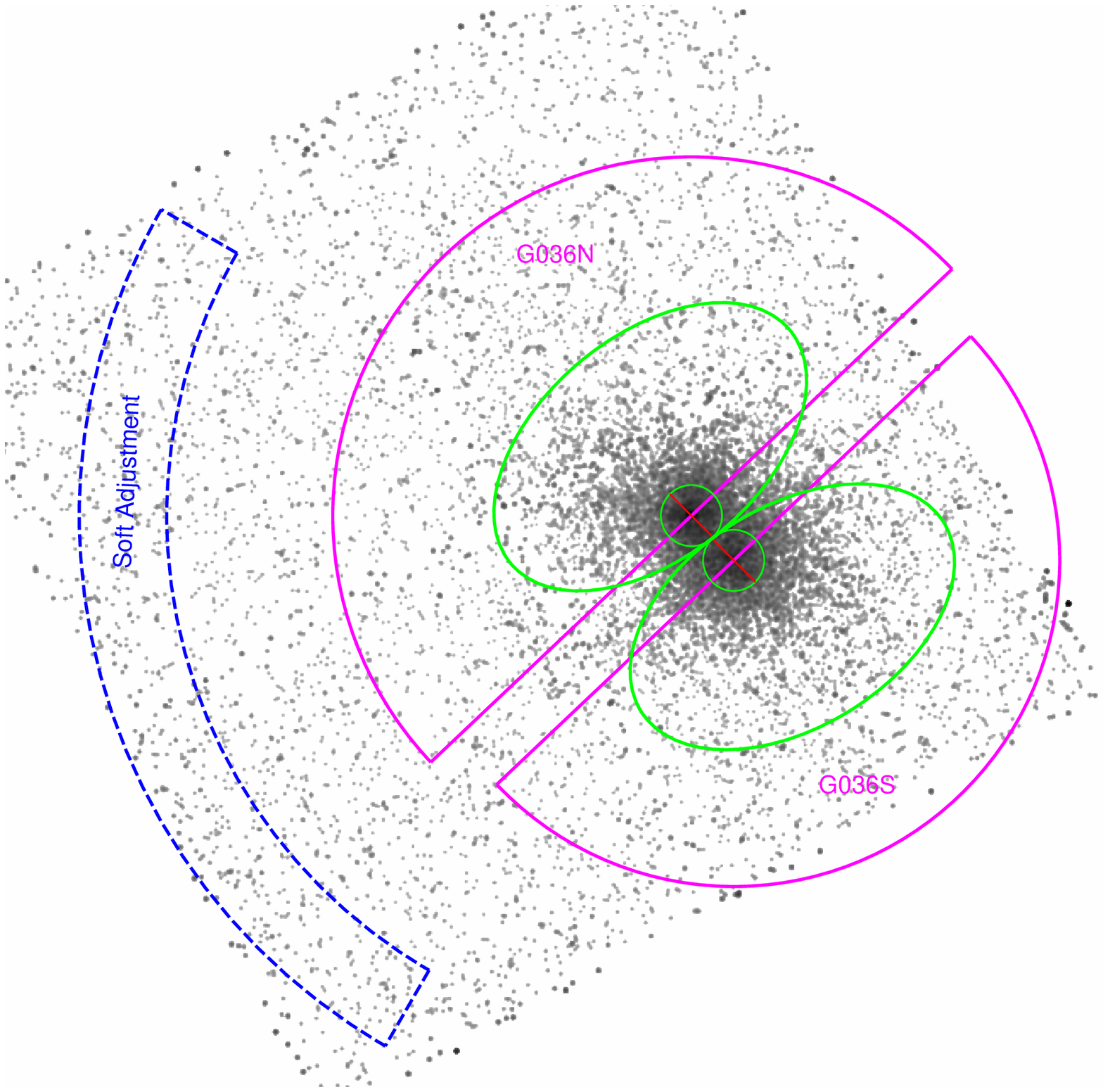}}
\caption{Regions used to perform spatial and spectral analyses. For either G036N or G036S, the central small green circle ($35''$ in radius) represents the core, while the outer green ellipse represents the bright part of the subcluster. A third region, the ellipse exclude the core, is also defined. These three regions are the same for both \xmm\ and \chandra . Spectra from these regions were extracted and fit together to determine the global properties (see Sections $4.1-4.3$) of G036N and G036S, with both \xmm\ and \chandra .  The magenta half circles mark $r_{500}$ for both G036N and G036S, and temperature and surface brightness profiles were obtained within these regions (note that the outmost radii for these profiles are different from $r_{500}$; see Figures 5 and 6). \textit{Left}: \xmm\ regions. For both G036N and G036S, $r_{500}=1.0$ Mpc (see Table 3). The outmost blue dashed annulus ($480''-720''$ in radii) is largely cluster emission free, and, together with the six regions defined above for both G036N and G036S, is used to determine the various components of the background (see Section 4.1.2 for details). \textit{Right}: \chandra\ regions. For G036N, $r_{500}=1.1$ Mpc, while $r_{500}=1.0$ Mpc for G036S (see Tables 3 and 5). The blue dashed region near the CCD chip is the region used to determine the parameters of the soft-band adjustment (see Section 4.1.1 for details).
}
\end{figure*}
%%%%%%%%%%%%%%%%%%%%%%%%%%%%%

%%%%%%%%%%%%%%%%%%%%%%%%%%%%%
\begin{figure*}[hbt!]
\centerline{
\includegraphics[width=0.3\textwidth, angle=-90]{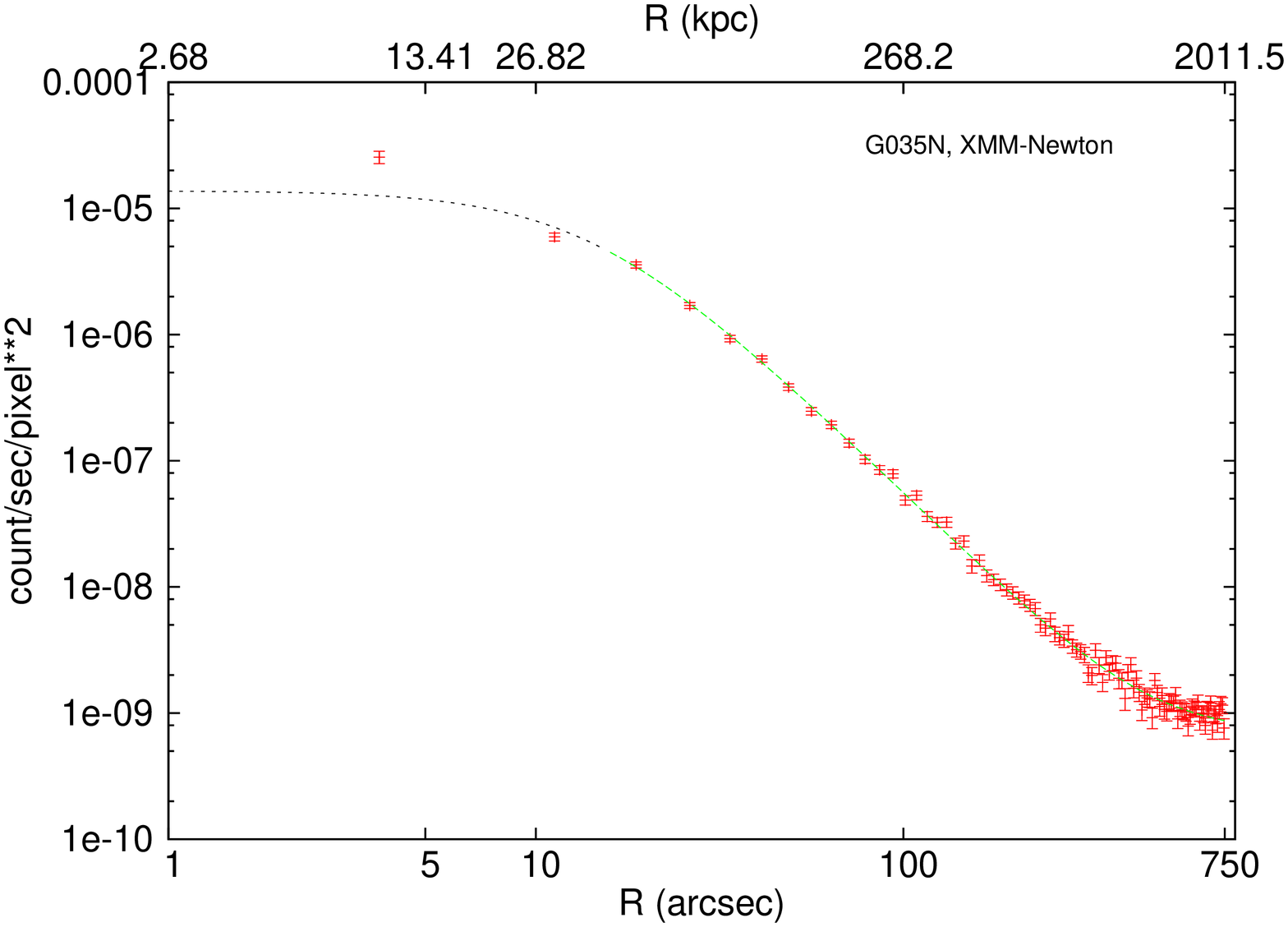}
\includegraphics[width=0.3\textwidth, angle=-90]{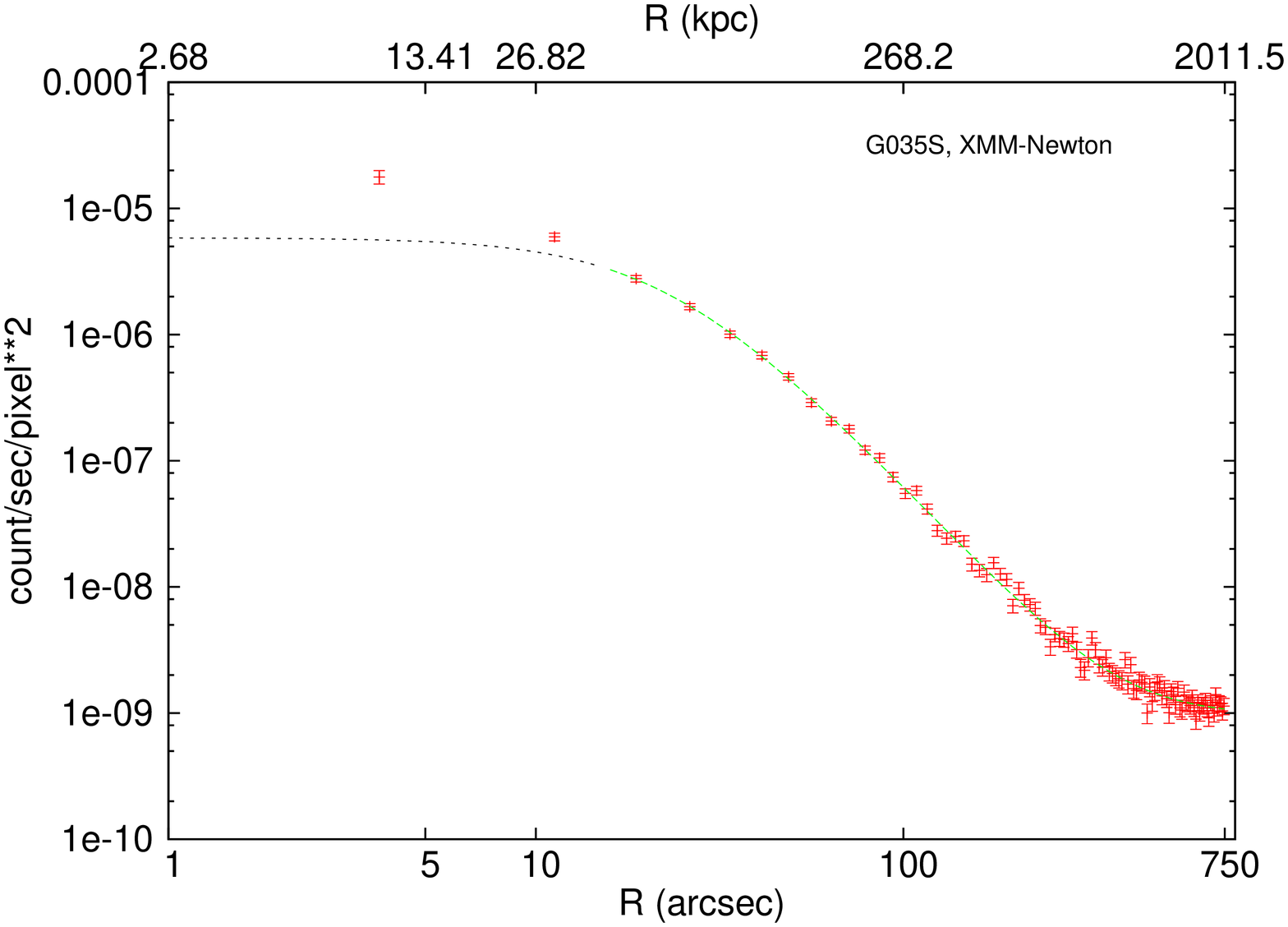}}
\centerline{
\includegraphics[width=0.3\textwidth, angle=-90]{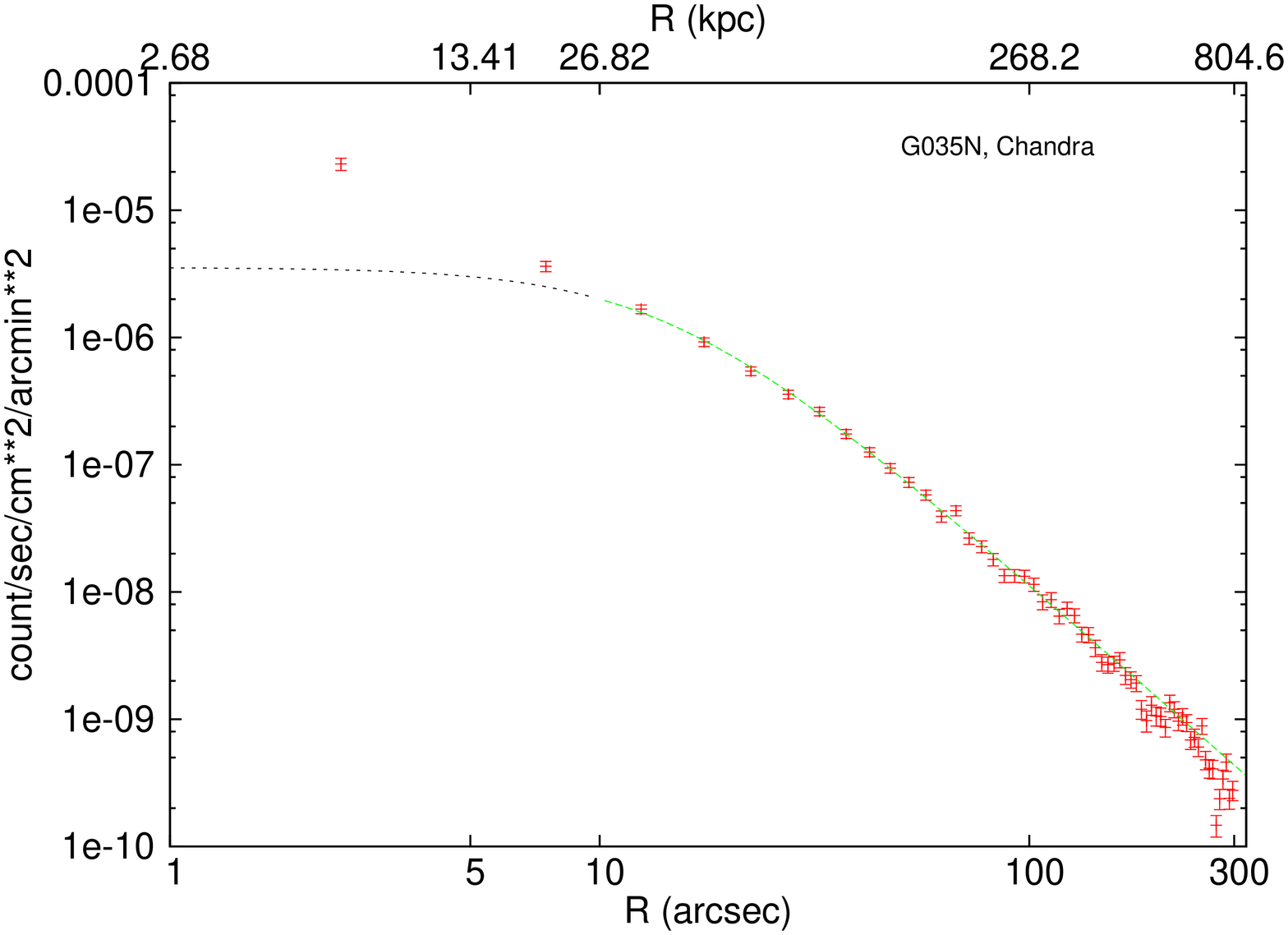}
\includegraphics[width=0.3\textwidth, angle=-90]{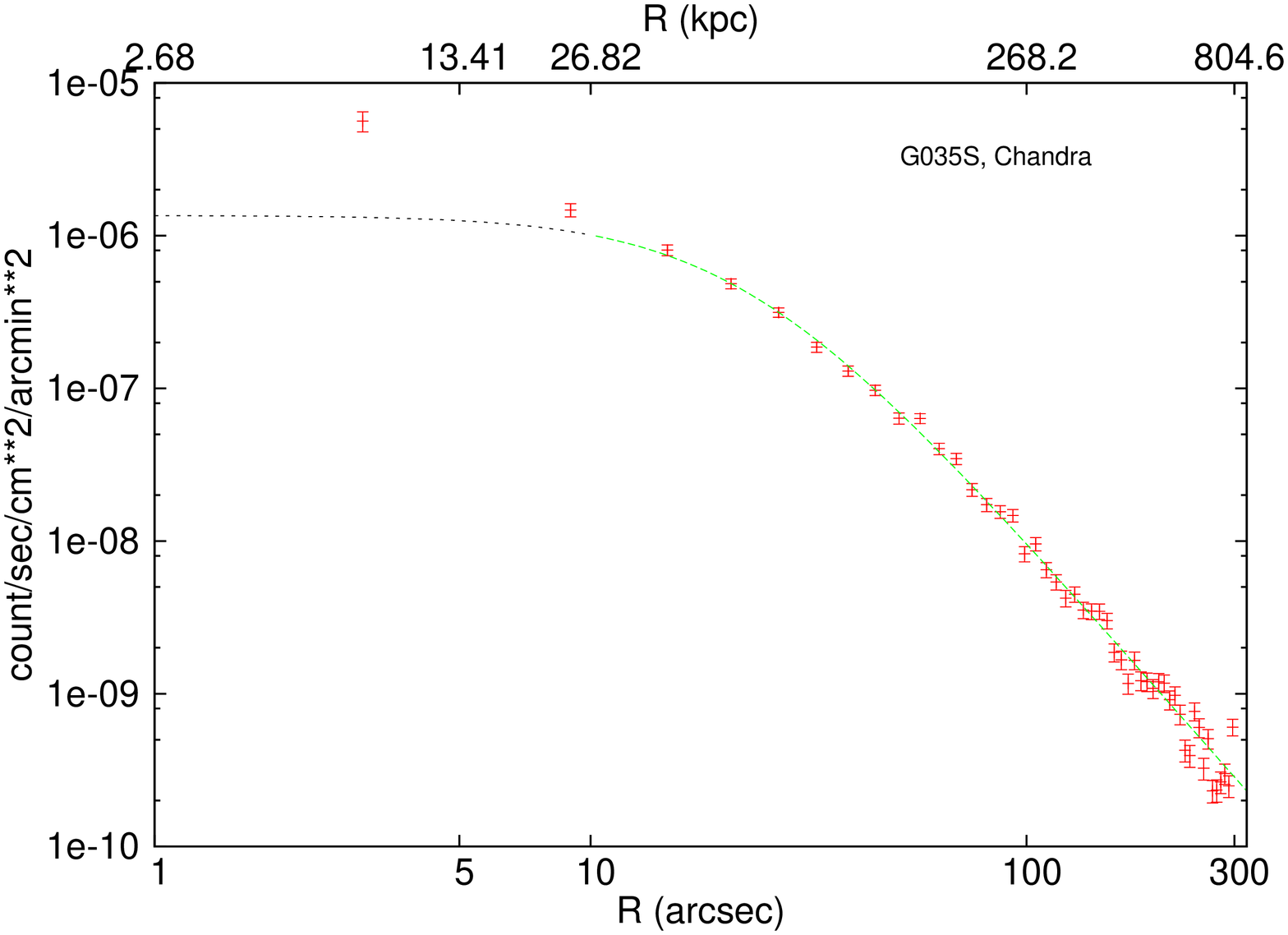}}
\caption{Surface brightness profiles extracted from the half circular regions in Figure 4. Red data points with errors are the measured values. The green dashed lines are the best fitting single $\beta$ models outside the central $\sim$$10''$ regions, where there is excess emission relative to the $\beta$ models. The $\beta$ models generally fit the data very well in these regions. The black dotted lines in the central $\sim$$10''$ regions are the interpolations of the $\beta$ models. The outer boundaries are shown by the largest values of the $x$-axis. \textit{Upper Left}: Surface brightness profile of G036N obtained by \xmm . \textit{Upper Right}: Surface brightness profile of G036S obtained by \xmm . \textit{Lower Left}: Surface brightness profile of G036N obtained by \chandra . \textit{Lower Right}: Surface brightness profile of G036S obtained by \chandra .
}
\end{figure*}
%%%%%%%%%%%%%%%%%%%%%%%%%%%%%

%%%%%%%%%%%%%%%%%%%%%%%%%%%%%
\begin{figure*}[hbt!]
\centerline{
\includegraphics[width=0.3\textwidth, angle=-90]{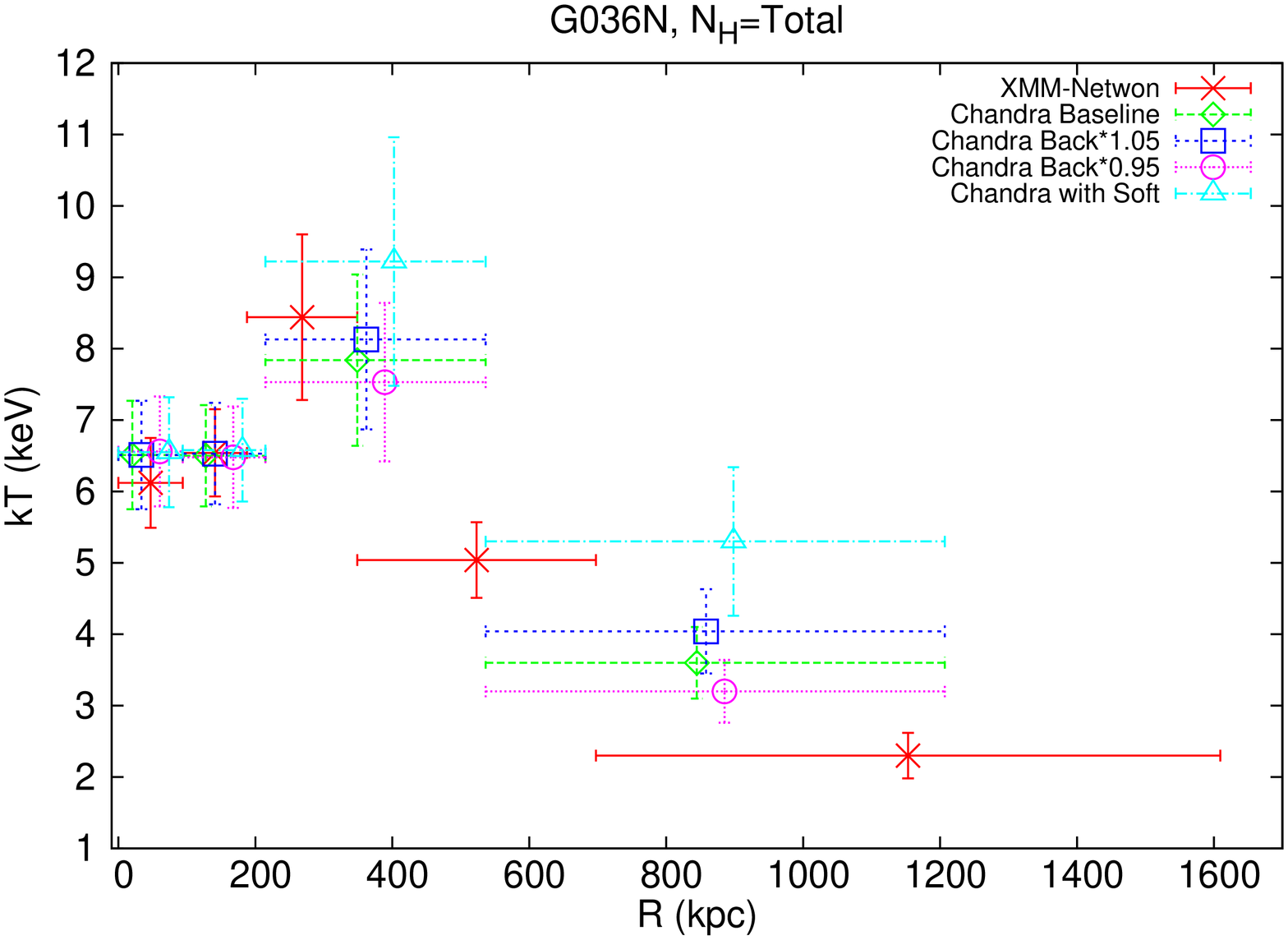}
\includegraphics[width=0.3\textwidth, angle=-90]{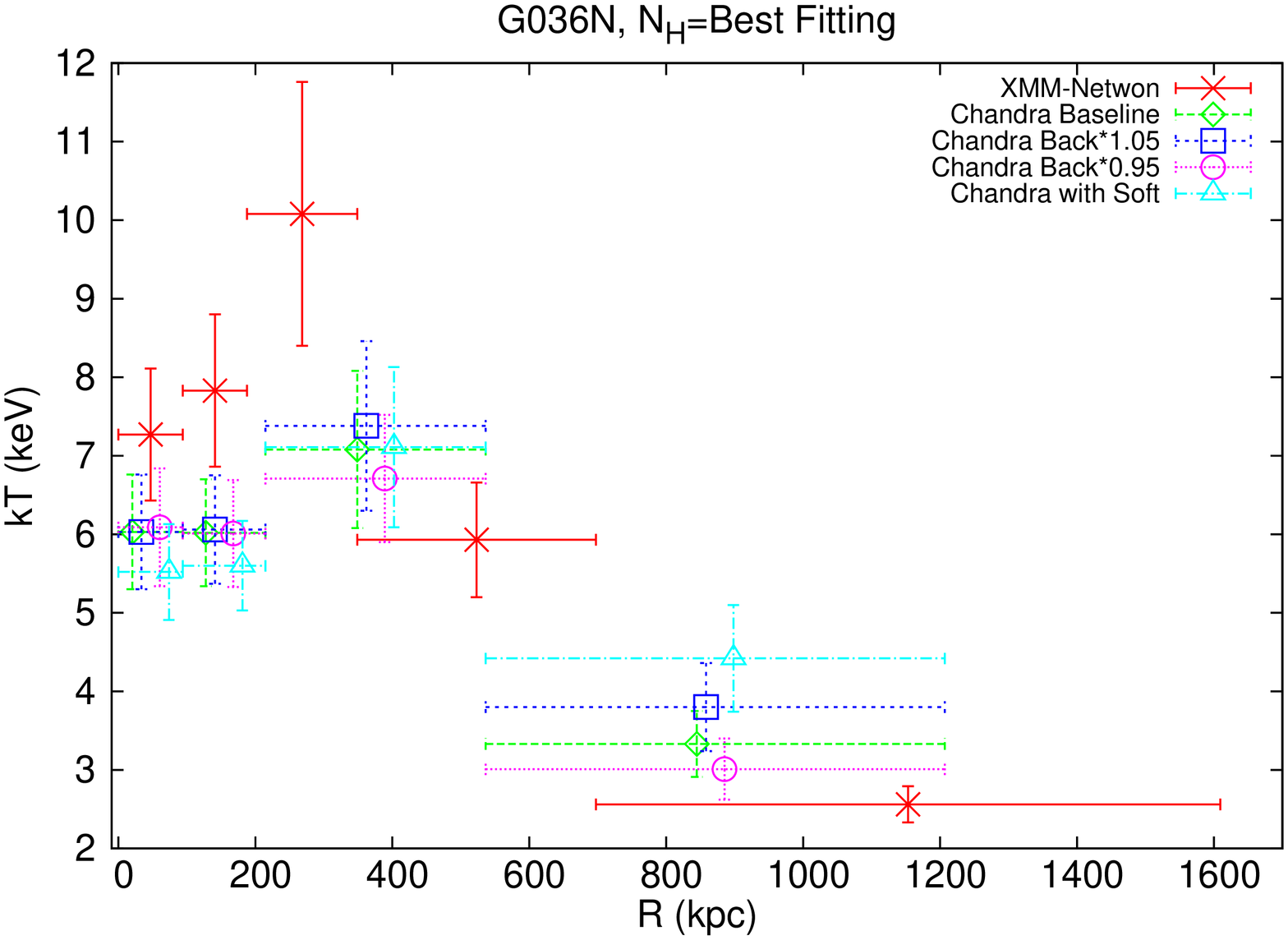}}
\centerline{
\includegraphics[width=0.3\textwidth, angle=-90]{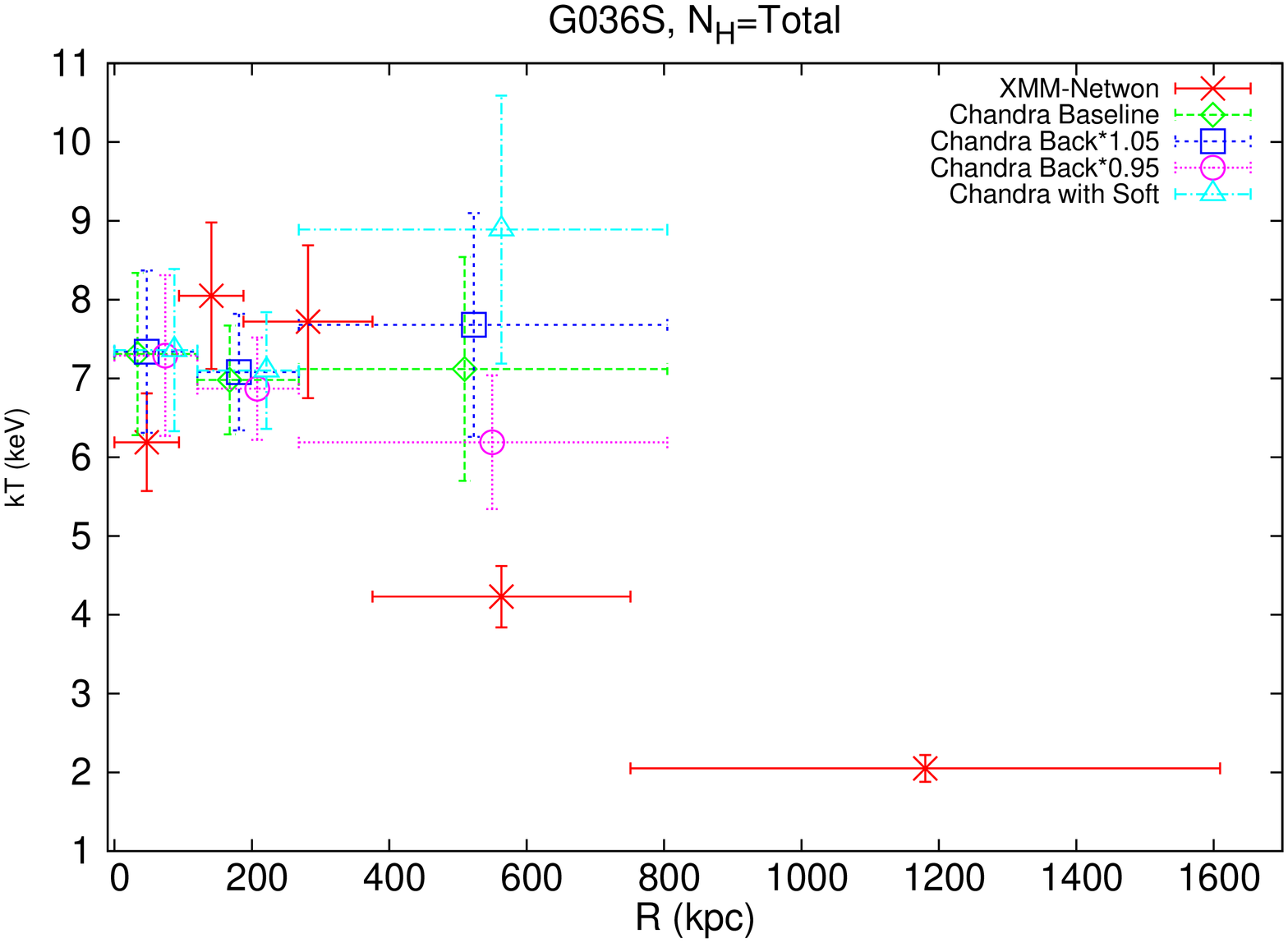}
\includegraphics[width=0.3\textwidth, angle=-90]{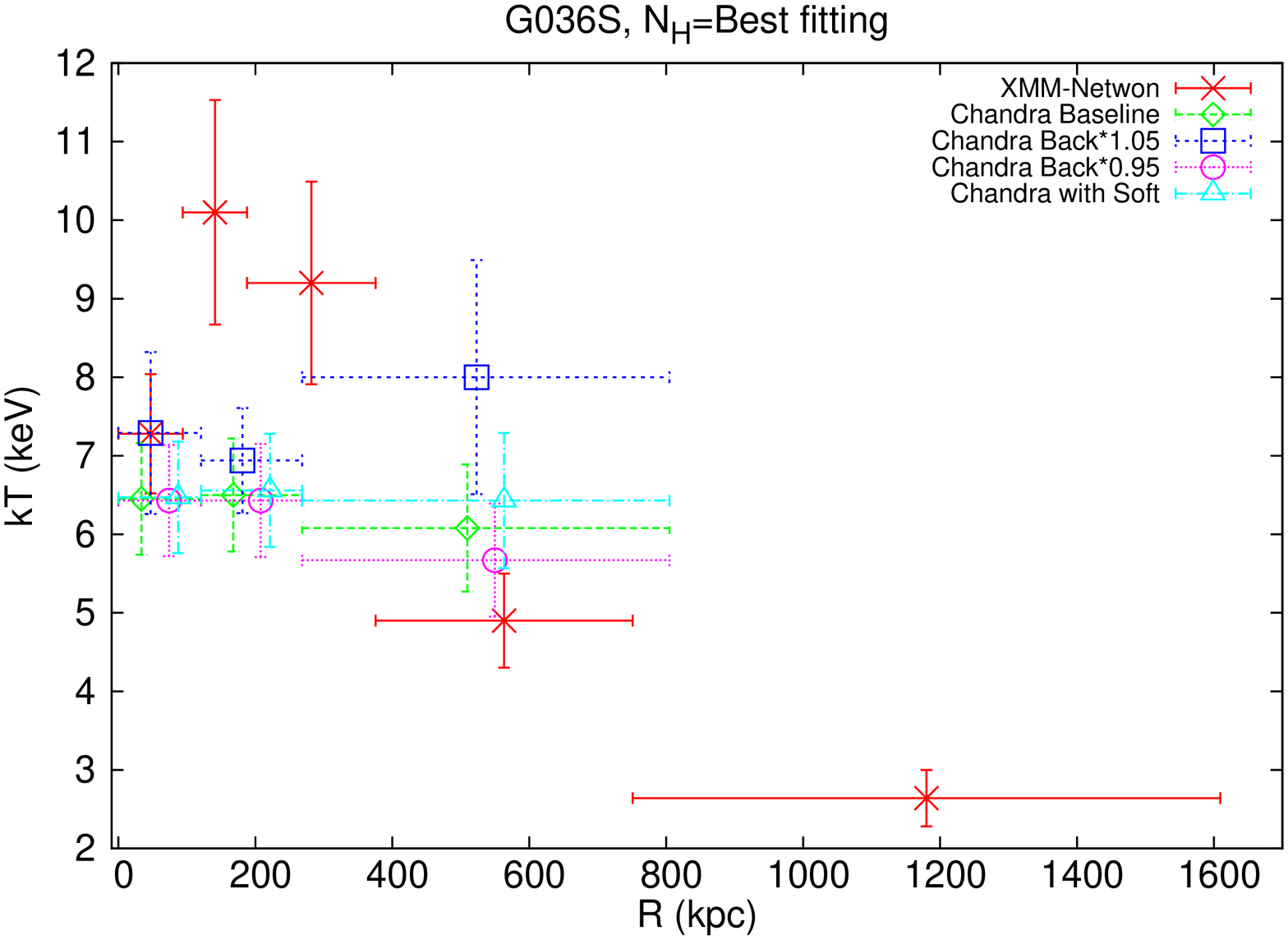}}
\caption{Temperature profiles extracted from the half circular regions in Figure 4. \xmm\ data are marked by red cross, \chandra\ baseline background model by green diamond, \chandra\ baseline background model varied by $-5\%$ by blue square, \chandra\ baseline background model varied by $+5\%$ by magenta circle, and \chandra\ baseline background model with a soft-band adjustment by cyan triangle. The horizontal error bars show the radial bin size. To show the data more clearly, we offset the \chandra\ data points by $\pm26.8$ and $\pm13.4$ kpc for the four \chandra\ background models. \textit{Upper Left}: Temperature profiles of G036N obtained with $N_H$ fixed at the total value, $0.136\times10^{22}$ cm$^{-2}$. \textit{Upper Right}: Temperature profiles of G036N obtained with $N_H$ fixed at the best fitting values, $0.092\times10^{22}$ cm$^{-2}$ for \xmm , $0.195\times10^{22}$ cm$^{-2}$ for the \chandra\ baseline background model with a soft-band adjustment, and $0.164\times10^{22}$ cm$^{-2}$ for the other three \chandra\ background models. \textit{Lower Left}: Temperature profiles of G036S obtained with $N_H$ fixed at the total value. \textit{Lower Right}: Temperature profiles of G036S obtained with $N_H$ fixed at the best fitting values.
}
\end{figure*}
%%%%%%%%%%%%%%%%%%%%%%%%%%%%%

%%%%%%%%%%%%%%%%%%%%%%%%%%%%%
\begin{figure*}[hbt!]
\centerline{
\includegraphics[width=0.5\textwidth]{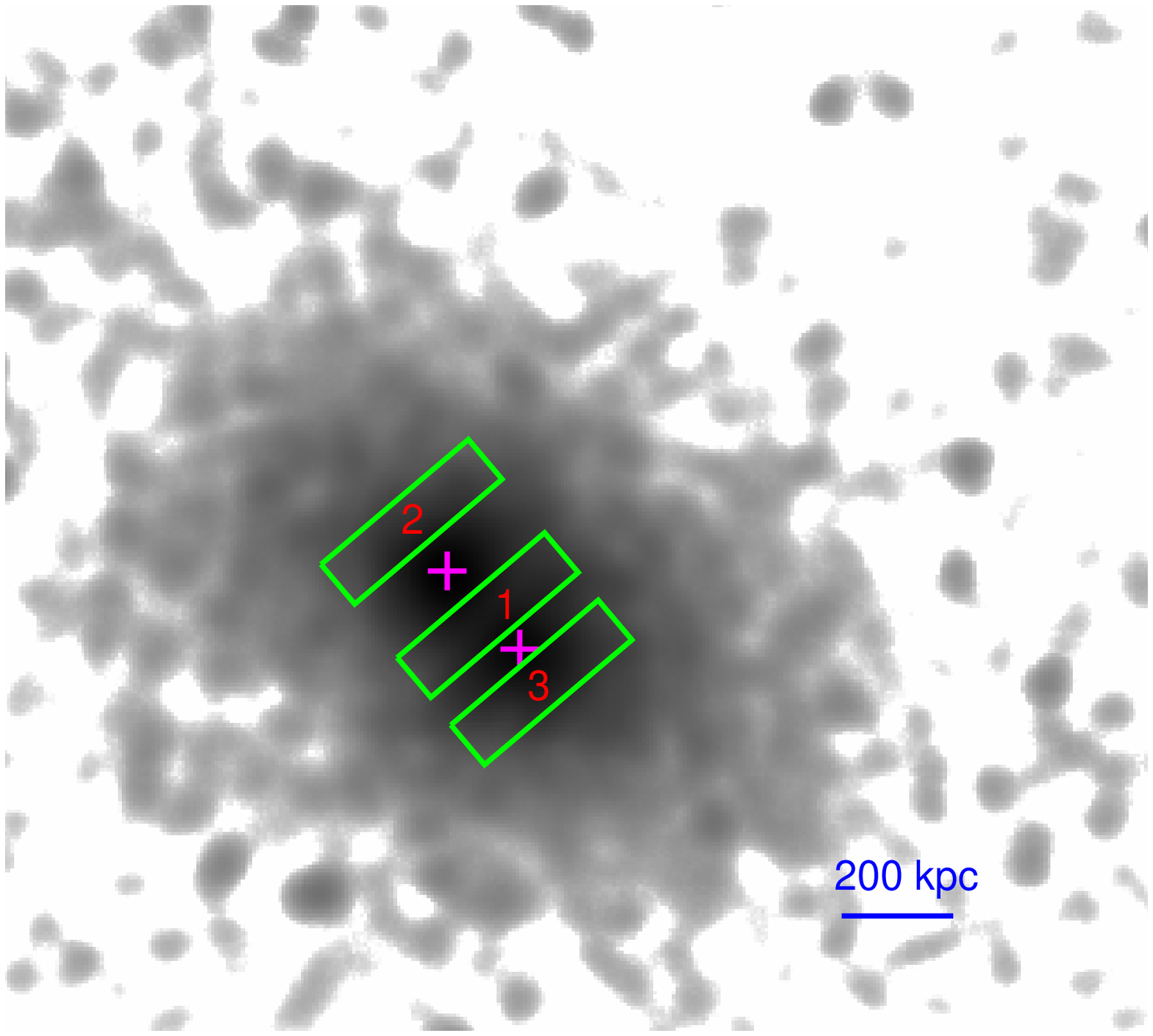}
\includegraphics[width=0.5\textwidth]{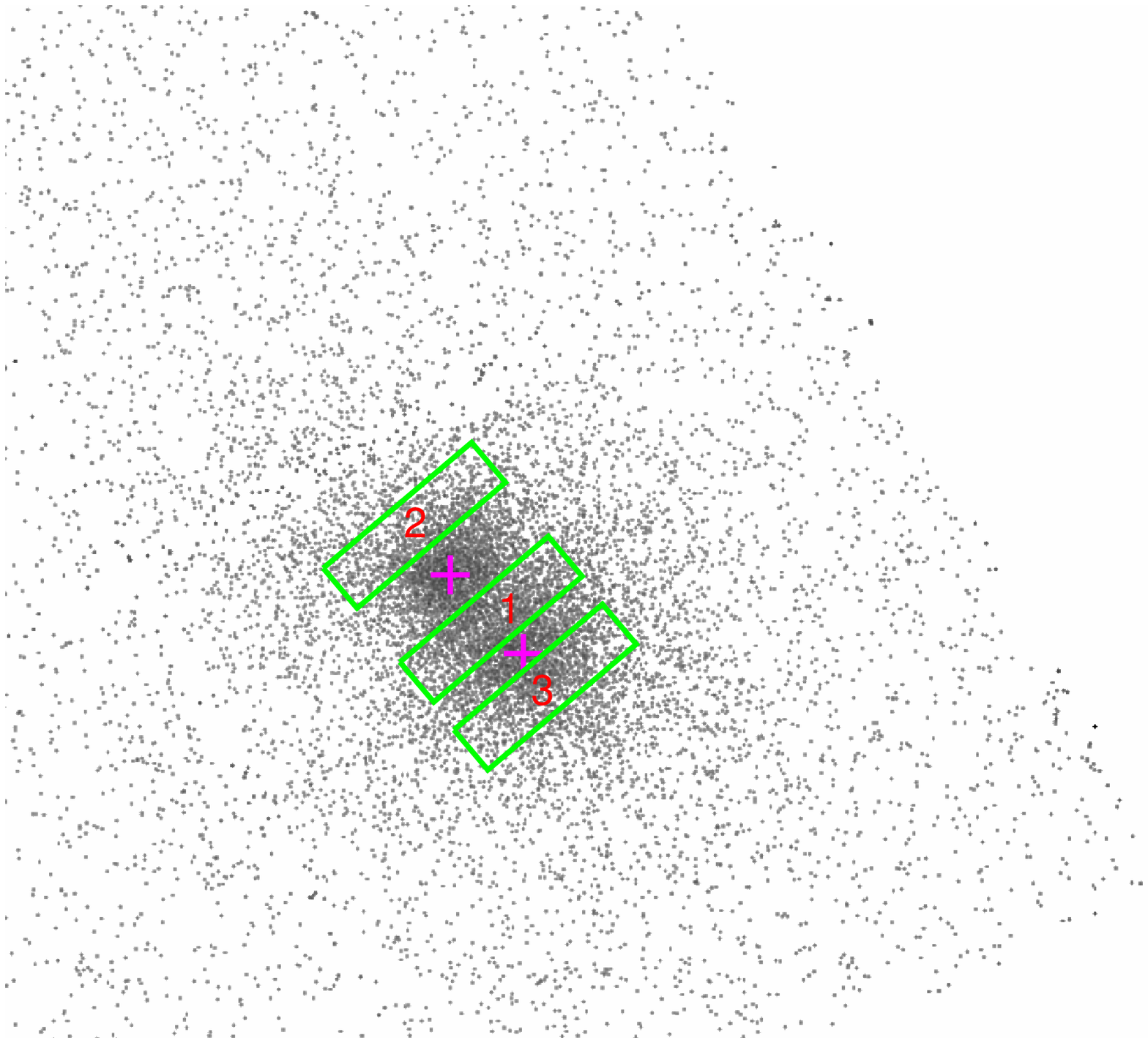}}
\caption{Regions used to determine the Mach number of the merger shock. Region 1 is the postshock region, and the symmetric region with respect to the subcluster center (magenta cross), Region 2 (3), is assumed to be the preshock region for G036N (G036S). \textit{Left}: Regions overlaid on the \xmm\ image. The lower right scale bar is 200 kpc. \textit{Right}: Regions overlaid on the \chandra\ image. This image has the same size as the left panel.
}
\end{figure*}
%%%%%%%%%%%%%%%%%%%%%%%%%%%%%

%%%%%%%%%%%%%%%%%%%%%%%%%%%%%
\begin{figure*}[hbt!]
\centerline{
\includegraphics[width=0.5\textwidth]{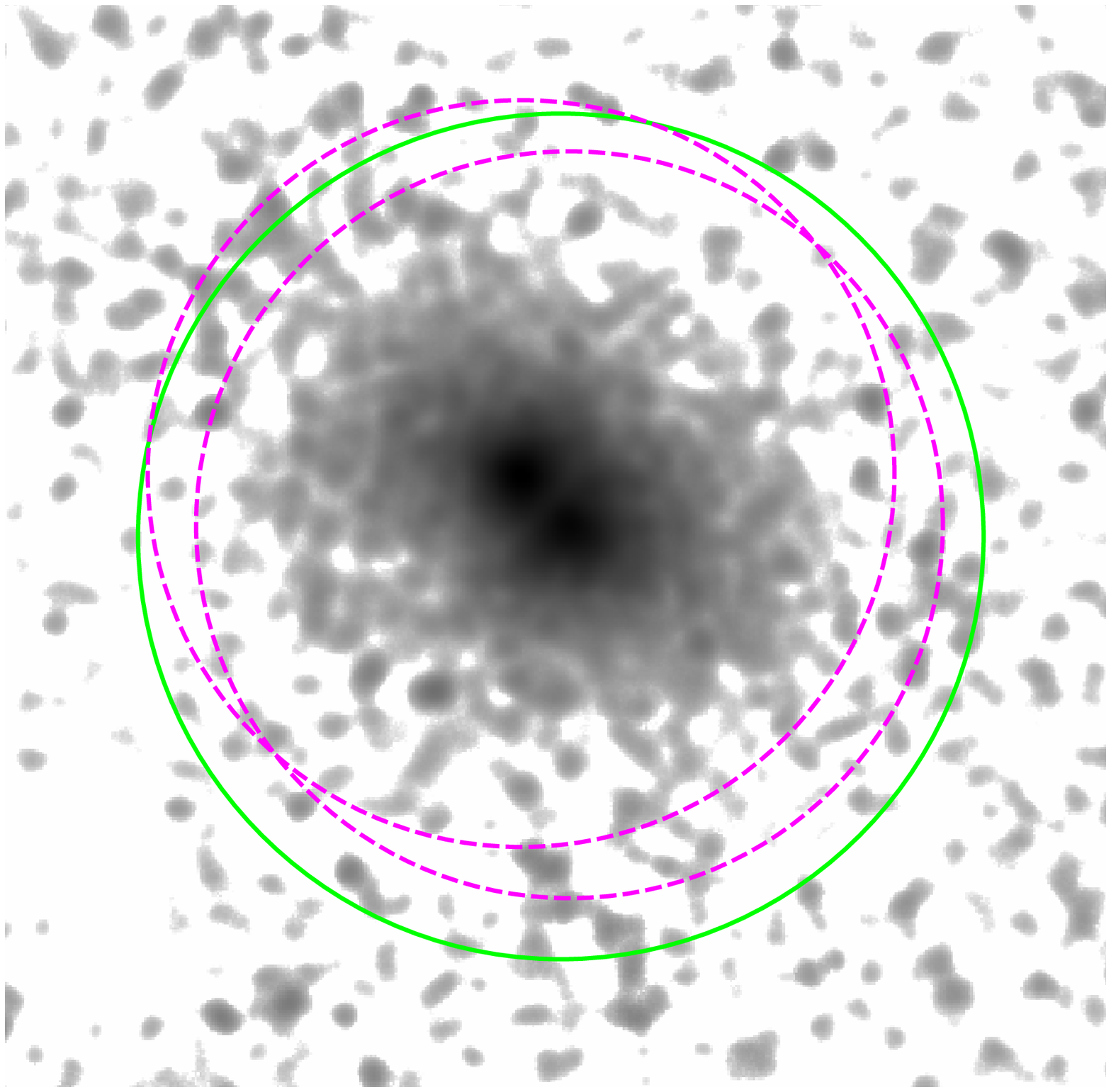}
\includegraphics[width=0.5\textwidth]{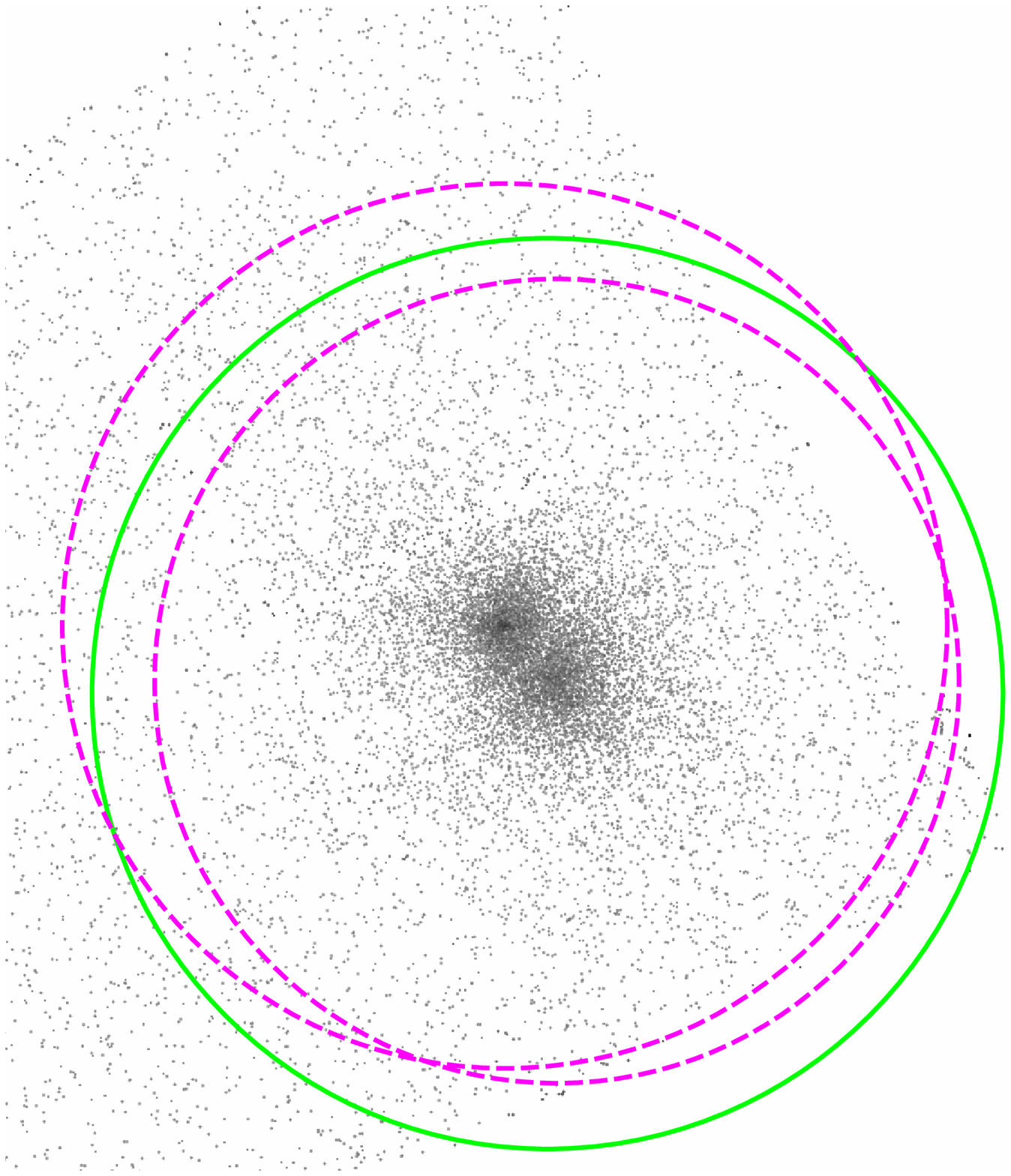}}
\caption{Regions used to determine the mass for both X-ray and SZ measurements. The green solid circle, with a radius of $r_{SZ,500}=1.12$ Mpc, is the region to measure the SZ mass by \planck , while the two magenta dashed circles, each with a radius of $r_{X,500}$, are the regions to ``measure'' (note that we only measure the mass in half circle and double it to obtain the total mass) the X-ray mass by \xmm\ or \chandra . The green solid circle is centered at the \planck\ position, while the magenta dashed circles are centered at the X-ray centroids for both G036N and G036S. \textit{Left}: \xmm\ regions. For both G036N and G036S, $r_{X,500}=1.0$ Mpc (see Table 3). The SZ region is larger by $\sim$$14\%$ than the X-ray region (the union of G036N and G036S). \textit{Right}: \chandra\ regions. For G036N, $r_{X,500}=1.1$ Mpc, while for G036S, $r_{X,500}=1.0$ Mpc (see Tables 3 and 5 for details). The SZ region is larger by $\sim$$3\%$ than the X-ray region.
}
\end{figure*}
%%%%%%%%%%%%%%%%%%%%%%%%%%%%%

\clearpage

\appendix
\section{TEMPERATURE, GAS MASS, TOTAL MASS, AND GAS MASS FRACTION PROFILES}

We present the model 2D and 3D temperature profiles, gas mass and total mass profiles, and gas mass fraction profiles in this Appendix. Sections 4.4 and 4.5 provide the details on how to obtain these profiles.

%%%%%%%%%%%%%%%%%%%%%%%%%%%%%
\begin{figure*}[hbt!]
\centerline{
\includegraphics[width=0.25\textwidth, angle=-90]{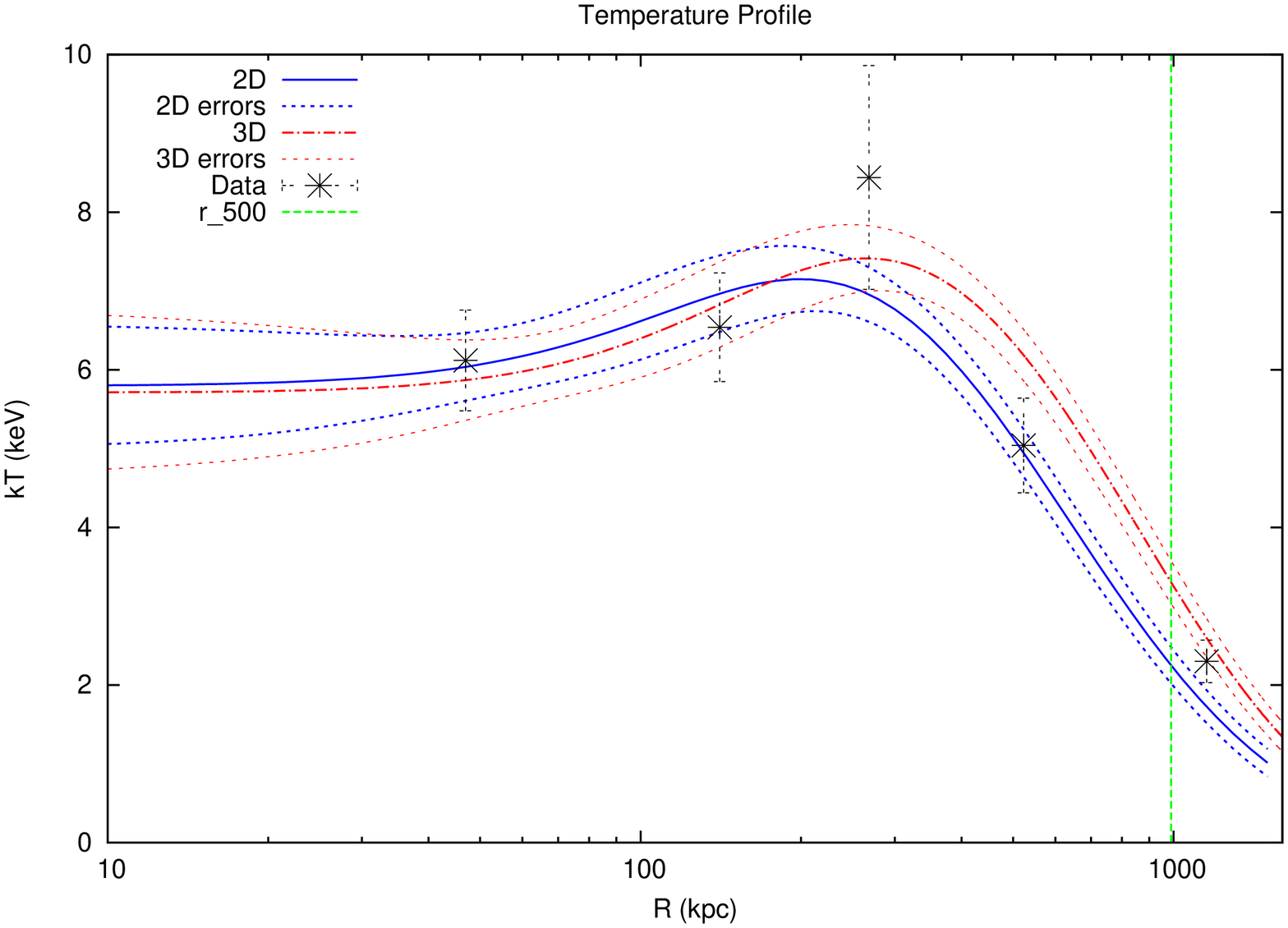}
\includegraphics[width=0.25\textwidth, angle=-90]{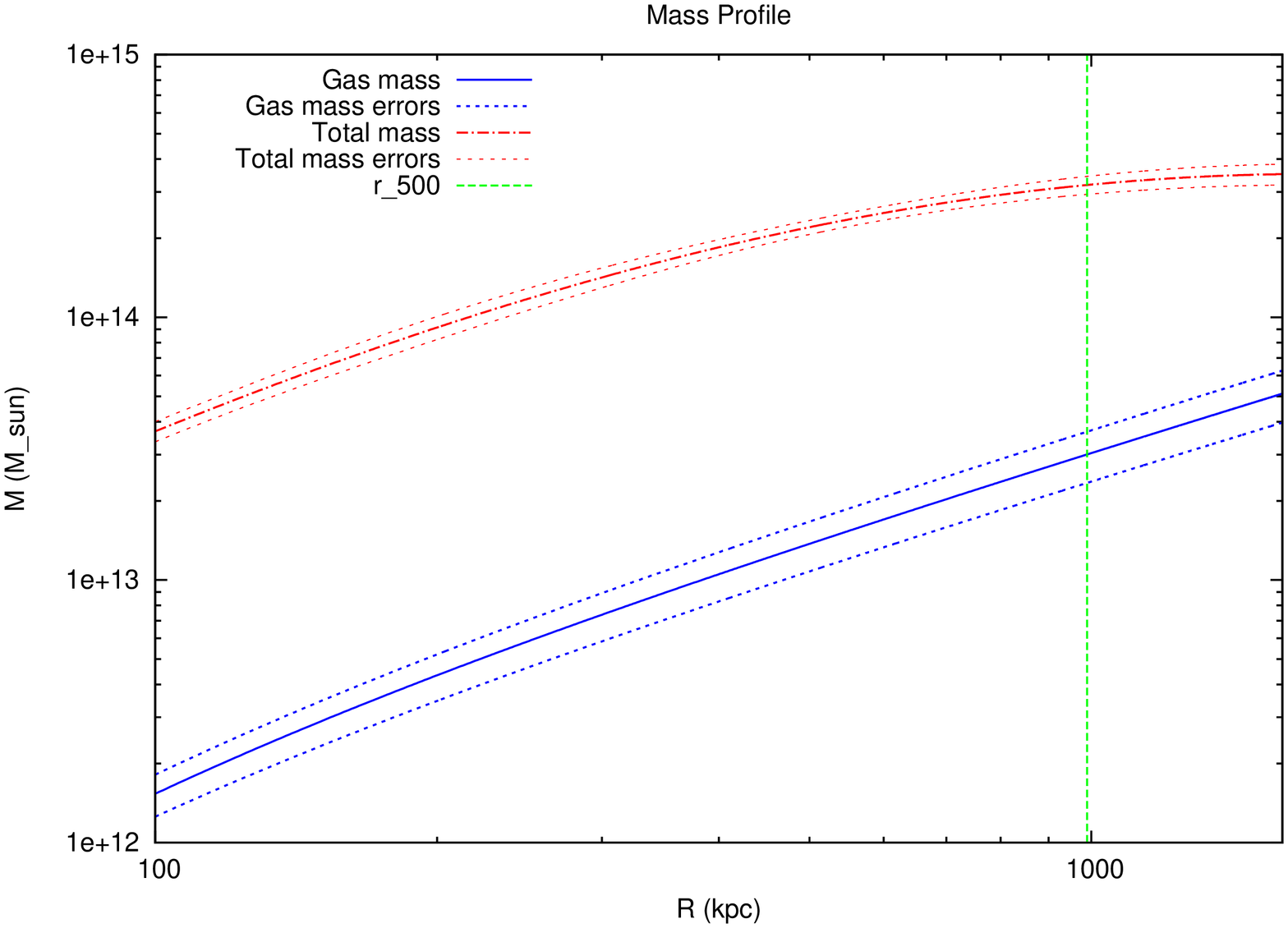}
\includegraphics[width=0.25\textwidth, angle=-90]{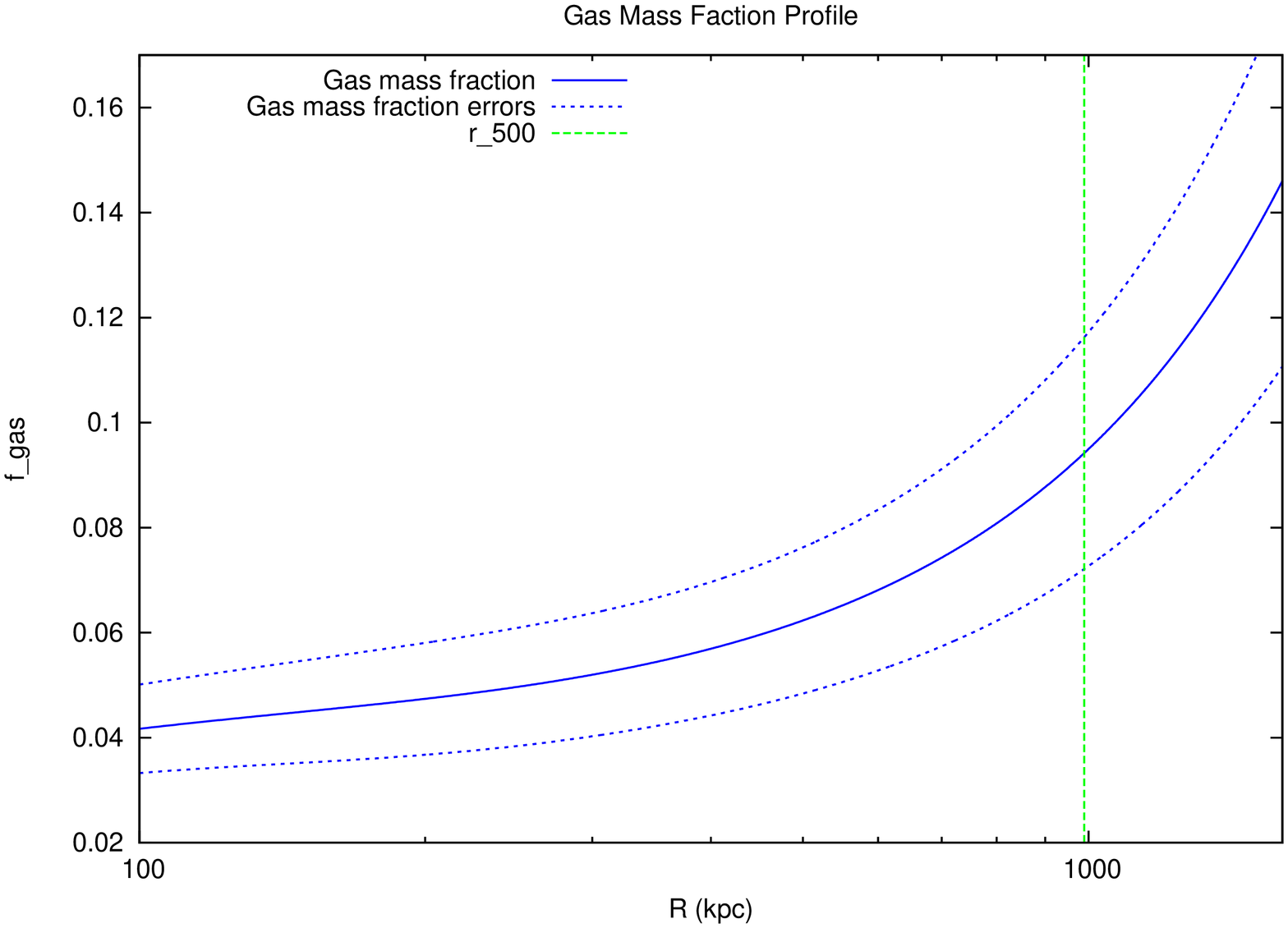}}
\caption{Temperature, gas mass and total mass, and gas mass fraction profiles for G036N obtained by \xmm , with $N_H=0.136\times10^{22}$ cm$^{-2}$ (total value). The vertical green dashed line marks $r_{500}$ for all three panels. \textit{Left}: Temperature profiles. The blue solid and blue dotted lines are the best fitting 2D temperature profile and the corresponding $1\sigma$ errors, while the red dash-dotted and red short-dashed lines are the best fitting 3D temperature profiles and the corresponding errors. The data points with errors are the measured 2D temperatures in each bin. \textit{Middle}: Gas mass and total mass profiles. The upper red dash-dotted and red short-dashed lines are the total mass profile and the corresponding errors, while the lower blue solid and blue dotted lines are the gas mass and the corresponding errors. \textit{Right}: Gas mass fraction profile. The blue solid and blue dotted lines are the gas mass fraction profile and the corresponding errors. 
}
\end{figure*}
%%%%%%%%%%%%%%%%%%%%%%%%%%%%%

%%%%%%%%%%%%%%%%%%%%%%%%%%%%%
\begin{figure*}[hbt!]
\centerline{
\includegraphics[width=0.25\textwidth, angle=-90]{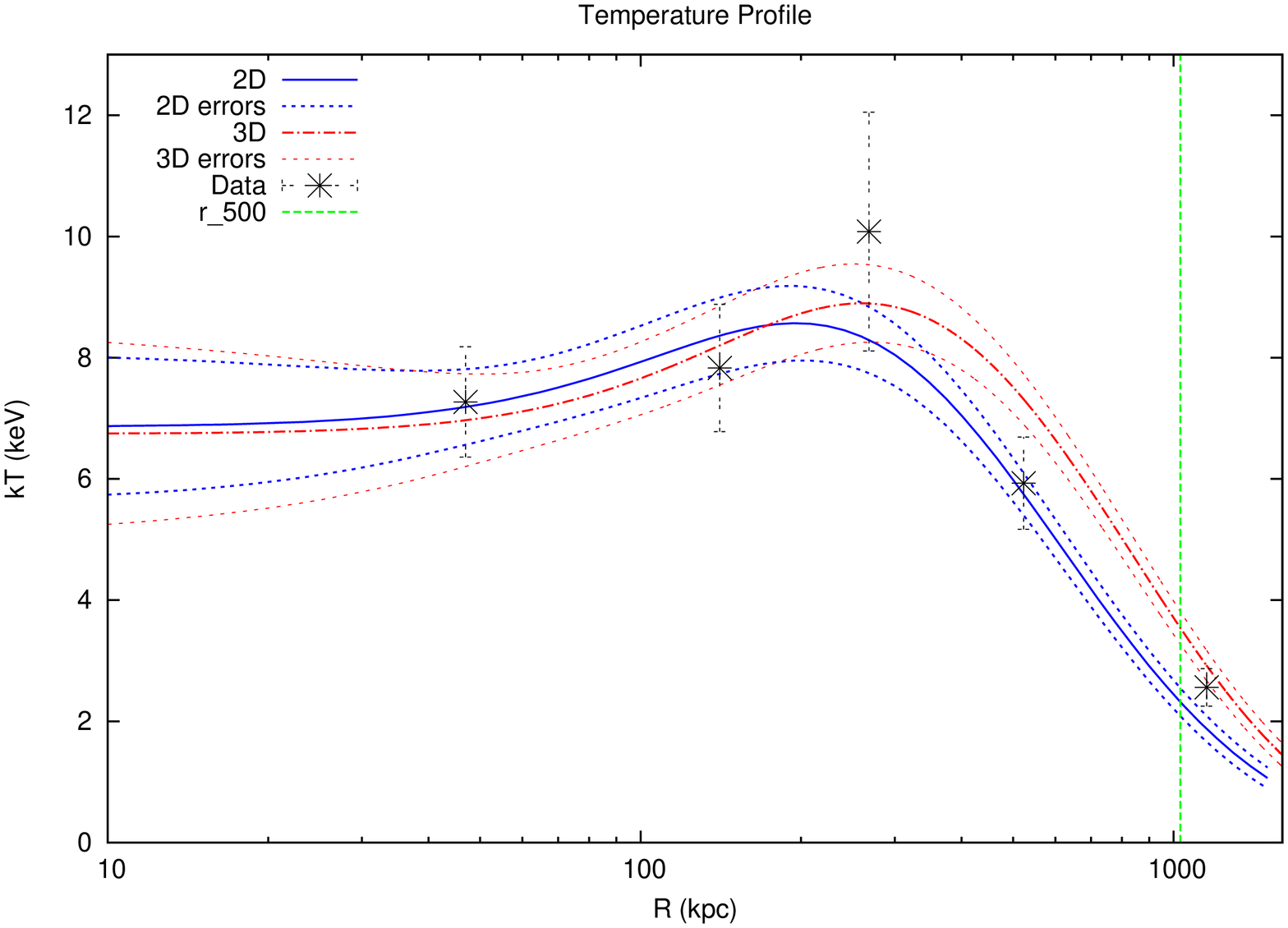}
\includegraphics[width=0.25\textwidth, angle=-90]{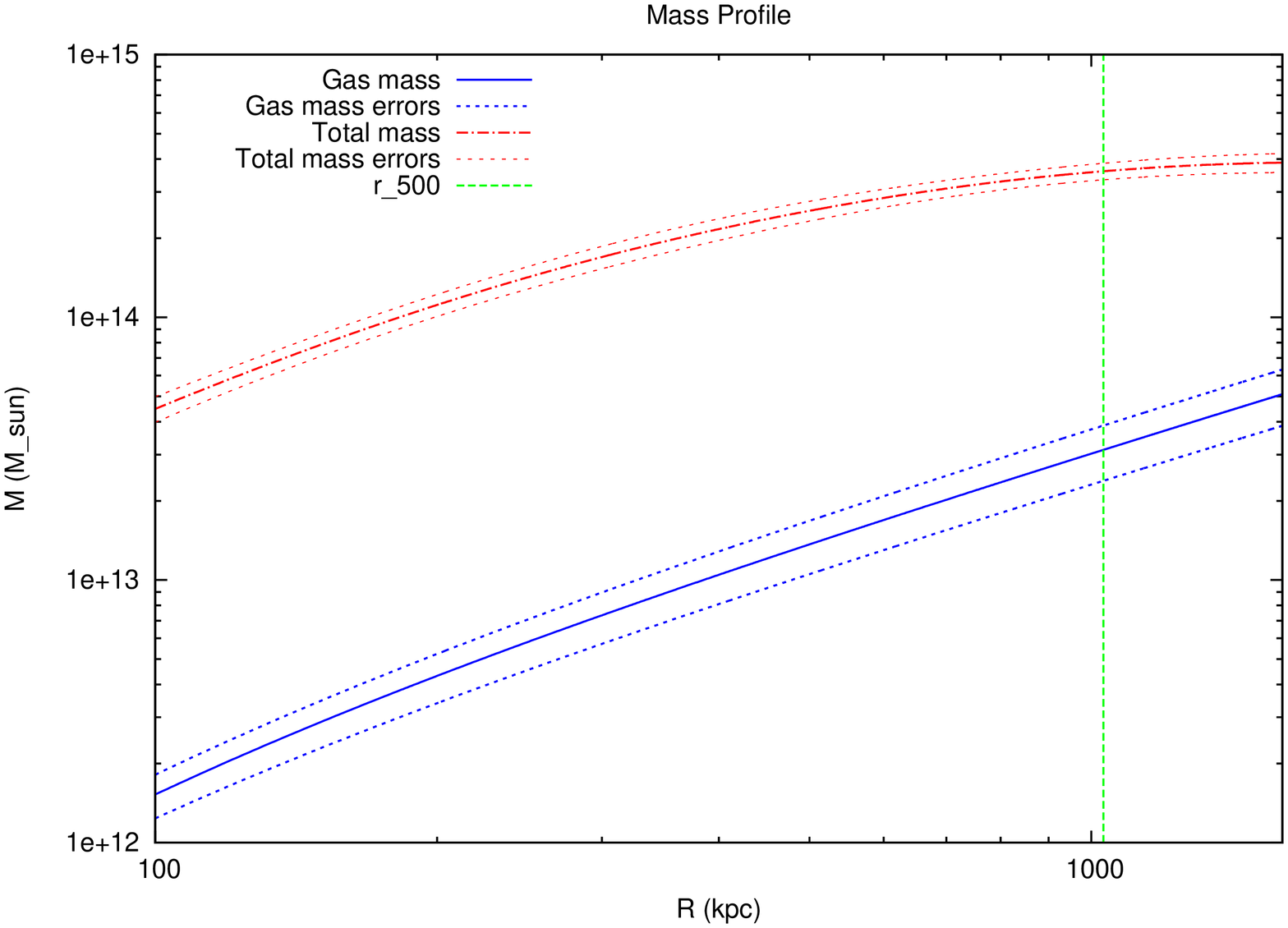}
\includegraphics[width=0.25\textwidth, angle=-90]{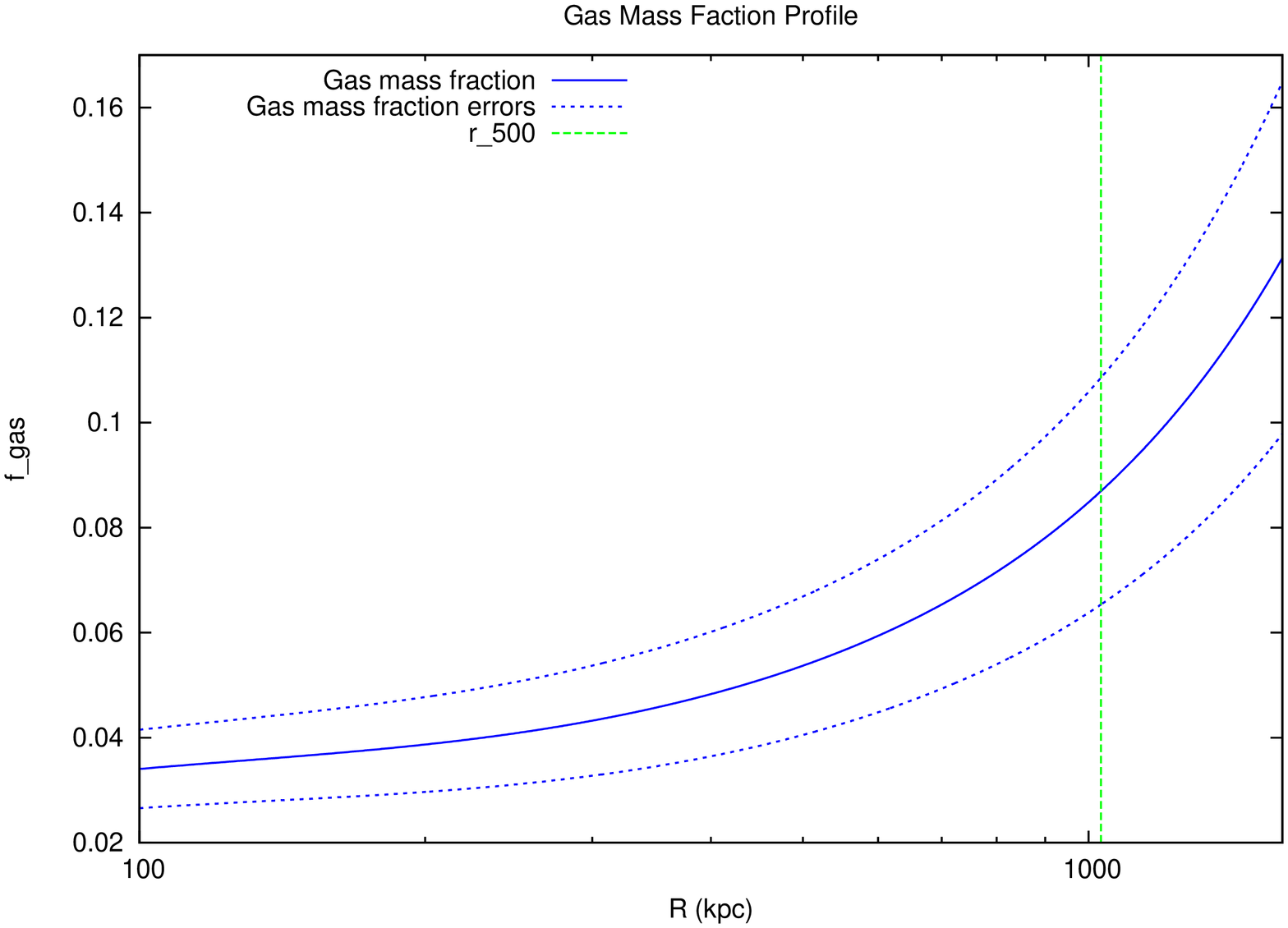}}
\caption{Same as Figure A9, except for $N_H=0.092\times10^{22}$ cm$^{-2}$ (best fitting value).
}
\end{figure*}
%%%%%%%%%%%%%%%%%%%%%%%%%%%%%

%%%%%%%%%%%%%%%%%%%%%%%%%%%%%
\begin{figure*}[hbt!]
\centerline{
\includegraphics[width=0.25\textwidth, angle=-90]{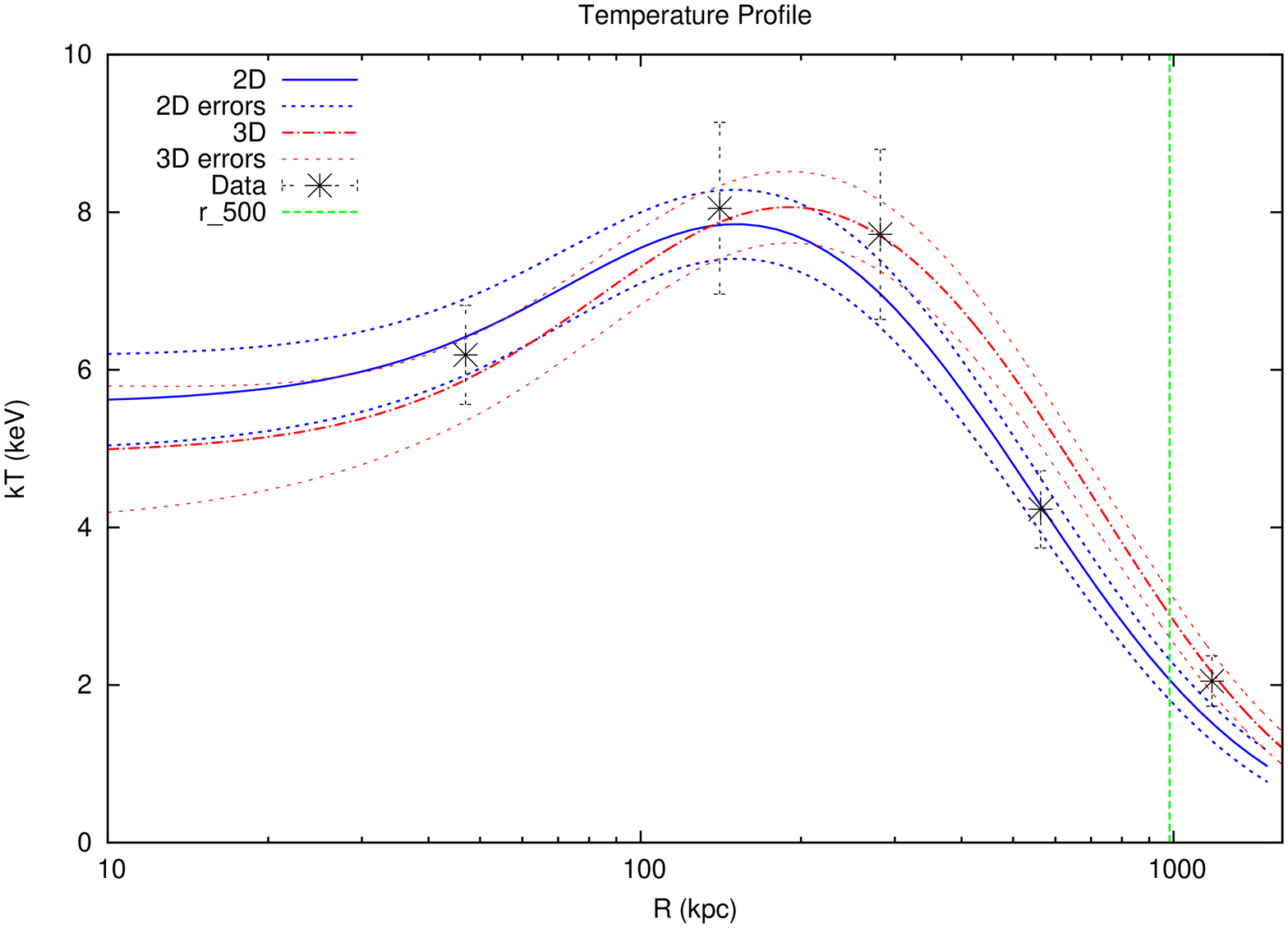}
\includegraphics[width=0.25\textwidth, angle=-90]{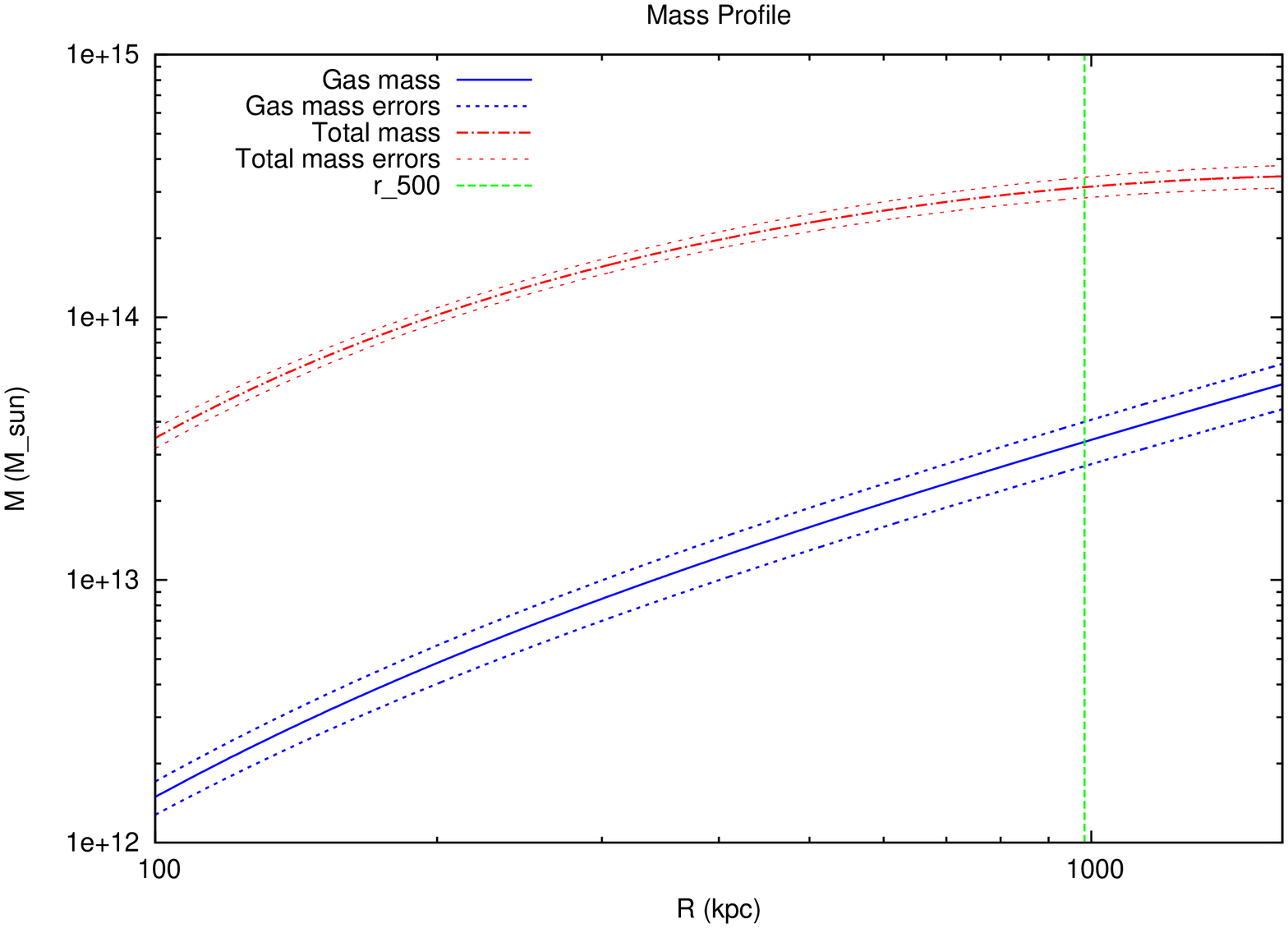}
\includegraphics[width=0.25\textwidth, angle=-90]{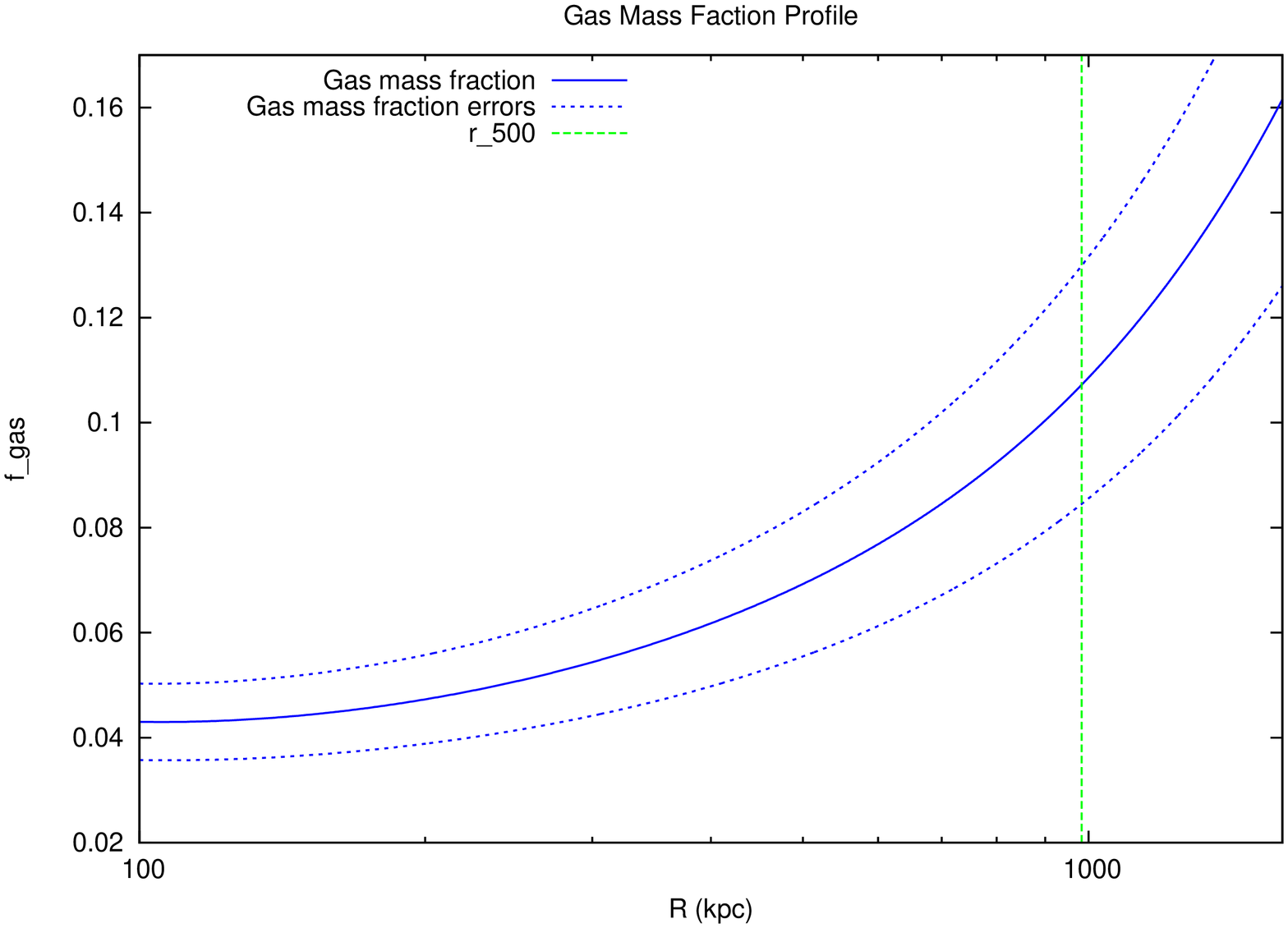}}
\caption{Same as Figure A9, except for G036S.
}
\end{figure*}
%%%%%%%%%%%%%%%%%%%%%%%%%%%%%

%%%%%%%%%%%%%%%%%%%%%%%%%%%%%
\begin{figure*}[hbt!]
\centerline{
\includegraphics[width=0.25\textwidth, angle=-90]{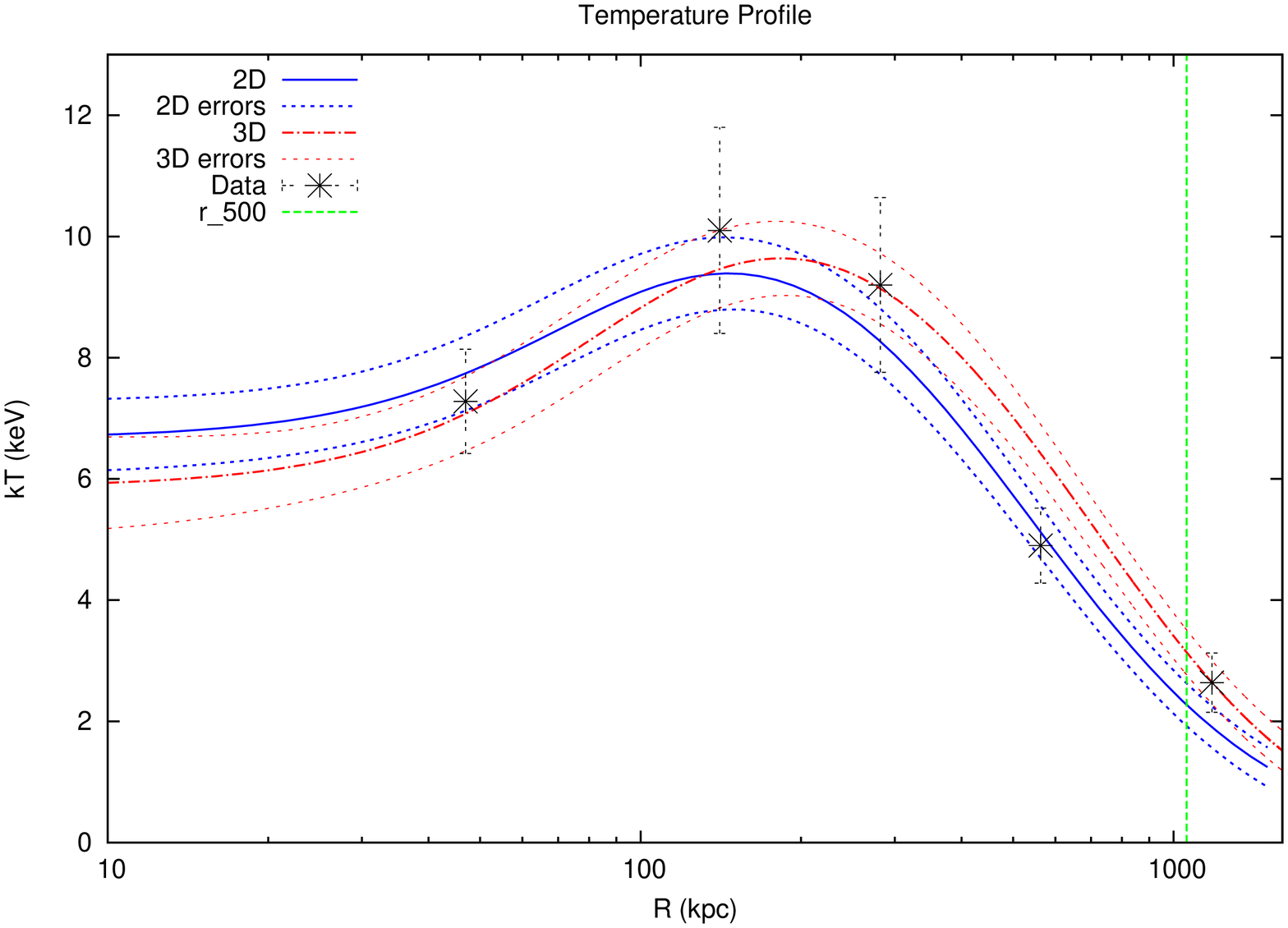}
\includegraphics[width=0.25\textwidth, angle=-90]{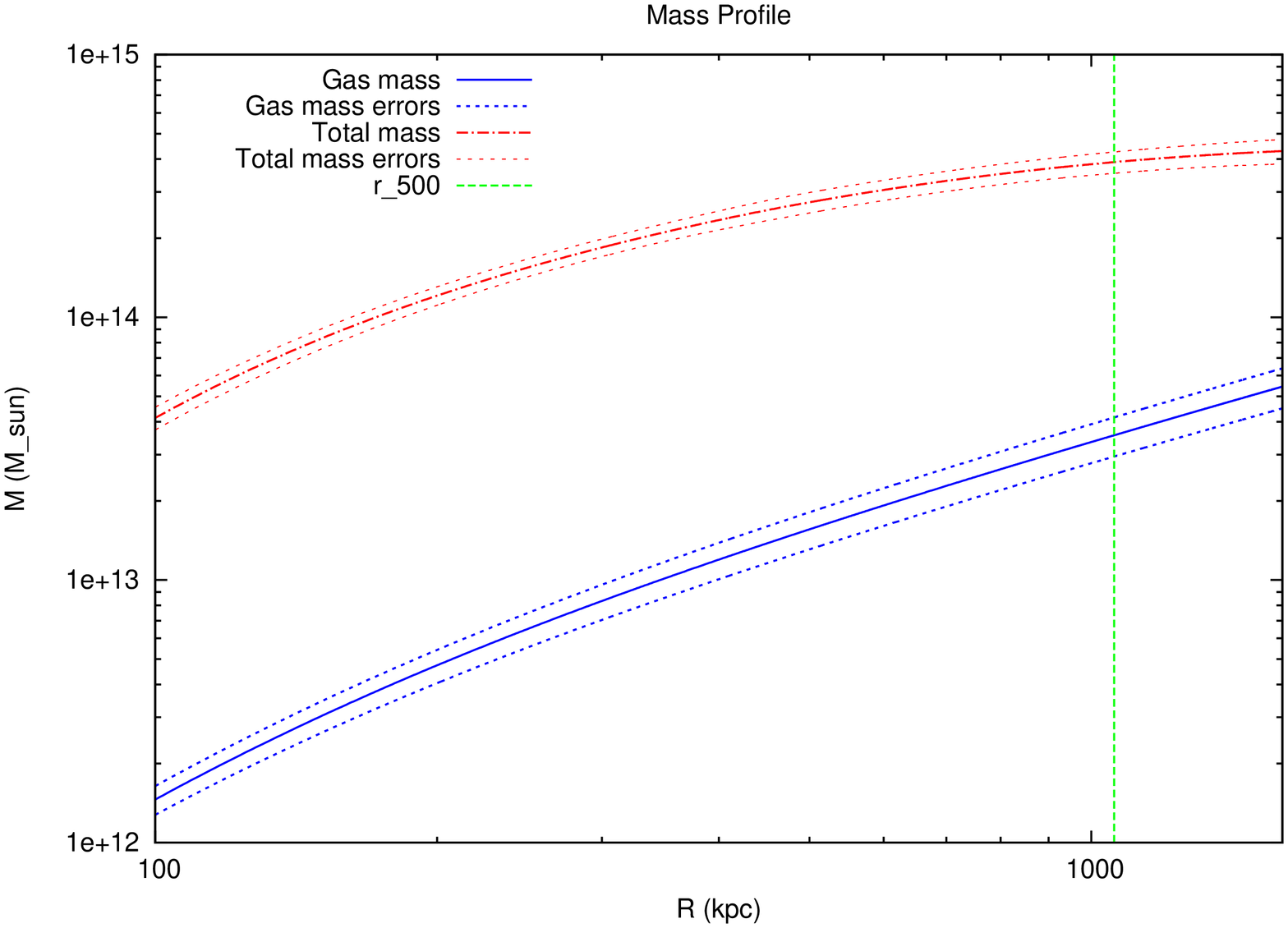}
\includegraphics[width=0.25\textwidth, angle=-90]{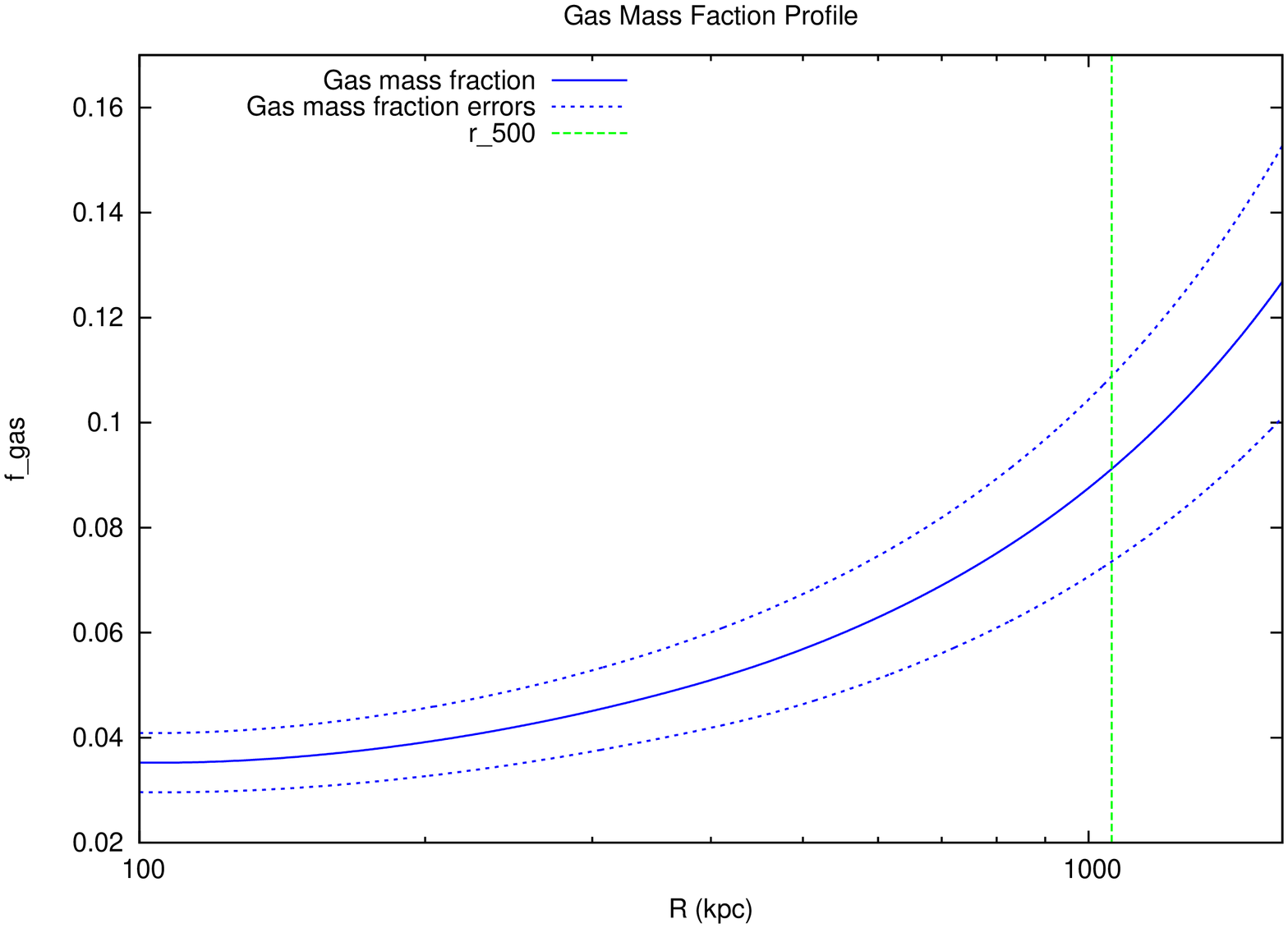}}
\caption{Same as Figure A9, except for G036S and $N_H=0.092\times10^{22}$ cm$^{-2}$ (best fitting value).
}
\end{figure*}
%%%%%%%%%%%%%%%%%%%%%%%%%%%%%

%%%%%%%%%%%%%%%%%%%%%%%%%%%%%
\begin{figure*}[hbt!]
\centerline{
\includegraphics[width=0.25\textwidth, angle=-90]{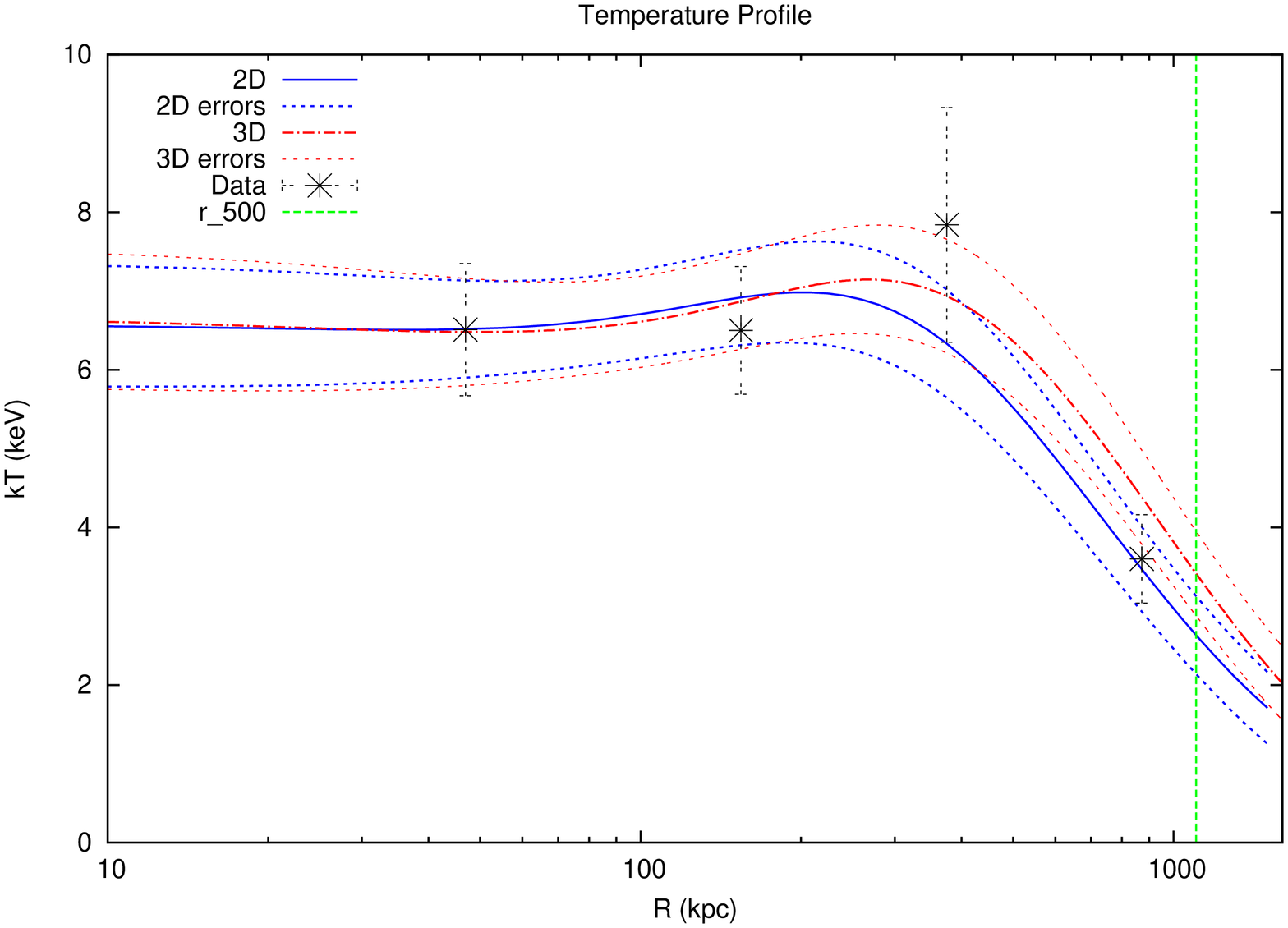}
\includegraphics[width=0.25\textwidth, angle=-90]{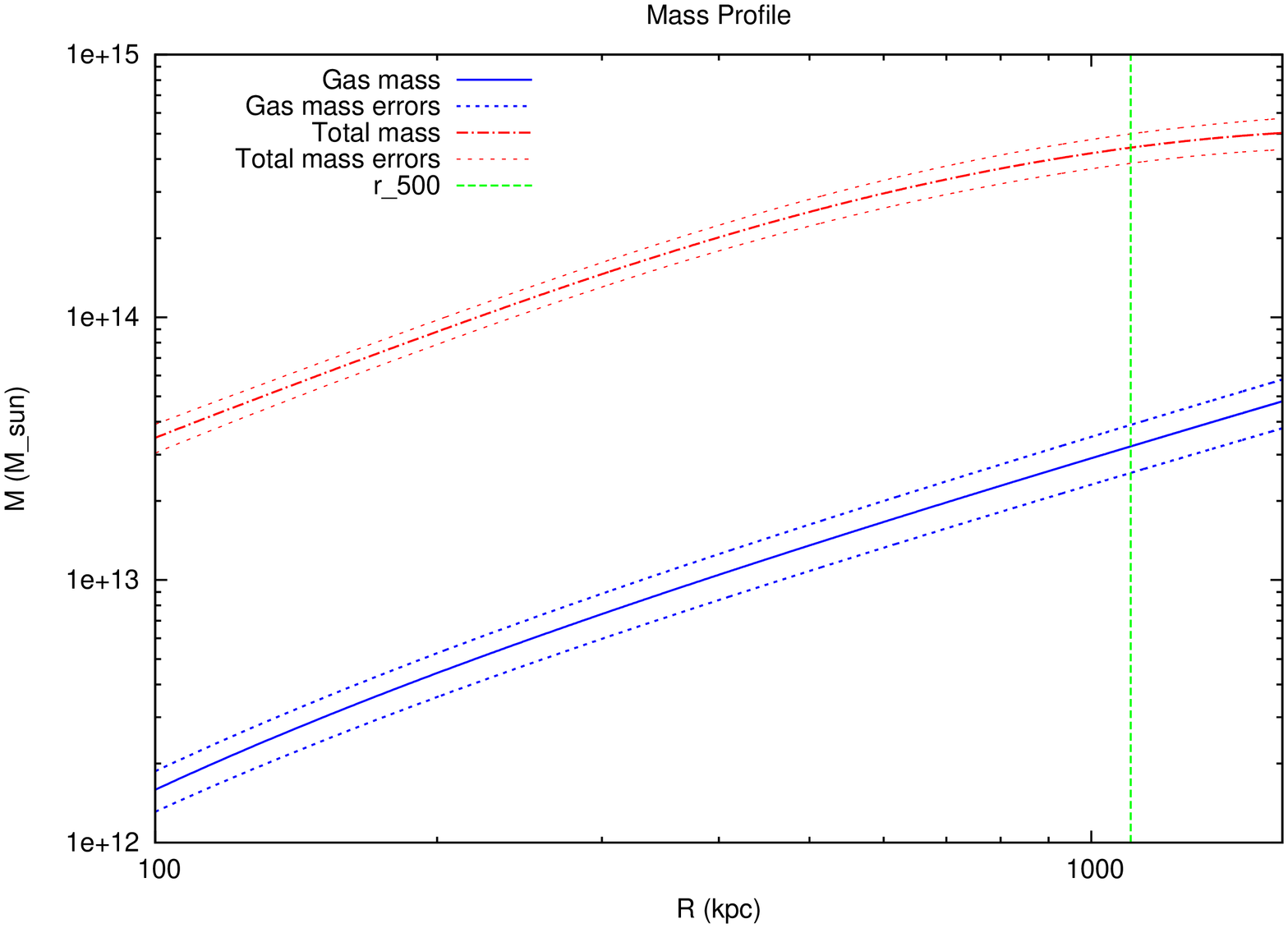}
\includegraphics[width=0.25\textwidth, angle=-90]{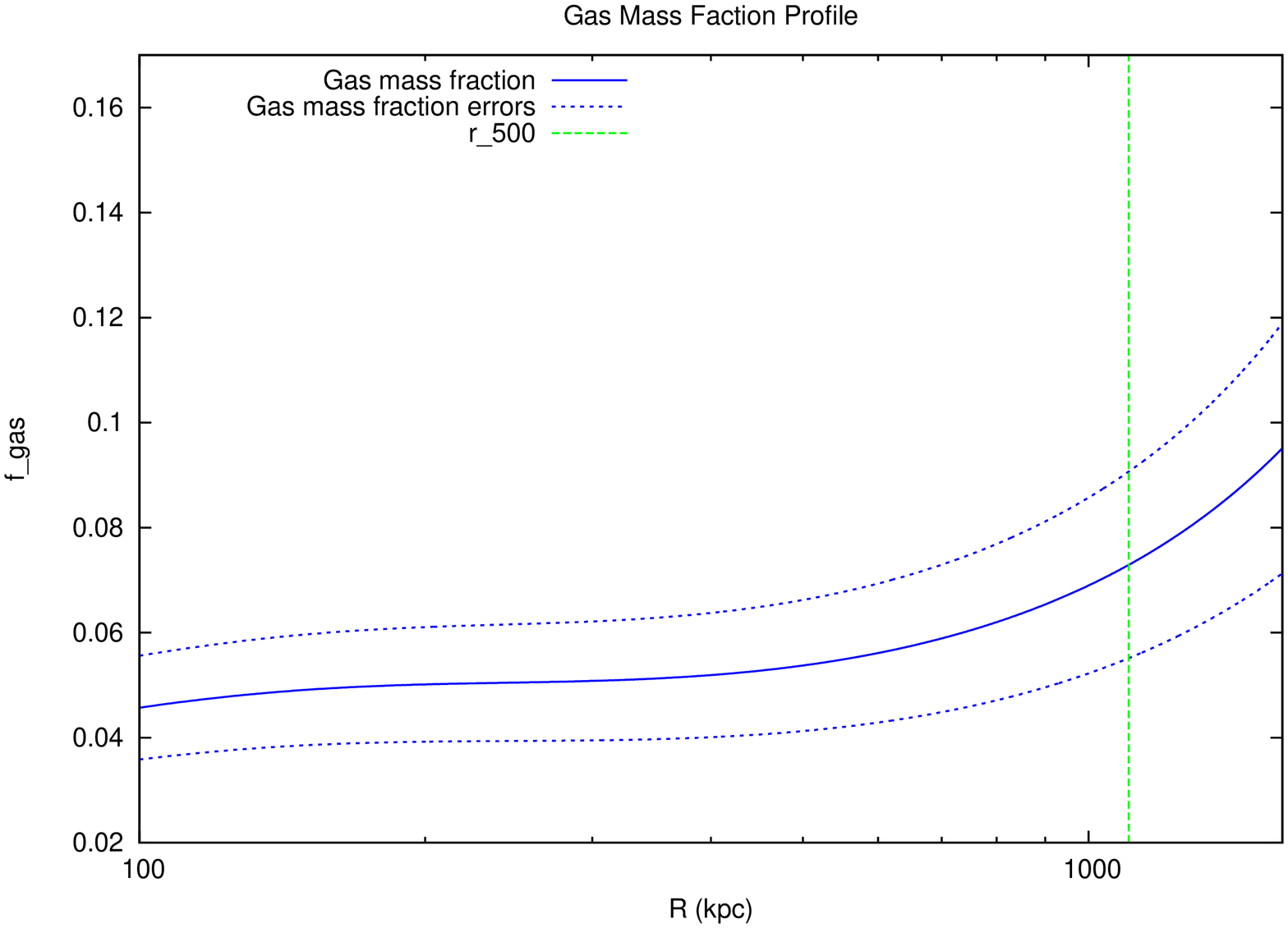}}
\caption{Same as Figure A9, except for \chandra\ baseline background model.
}
\end{figure*}
%%%%%%%%%%%%%%%%%%%%%%%%%%%%%

%%%%%%%%%%%%%%%%%%%%%%%%%%%%%
\begin{figure*}[hbt!]
\centerline{
\includegraphics[width=0.25\textwidth, angle=-90]{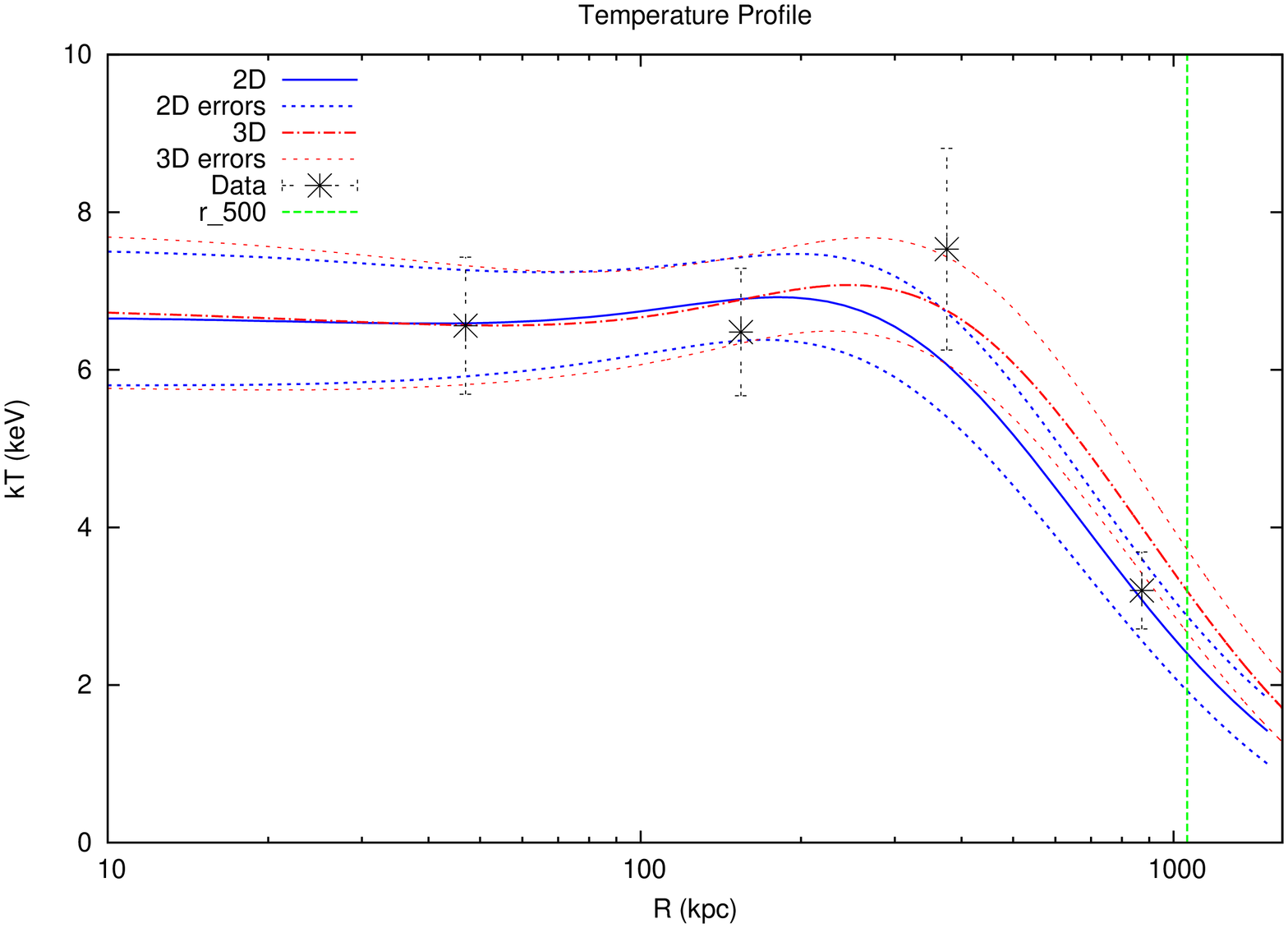}
\includegraphics[width=0.25\textwidth, angle=-90]{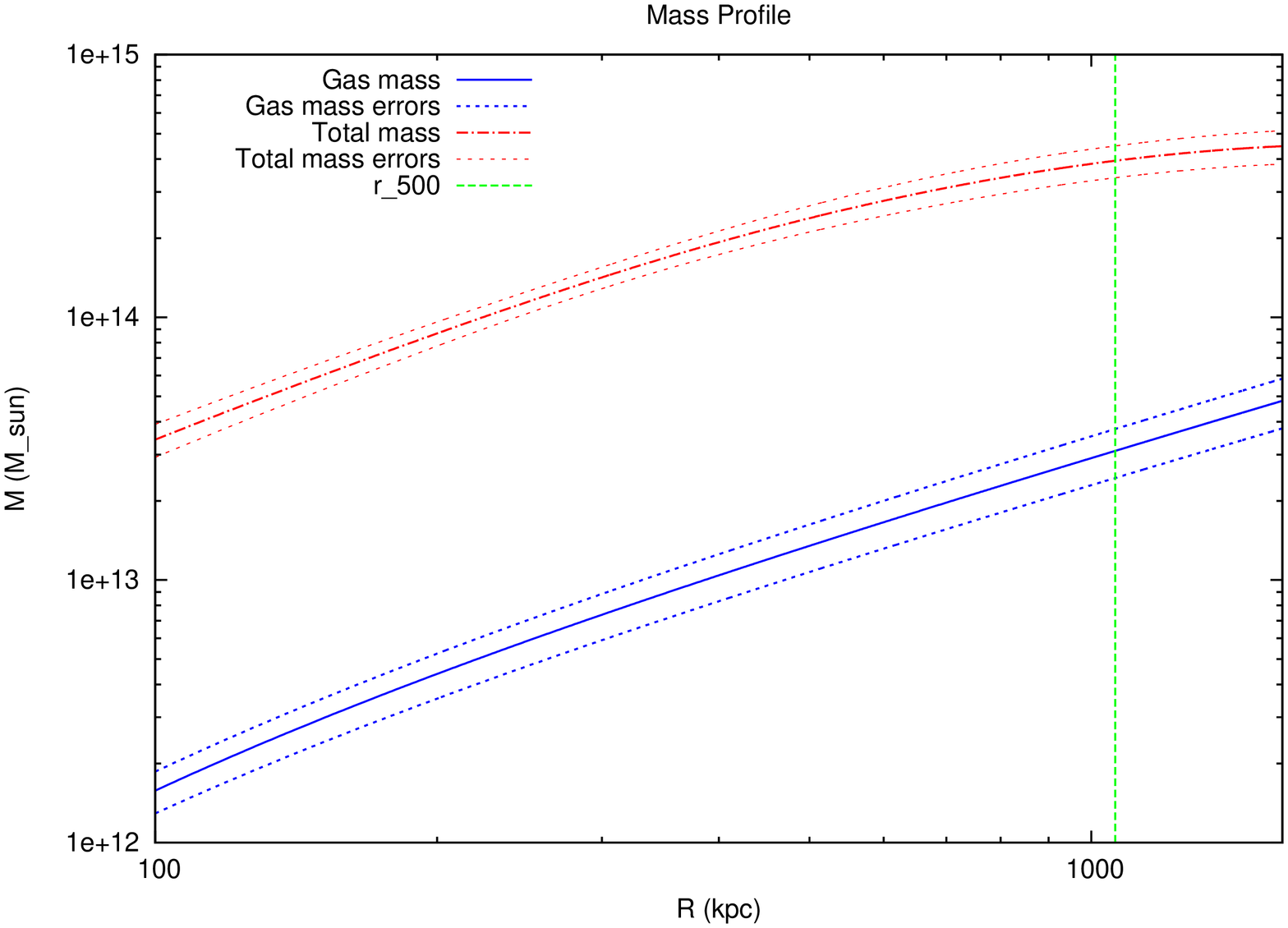}
\includegraphics[width=0.25\textwidth, angle=-90]{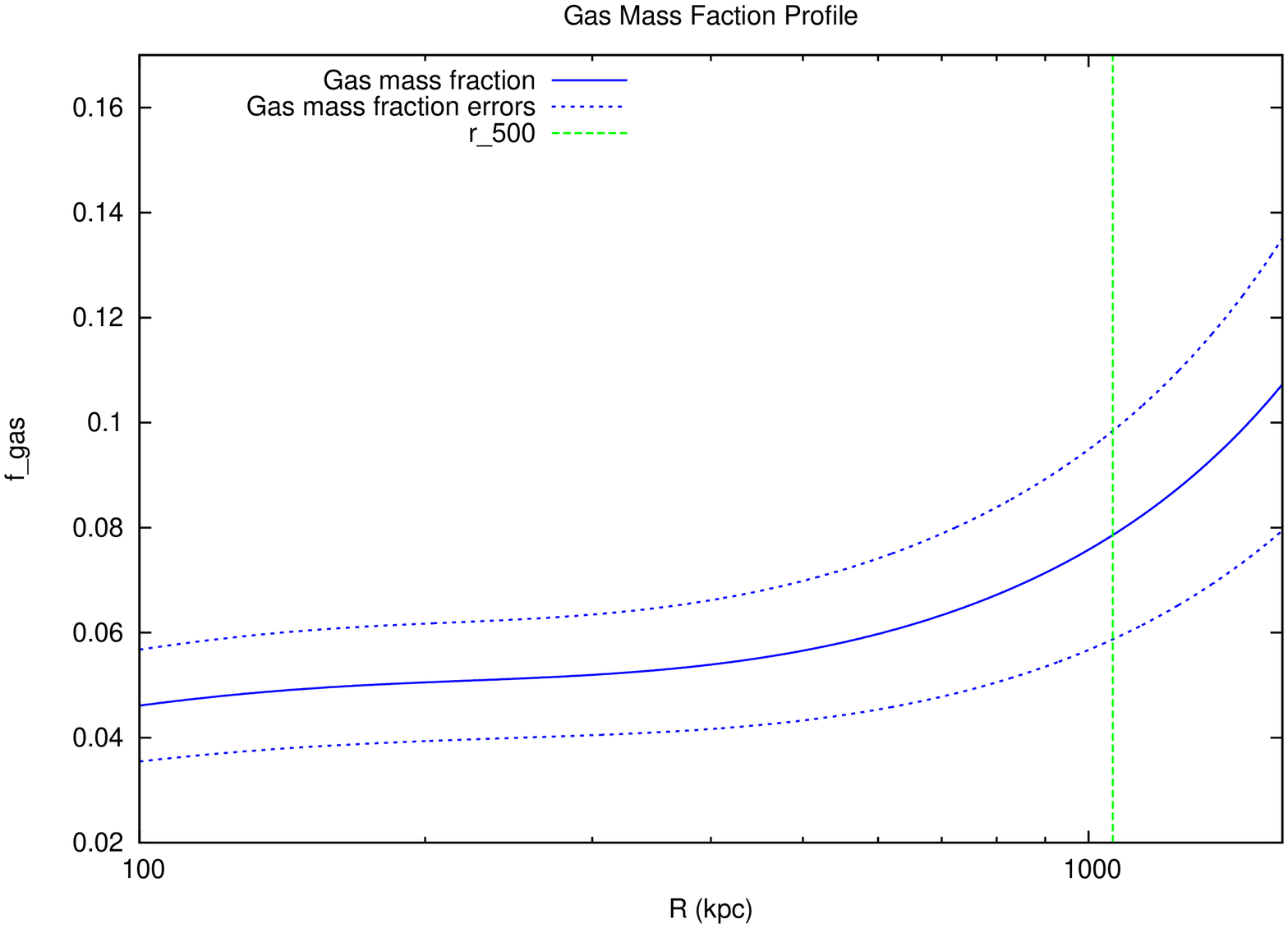}}
\caption{Same as Figure A9, except for \chandra\ baseline background model varying by $+5\%$. 
}
\end{figure*}
%%%%%%%%%%%%%%%%%%%%%%%%%%%%%

%%%%%%%%%%%%%%%%%%%%%%%%%%%%%
\begin{figure*}[hbt!]
\centerline{
\includegraphics[width=0.25\textwidth, angle=-90]{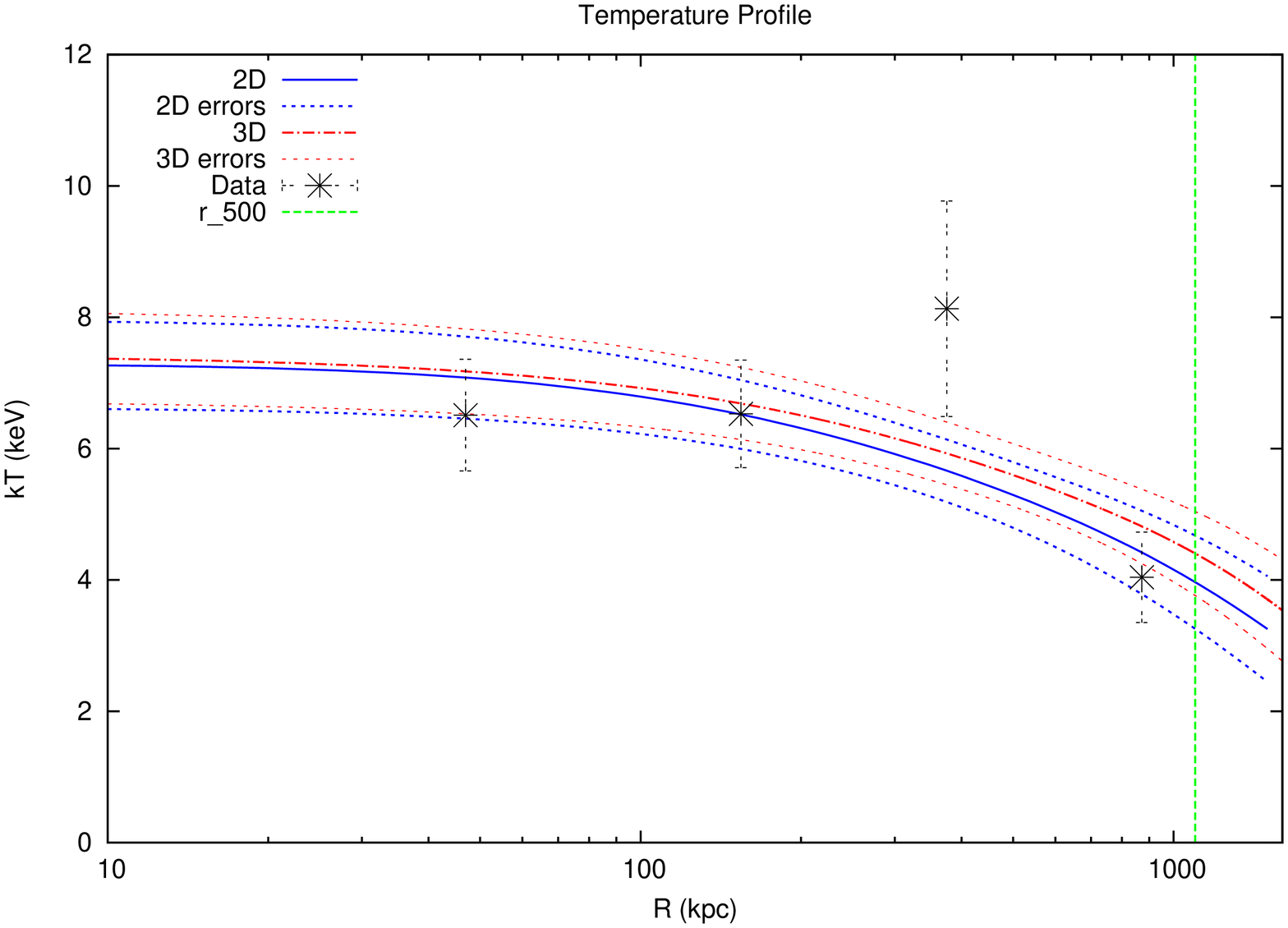}
\includegraphics[width=0.25\textwidth, angle=-90]{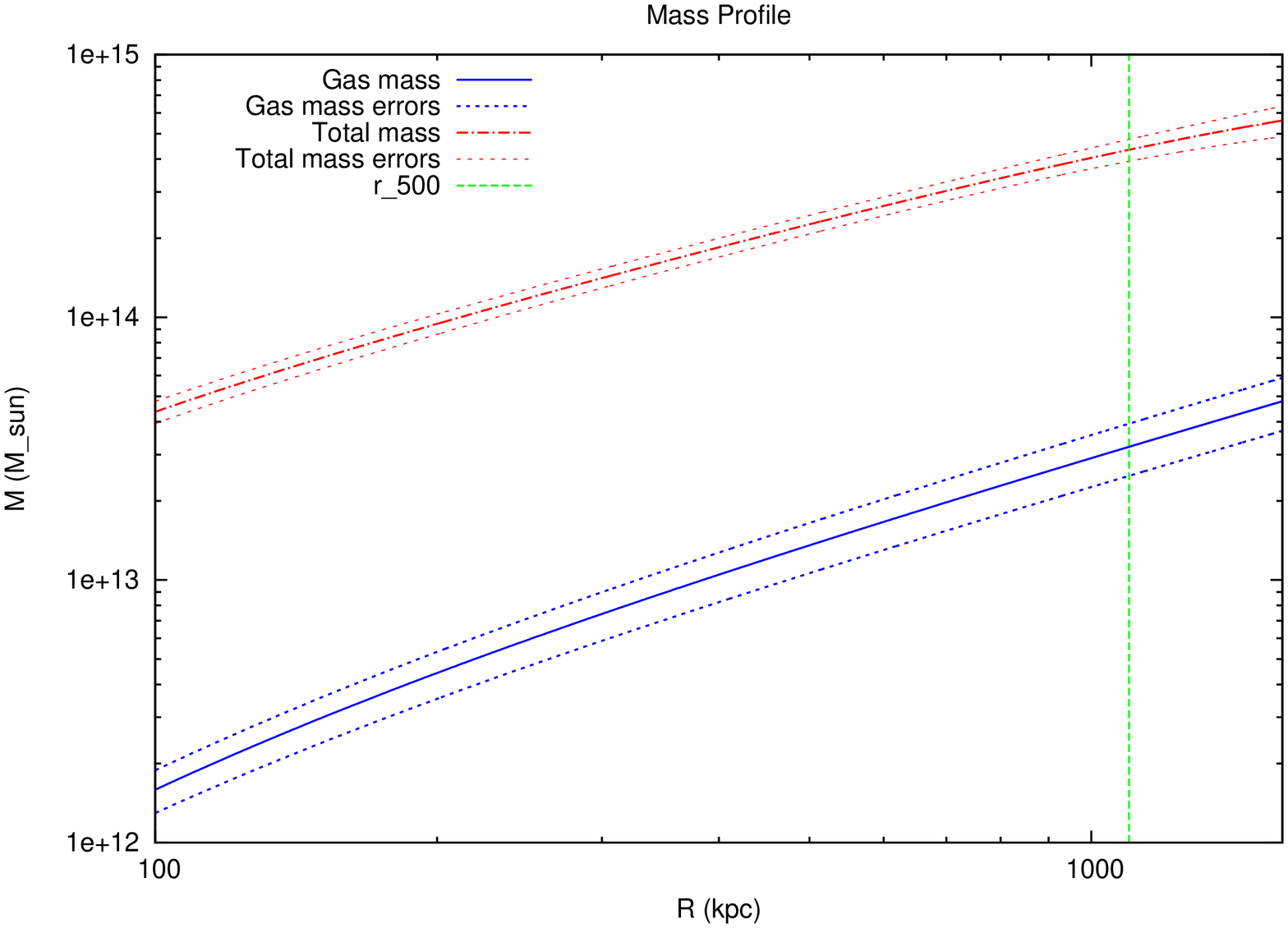}
\includegraphics[width=0.25\textwidth, angle=-90]{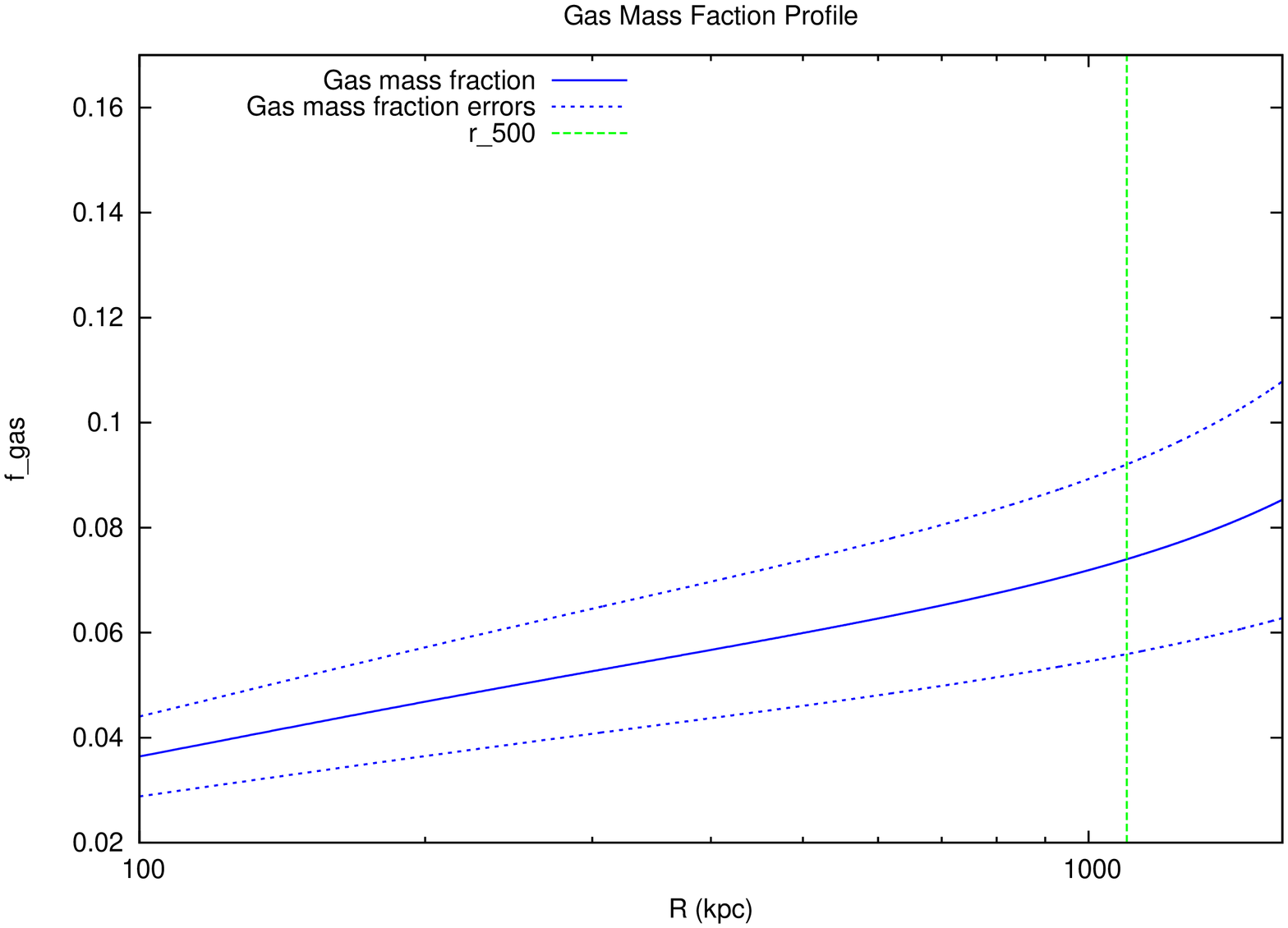}}
\caption{Same as Figure A9, except for \chandra\ baseline background model varying by $-5\%$. 
}
\end{figure*}
%%%%%%%%%%%%%%%%%%%%%%%%%%%%%

%%%%%%%%%%%%%%%%%%%%%%%%%%%%%
\begin{figure*}[hbt!]
\centerline{
\includegraphics[width=0.25\textwidth, angle=-90]{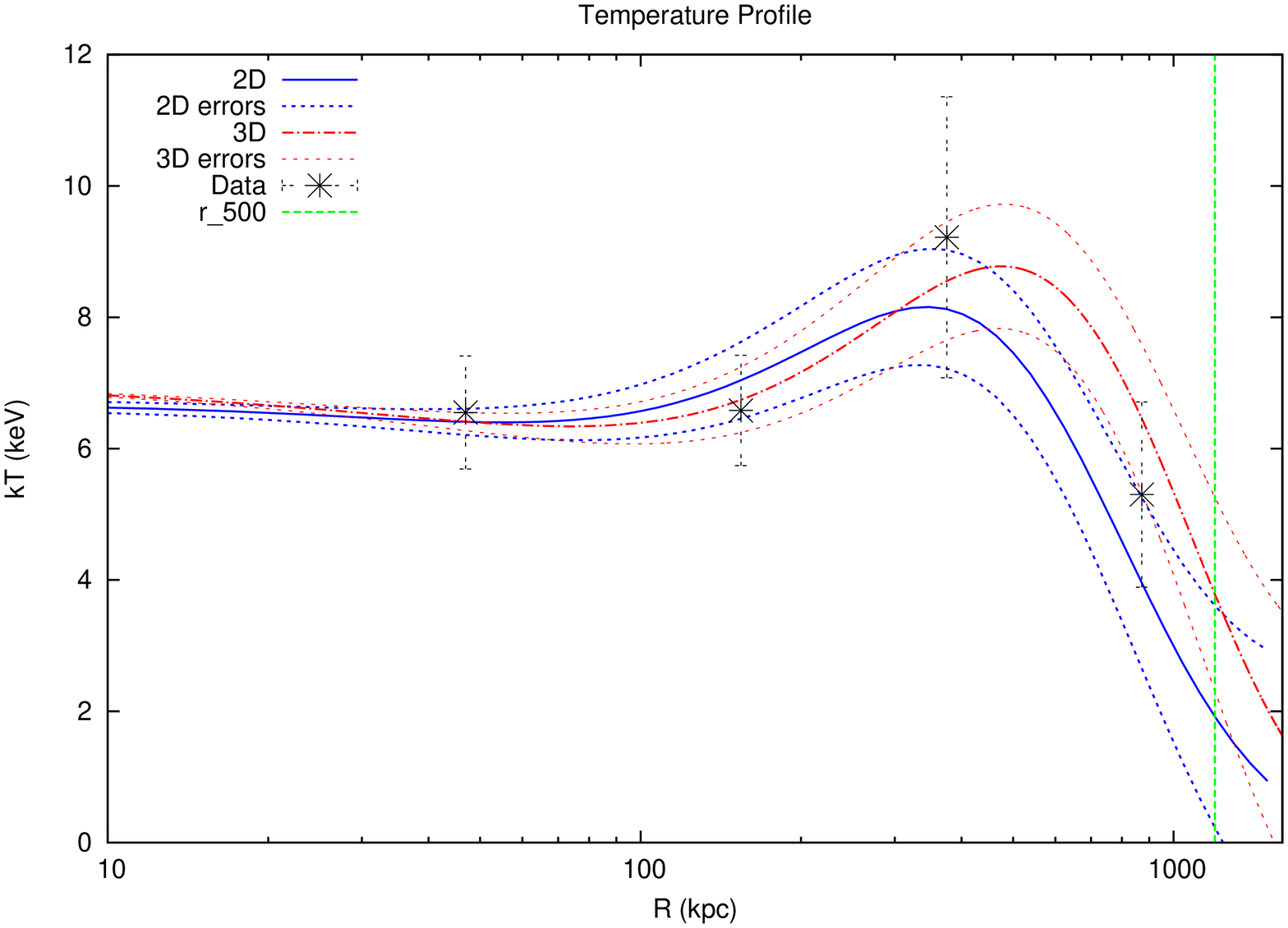}
\includegraphics[width=0.25\textwidth, angle=-90]{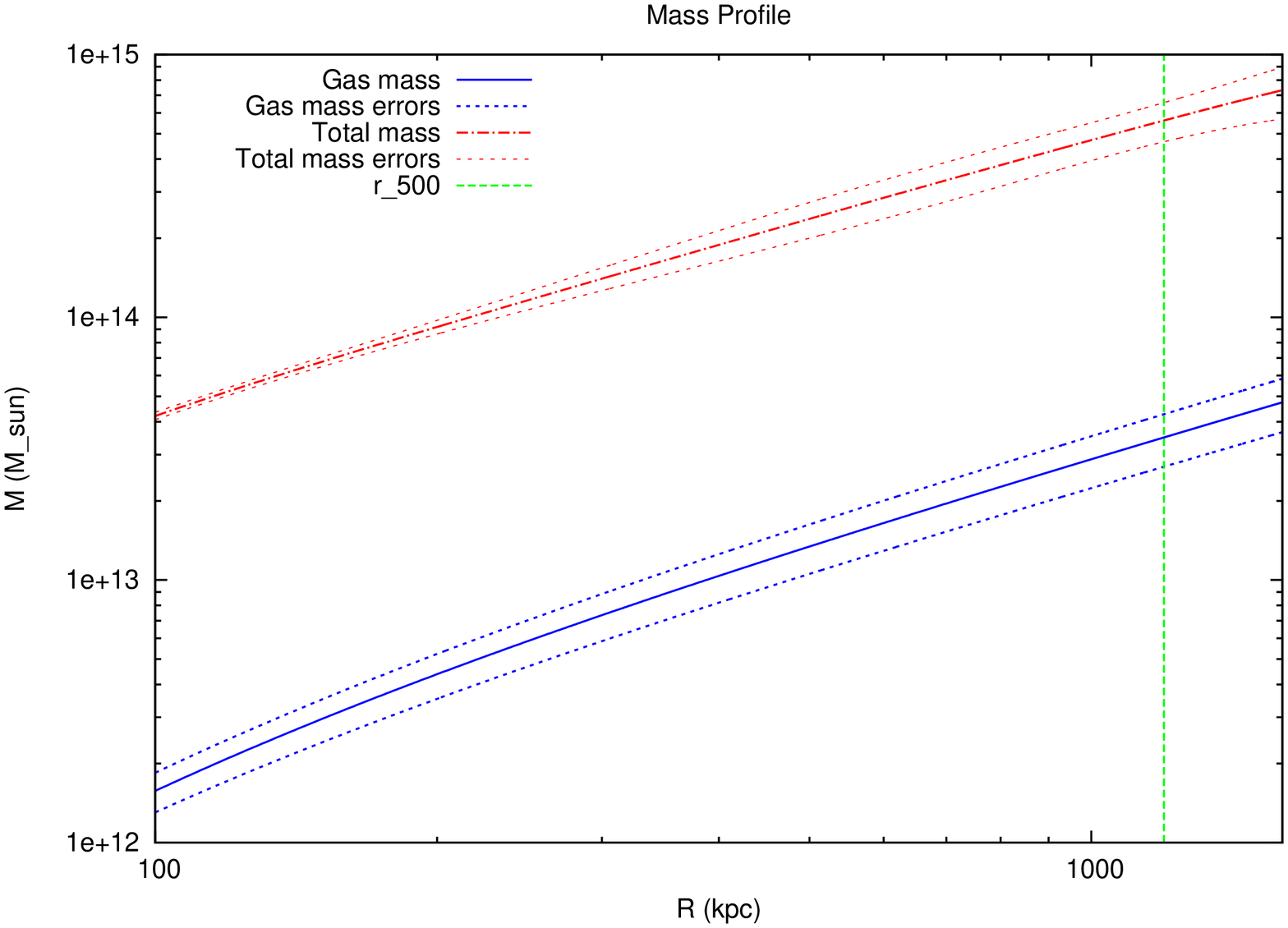}
\includegraphics[width=0.25\textwidth, angle=-90]{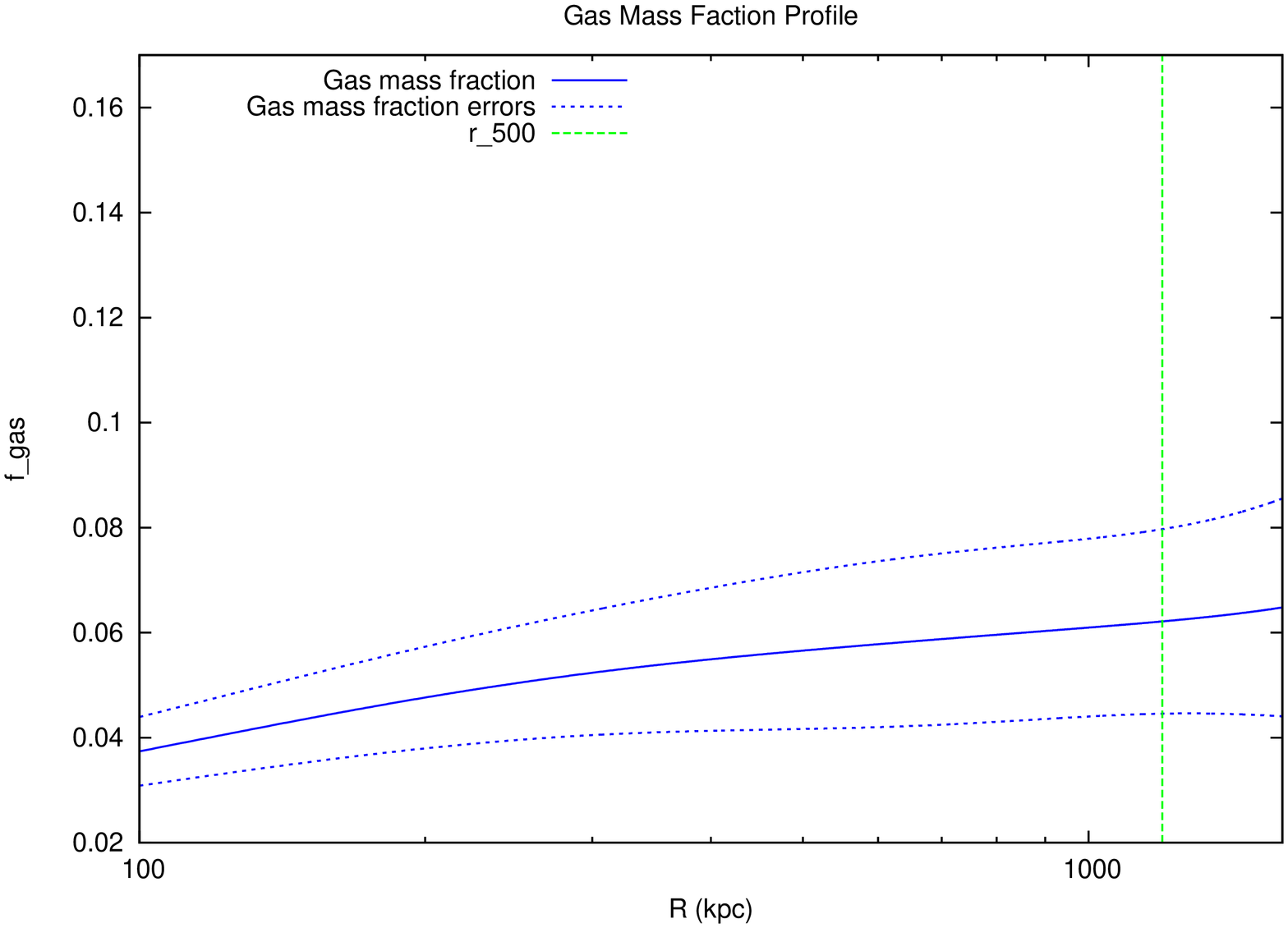}}
\caption{Same as Figure A9, except for \chandra\ baseline background model with a soft band adjustment.
}
\end{figure*}
%%%%%%%%%%%%%%%%%%%%%%%%%%%%%

%%%%%%%%%%%%%%%%%%%%%%%%%%%%%
\begin{figure*}[hbt!]
\centerline{
\includegraphics[width=0.25\textwidth, angle=-90]{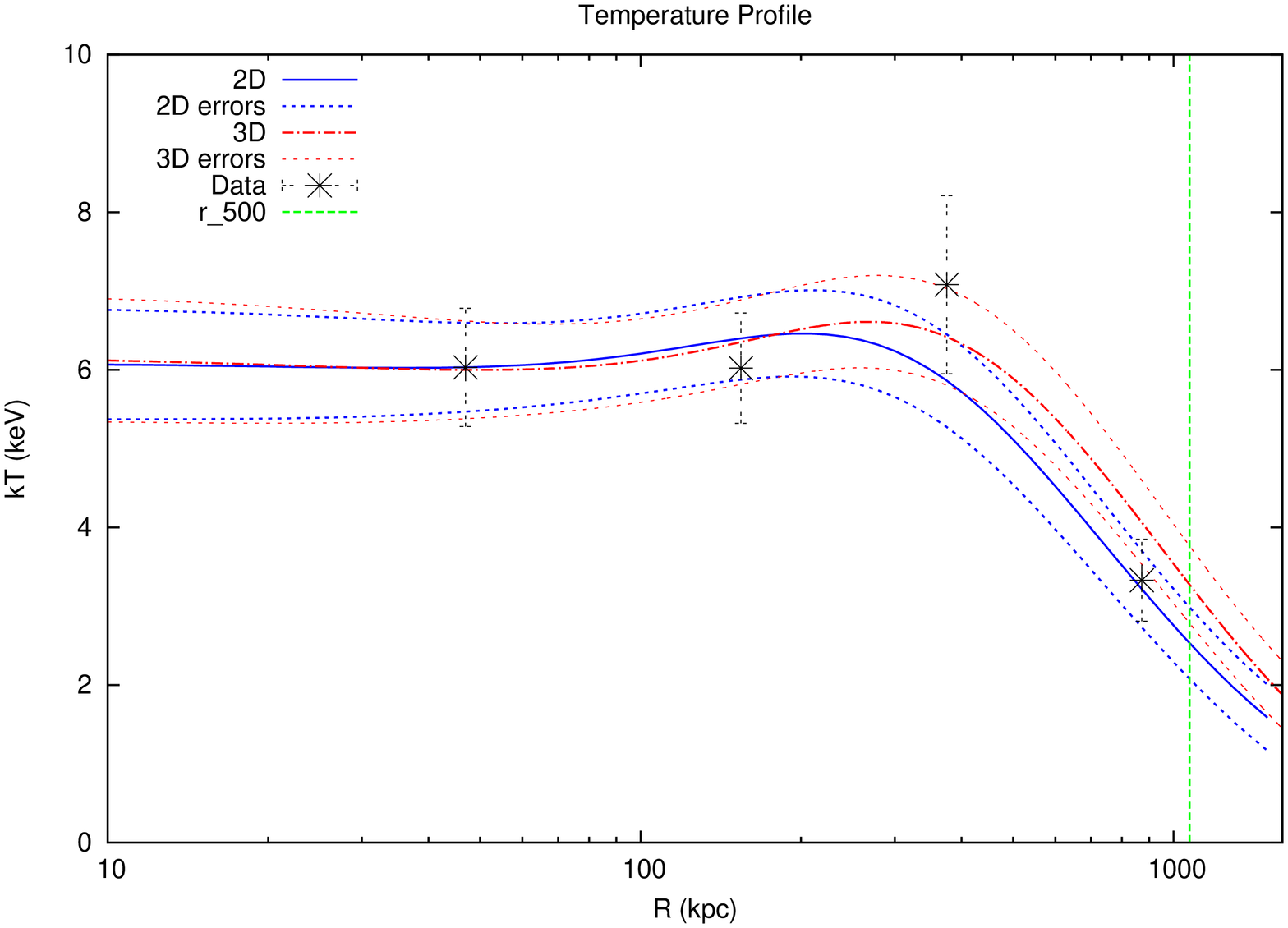}
\includegraphics[width=0.25\textwidth, angle=-90]{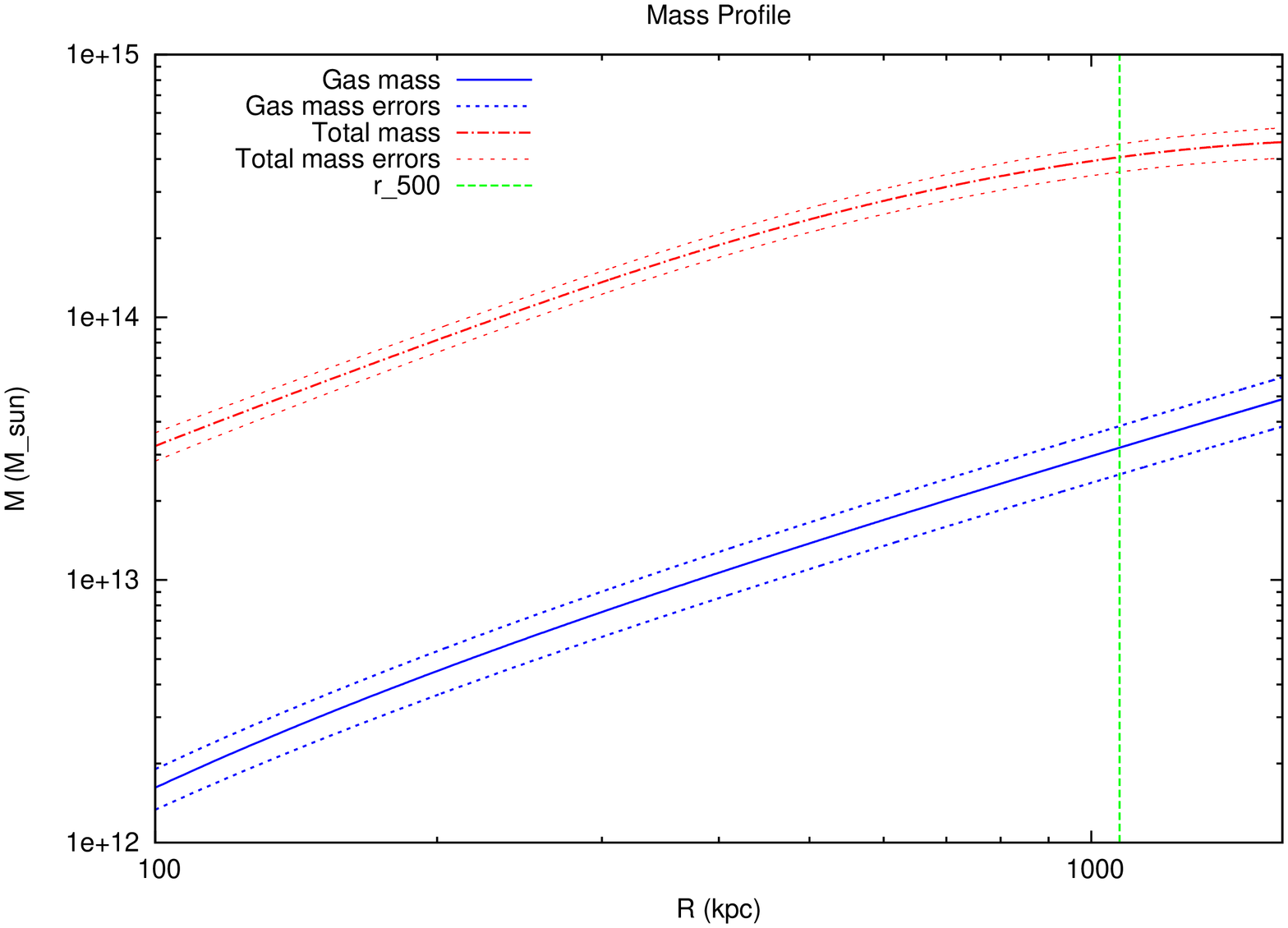}
\includegraphics[width=0.25\textwidth, angle=-90]{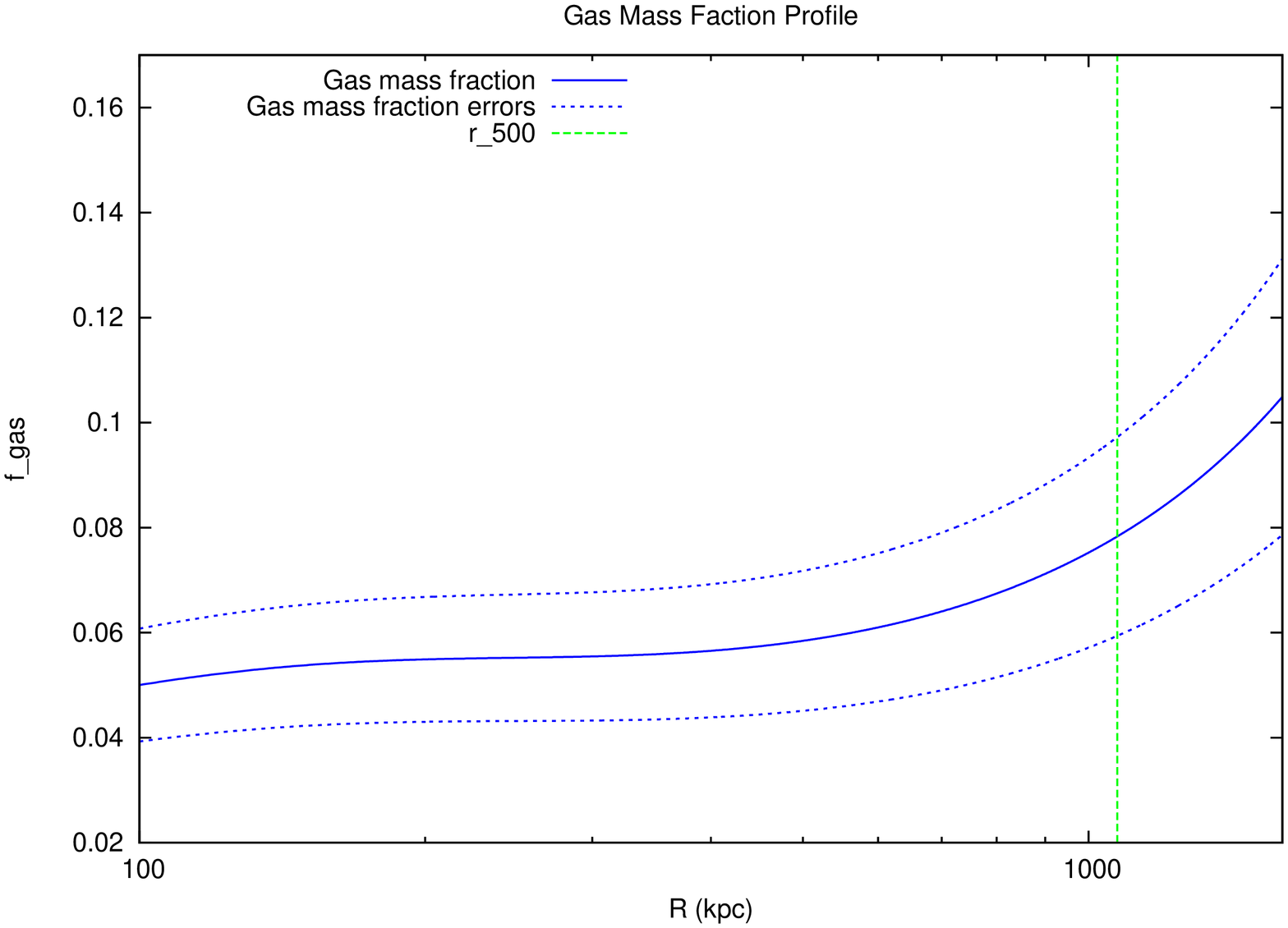}}
\caption{Same as Figure A9, except for \chandra\ baseline background model and $N_H=0.164\times10^{22}$ cm$^{-2}$ (best fitting value).
}
\end{figure*}
%%%%%%%%%%%%%%%%%%%%%%%%%%%%%

%%%%%%%%%%%%%%%%%%%%%%%%%%%%%
\begin{figure*}[hbt!]
\centerline{
\includegraphics[width=0.25\textwidth, angle=-90]{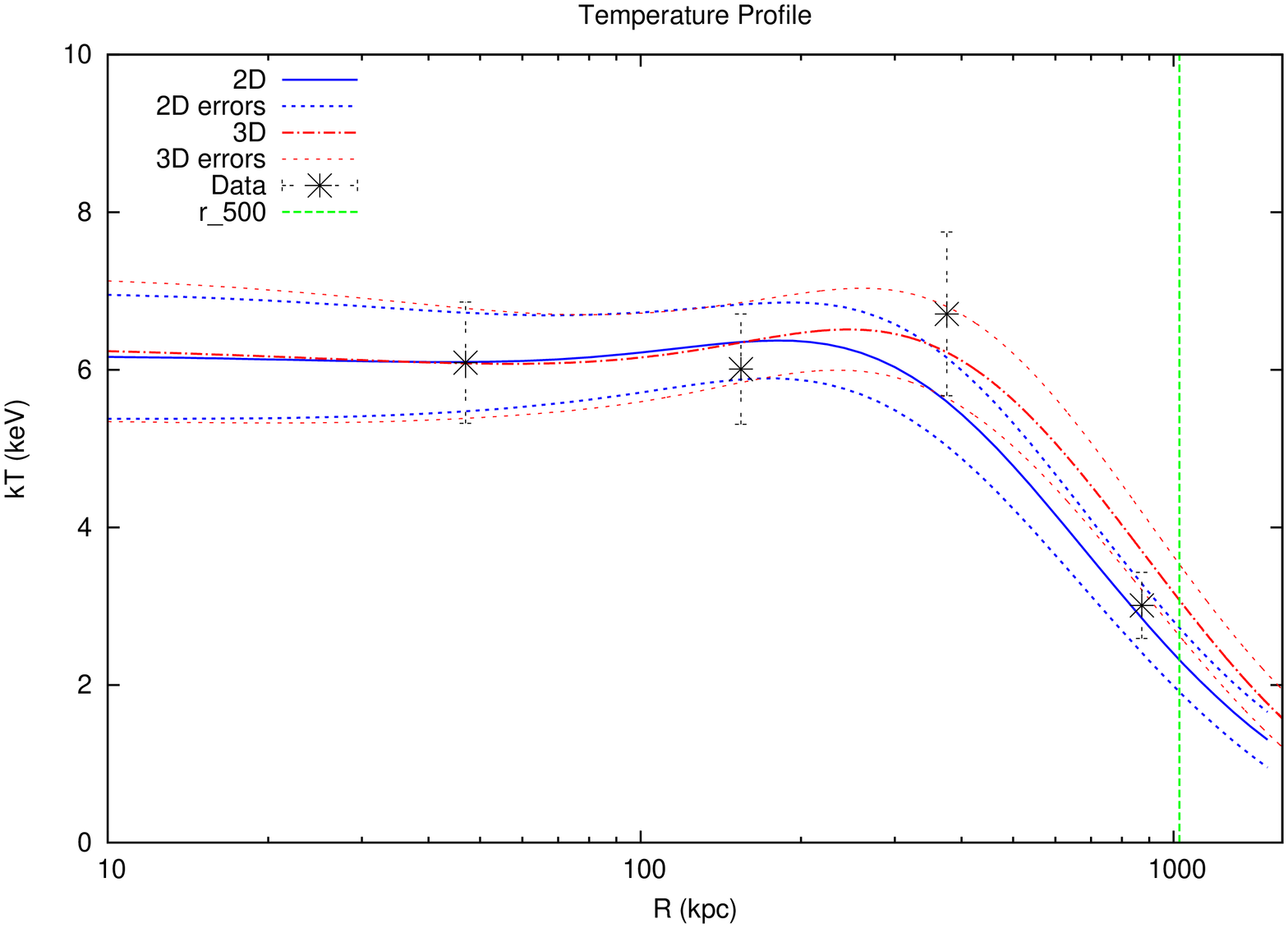}
\includegraphics[width=0.25\textwidth, angle=-90]{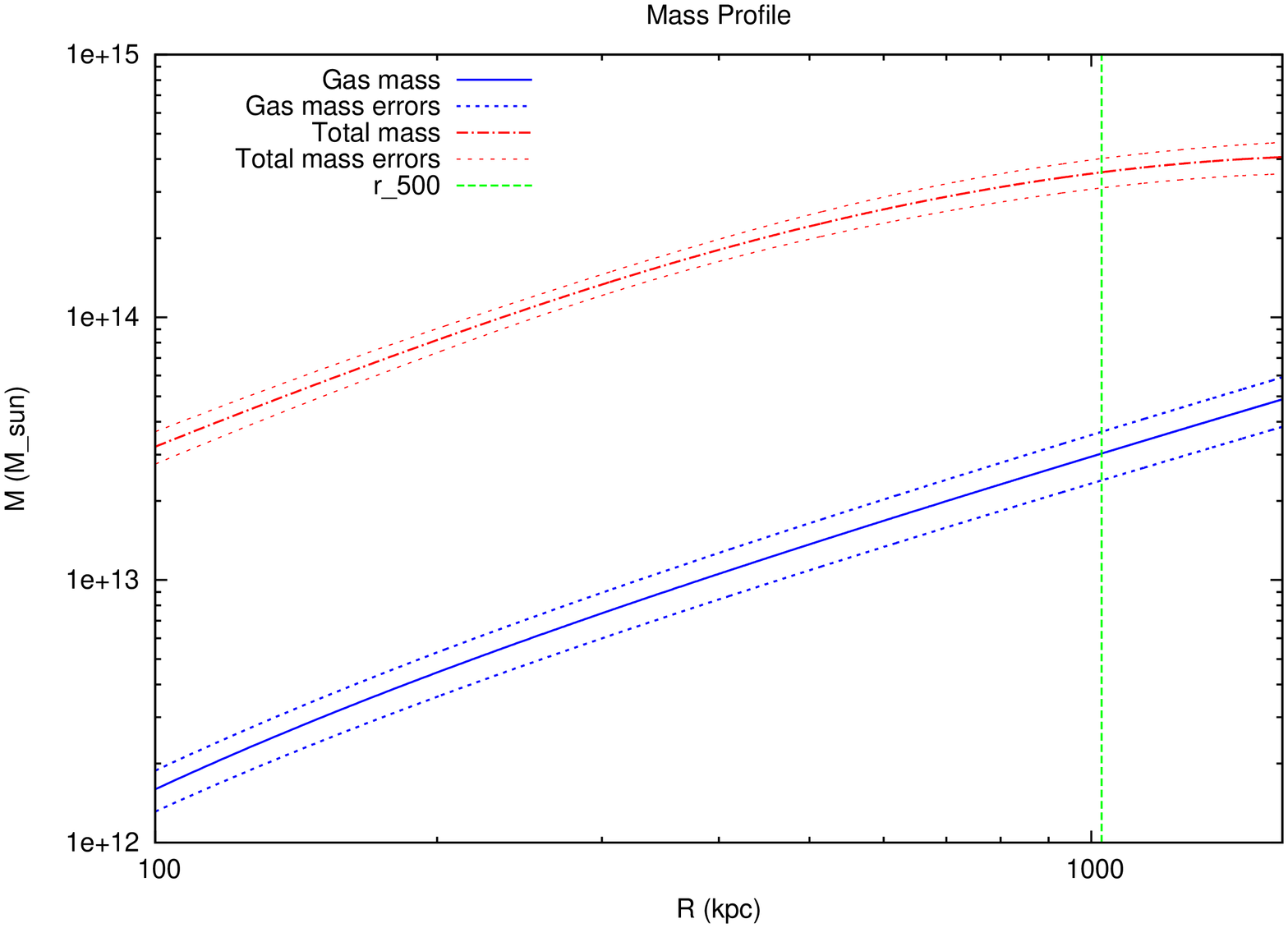}
\includegraphics[width=0.25\textwidth, angle=-90]{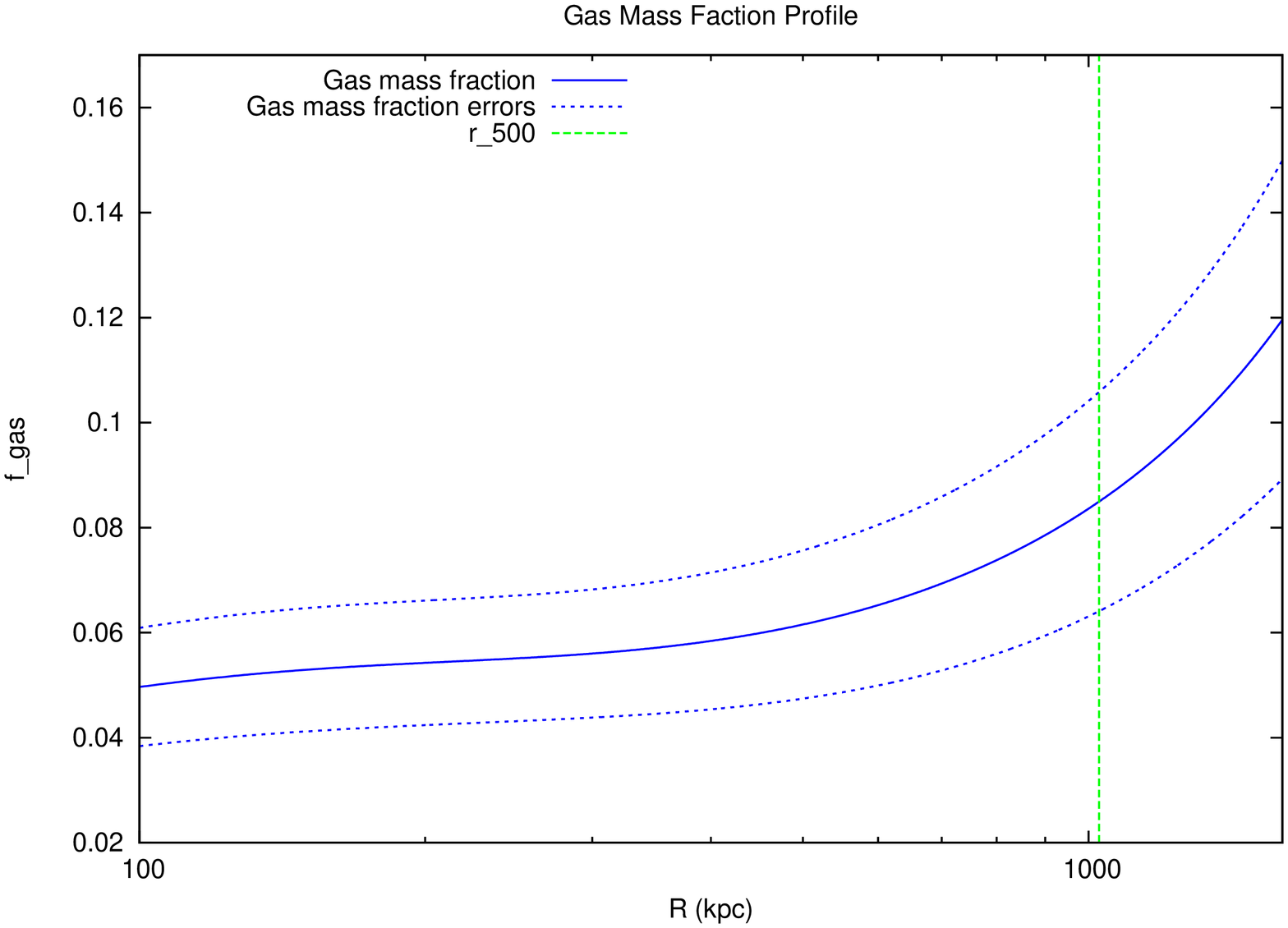}}
\caption{Same as Figure A9, except for \chandra\ baseline background model varying by $+5\%$ and $N_H=0.164\times10^{22}$ cm$^{-2}$ (best fitting value).
}
\end{figure*}
%%%%%%%%%%%%%%%%%%%%%%%%%%%%%

%%%%%%%%%%%%%%%%%%%%%%%%%%%%%
\begin{figure*}[hbt!]
\centerline{
\includegraphics[width=0.25\textwidth, angle=-90]{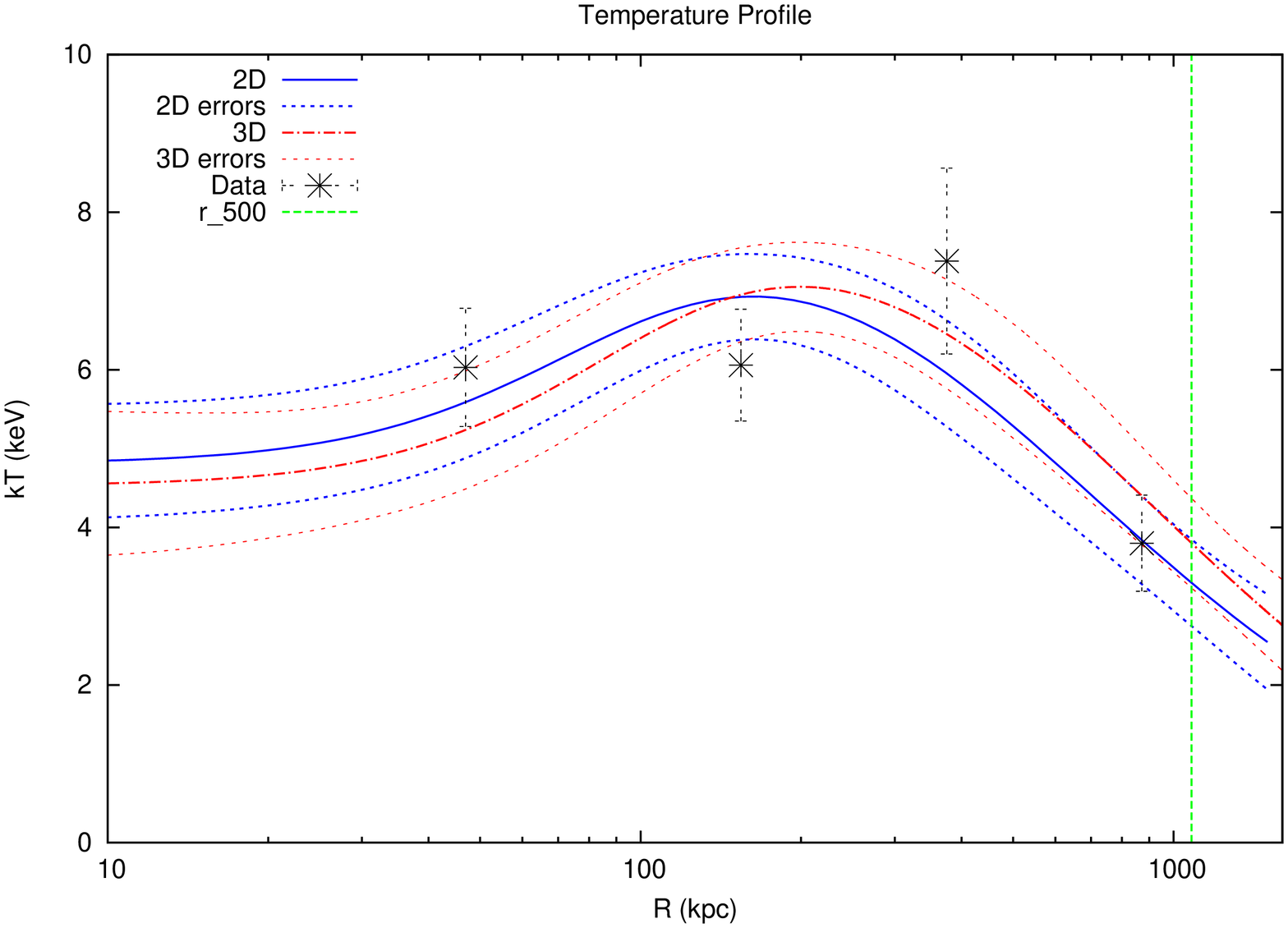}
\includegraphics[width=0.25\textwidth, angle=-90]{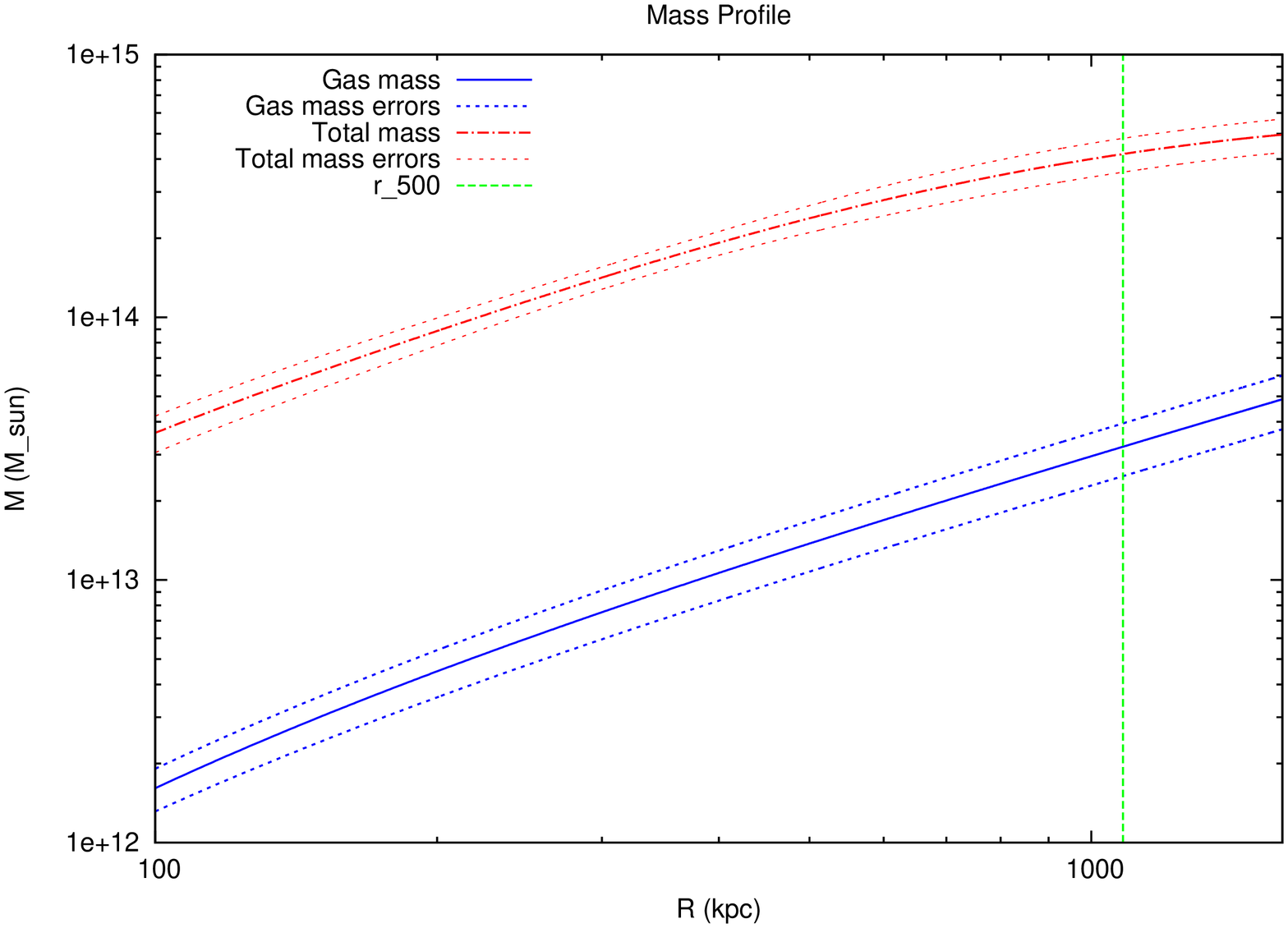}
\includegraphics[width=0.25\textwidth, angle=-90]{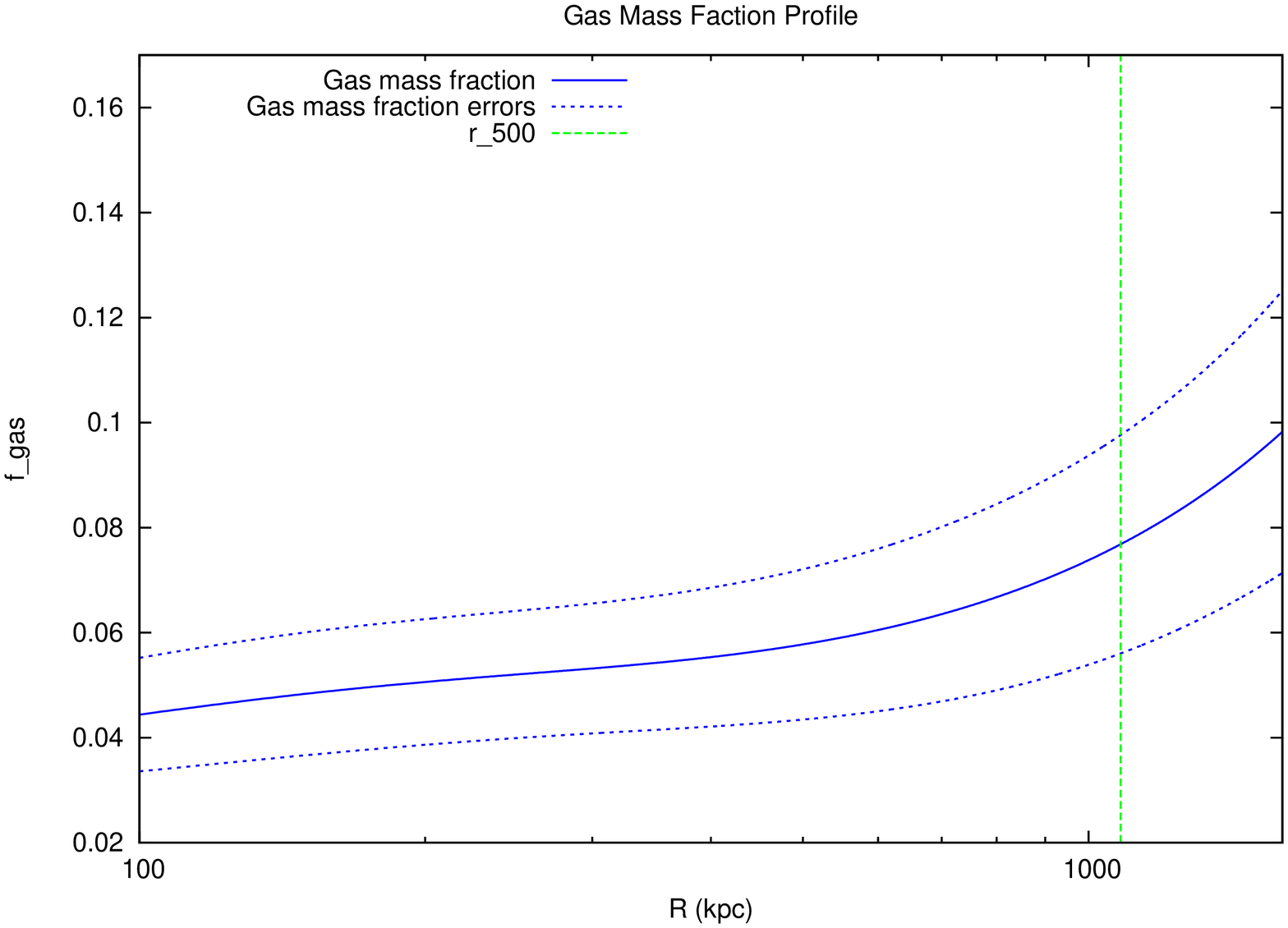}}
\caption{Same as Figure A9, except for \chandra\ baseline background model varying by $-5\%$ and $N_H=0.164\times10^{22}$ cm$^{-2}$ (best fitting value).
}
\end{figure*}
%%%%%%%%%%%%%%%%%%%%%%%%%%%%%

%%%%%%%%%%%%%%%%%%%%%%%%%%%%%
\begin{figure*}[hbt!]
\centerline{
\includegraphics[width=0.25\textwidth, angle=-90]{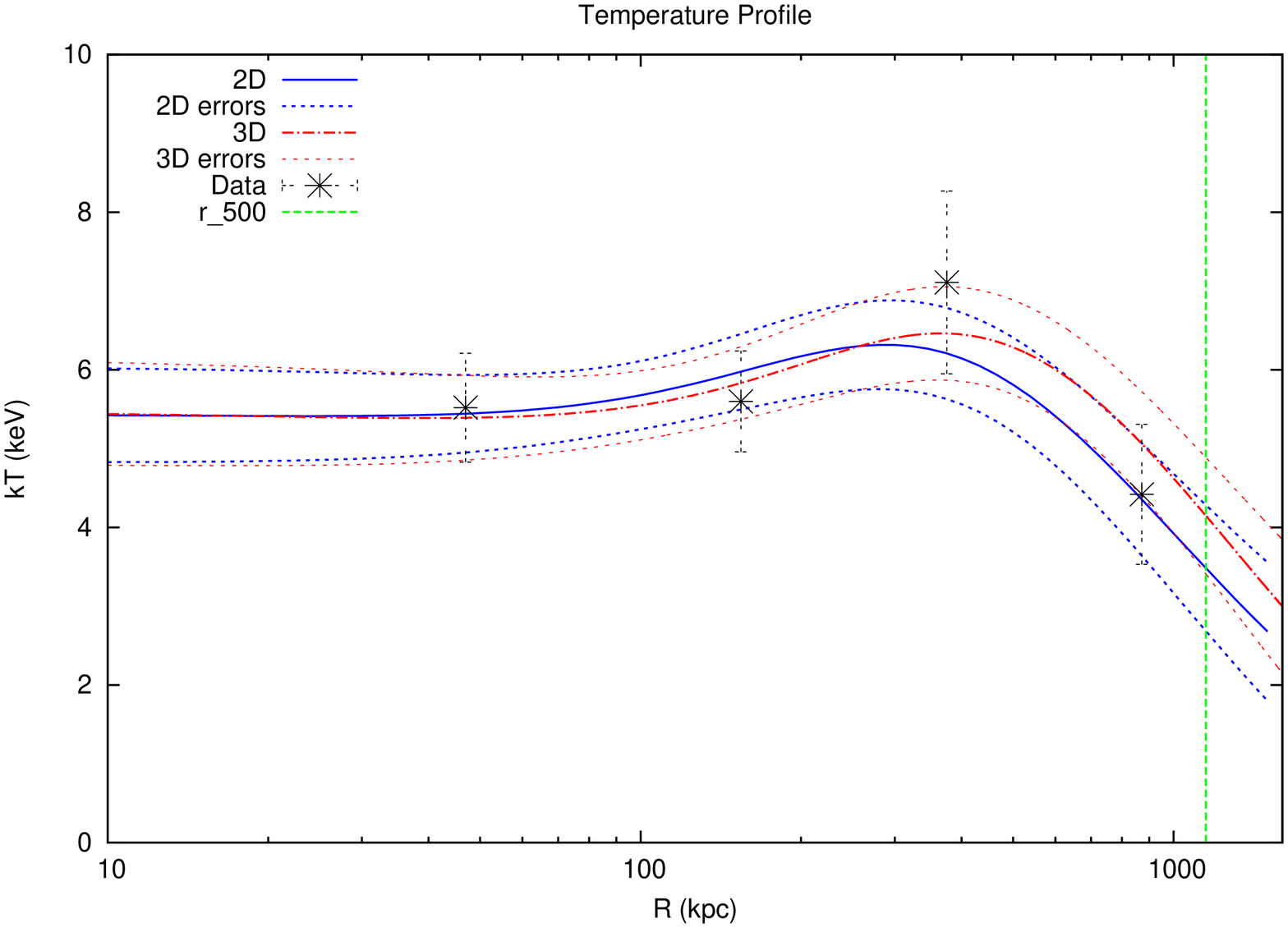}
\includegraphics[width=0.25\textwidth, angle=-90]{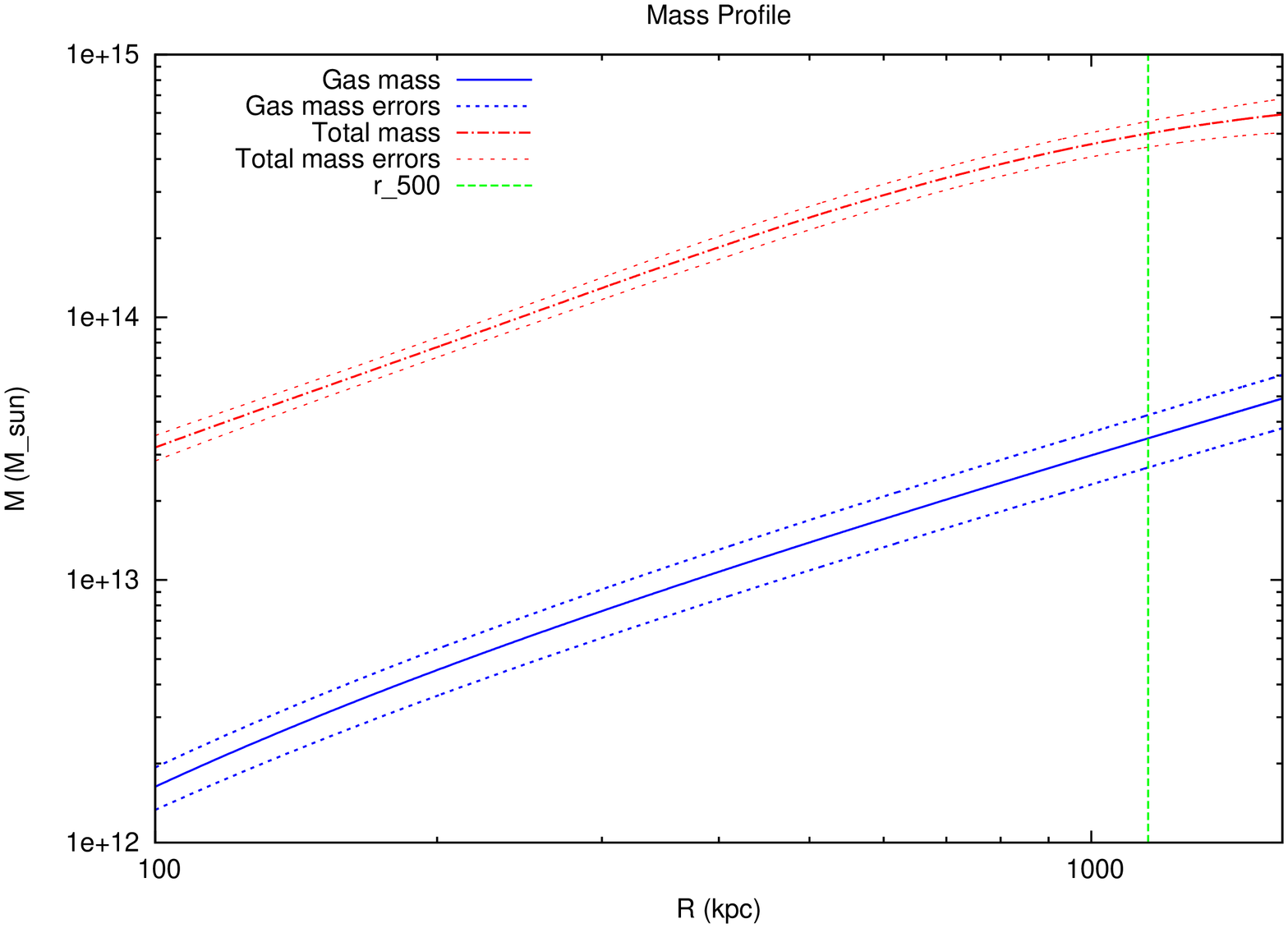}
\includegraphics[width=0.25\textwidth, angle=-90]{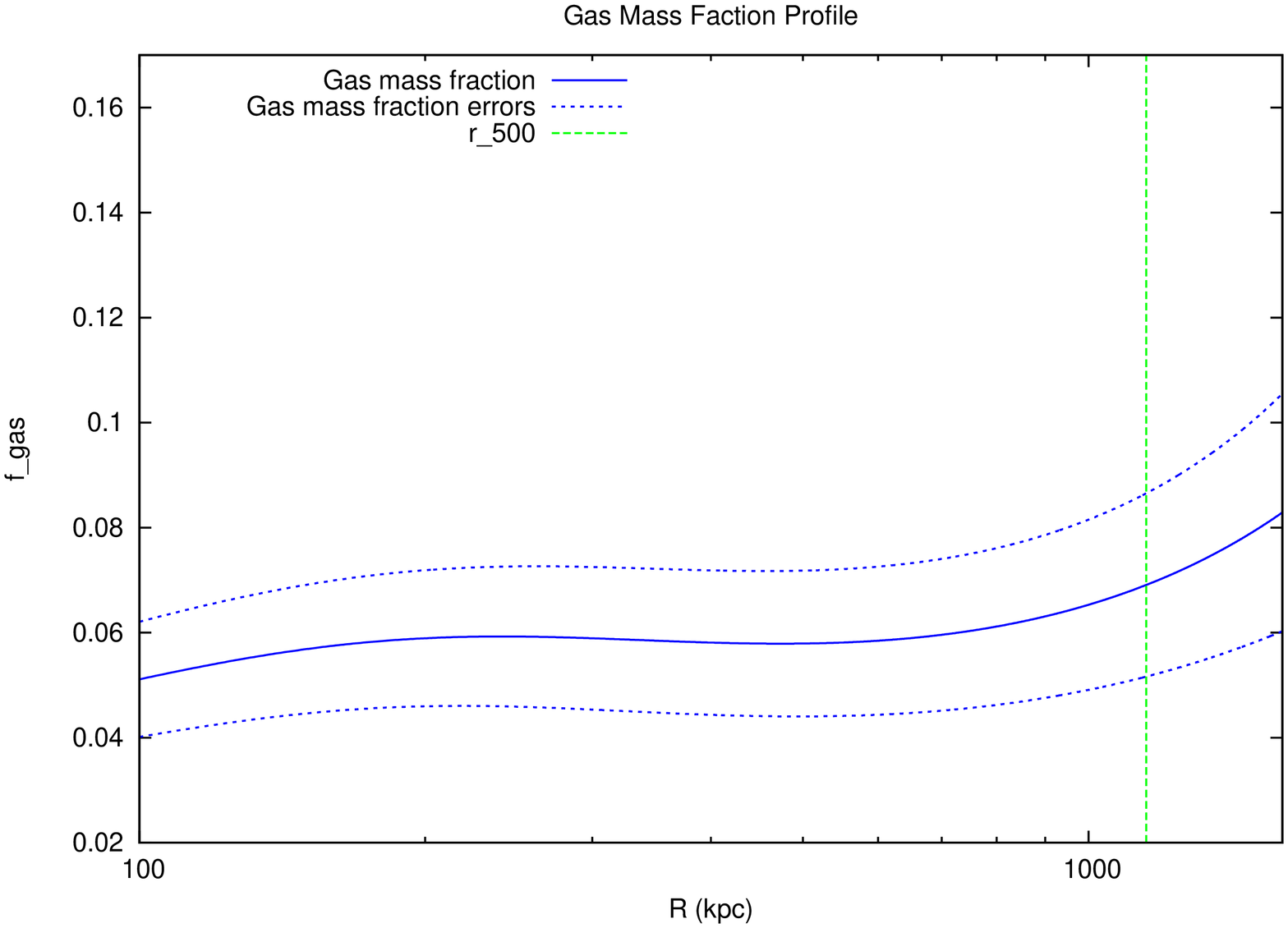}}
\caption{Same as Figure A9, except for \chandra\ baseline background model with a soft band adjustment and $N_H=0.195\times10^{22}$ cm$^{-2}$ (best fitting value).
}
\end{figure*}
%%%%%%%%%%%%%%%%%%%%%%%%%%%%%

%%%%%%%%%%%%%%%%%%%%%%%%%%%%%%%%%%%%%%%%%%%%%%%%%%%%%%%%%%%%%%%%%%%%%%%%%%

\end{document}